\documentclass[12pt,singlespace]{puthesis}

\newcommand{\proquestmode}{}
          
\usepackage[top=1.4in, bottom=1in, left=1in, right=1in]{geometry}

\title{Variational Integration for Ideal Magnetohydrodynamics and Formation of Current Singularities}

\submitted{September 2017}  
\copyrightyear{2017}  
\author{Yao Zhou}
\adviser{Hong Qin, Amitava Bhattacharjee} 
\department{Astrophysical Sciences \linebreak Program in Plasma Physics}

\usepackage{amsfonts}
\usepackage{amsmath}
\usepackage{amssymb}
\usepackage{mathtools}
\usepackage{framed}
\usepackage{bm}
\usepackage{natbib}

\usepackage{siunitx}

\usepackage{graphicx}
\usepackage{caption}

\usepackage{verbatim}

\usepackage{multirow}
\usepackage{longtable}

\usepackage{booktabs}

\usepackage{fancyhdr}
\ifdefined\printmode

\usepackage{url}

\else

\ifdefined\proquestmode

\usepackage{hyperref}
\hypersetup{bookmarksnumbered}

\makeatletter
\hypersetup{pdftitle=\@title,pdfauthor=\@author}
\makeatother

\else

\usepackage{hyperref}
\hypersetup{colorlinks,bookmarksnumbered}

\makeatletter
\hypersetup{pdftitle=\@title,pdfauthor=\@author}
\makeatother

\fi 
\fi 

\ifodd 0
\renewcommand{\maketitlepage}{}

\else

\abstract{
Coronal heating has been a long-standing conundrum in solar physics. Parker's conjecture that spontaneous current singularities lead to nanoflares that heat the corona has been controversial. In ideal magnetohydrodynamics (MHD), can genuine current singularities emerge from a smooth 3D line-tied magnetic field? To numerically resolve this issue, the schemes employed must preserve magnetic topology exactly to avoid artificial reconnection in the presence of (nearly) singular current densities. 

Structure-preserving numerical methods are favorable for mitigating numerical dissipation, and variational integration is a powerful machinery for deriving them. However, successful applications of variational integration to ideal MHD have been scarce. In this thesis, we develop variational integrators for ideal MHD in Lagrangian labeling by discretizing Newcomb's Lagrangian on a moving mesh using discretized exterior calculus. With the built-in  frozen-in equation, the schemes are free of artificial reconnection, hence optimal for studying current singularity formation.

Using this method, we first study a fundamental prototype problem in 2D, the Hahm-Kulsrud-Taylor (HKT) problem. It considers the effect of boundary perturbations on a 2D plasma magnetized by a sheared field, and its linear solution is singular. We find that with increasing resolution, the nonlinear solution converges to one with a current singularity. The same signature of current singularity is also identified in other 2D cases with more complex magnetic topologies, such as the coalescence instability of magnetic islands. 

We then extend the HKT problem to 3D line-tied geometry, which models the solar corona by anchoring the field lines in the boundaries. The effect of such geometry is crucial in the controversy over Parker's conjecture. The linear solution, which is singular in 2D, is found to be smooth. However, with finite amplitude, it can become pathological above a critical system length. The nonlinear solution turns out smooth for short systems. Nonetheless, the scaling of peak current density vs.~system length suggests that the nonlinear solution may become singular at a finite length. With the results in hand, we cannot confirm or rule out this possibility conclusively, since we cannot obtain solutions with system lengths near the extrapolated critical value. 
}

\acknowledgements{
First and foremost, I would like to thank my advisors, Hong Qin and Amitava Bhattacharjee, who made this thesis possible. Hong led me to Princeton, and introduced me to geometric mechanics and structure-preserving numerical methods. When I was stuck with discretizing in Eulerian labeling for months, it was him that reminded me to try Lagrangian labeling instead. Amitava first recognized that my method can be most suitable for studying current singularity formation, and has since supported my project with the utmost enthusiasm and more than enough (computational) resources that I deserve to squander. Above all, both of them have trusted me with complete freedom to pursue my own interests, which can be annoyingly peculiar, at my own pace, which can be embarrassingly slow.

I thank Yi-Min Huang, to whom I turn whenever I run into technical difficulties, for being the most patient mentor and most pleasant collaborator. I have enjoyed working with and learning from him, and I look forward to much more of those. I appreciate the continuous interest in my research and support to my career from Ellen Zweibel. I am grateful to Stuart Hudson for reading my thesis, and Jim Stone for serving on my committee.

Looking back at my six years at Princeton, I must thank Sam Cohen and Bill Tang for tolerating how poorly I did on my first and second year projects. I enjoyed the curriculum and the research atmosphere provided by Nat Fisch and all the faculties of the Program in Plasma Physics. Meanwhile, I appreciate the help from Barbara Sarfaty, Beth Leman, and Jen Jones on all sorts of administrative issues.

Still, it was with my peers that I spent most of my time at Princeton. Chang, you and I are so close that I do not even know what to thank you for. Josh, thanks for being a great math teacher and office mate. Lee and Jono, it was a lot of fun hanging out with you guys, particularly on our trip to China. Daniel, thanks for leaving me some big shoes to fill. Eric, Jon, Jack, and Ben, being in the same class as you guys has been a unique experience. Seth, Mike, Jeff, Eero, Luca, David, Manasvi, and all the fellow students and postdocs in and out of the theory nub, you made the lab a delightful place to work at. Members of the A-Team, it has been a pleasure of mine. Lei, Yuan, Qian, Ge, Hongxuan, Qing, Kenan, Chuanfei, and all my Chinese friends at Princeton, with your company I am rarely homesick.

Overall, my life at Princeton has been so comfortable that I decided to spend a few more years here. For the opportunity to do that, I thank Ilya Dodin. I am very excited about the research endeavors that we are about to take on. In addition, I am indebted to all those who helped me in one way or another during my job hunt, including Hantao Ji, Matt Kunz, Phil Morrison, Michael Kraus, Nick Murphy, Antoine Cerfon, Harold Weitzner, and all those who considered hiring me at some point.

Finally, it is time for family. To my mother, thank you for being an excellent parent in every aspect, which made me what I am today. To my in-laws, your open-mindedness and generosity have made the life of Ningshan and me in the U.S. so much easier. To Ningshan, thank you for supporting me to be who I am and do what I want during and beyond my PhD study; I can only return the favor for the rest of your PhD study, and the many, many years ahead of us.

This work in this dissertation was supported by the U.S.~Department of Energy under Contract No.~DE-AC02-09CH11466, and used resources of the Oak Ridge Leadership Computing Facility at the Oak Ridge National Laboratory, which is supported by the Office of Science of the U.S.~Department of Energy under Contract No.~DE-AC05-00OR22725.

}

\fi  

\begin{document}

\makefrontmatter


\chapter{Introduction\label{ch:intro}}
The work in this thesis consists of two parts: the first is the development of a novel numerical method, variational integration for ideal magnetohydrodynamics (MHD); the second is to apply this method to studying a computationally challenging problem, formation of current singularities. This chapter serves the purpose of briefly surveying the background literatures for these two projects. In Sec.\,\ref{intro:structure}, we identify how our method fits into the lively landscape of structure-preserving numerical methods (for plasma physics). In Sec.\,\ref{intro:formation}, we review what has previously been accomplished on the problem of current singularity formation, in the contexts of both toroidal fusion plasmas and solar coronal heating, and overview what progress is made in this thesis. Readers that are primarily interested in the new results can safely skip this chapter and proceed to the subsequent ones, which are meant to be self-contained.

\section{Structure-preserving numerical methods}\label{intro:structure}

Computation has become widely acknowledged as the third pillar of science. Still, the pursuit for better computational methods has never ceased. A specific motivation in the development of numerical methods for computational physics is to respect the physical properties of the original systems, which are often described in terms of conservation laws. In other words, it is preferable that the accumulation of the errors embedded in the numerical schemes do not violate the conservation laws of the systems. Otherwise, the fidelity of the numerical results is undermined, particularly in long-term simulations. For example, using the standard Runge-Kutta methods to simulate Hamiltonian systems (as simple as a pendulum) often results in coherent drifts in energy, which eventually become significant and make the solutions unphysical and unreliable.

Many have come up with various manipulations to enforce conservation laws for different systems when developing numerical schemes. The caveat with these manipulations is that they are often \textit{ad hoc} and lack generality, and therefore can be troublesome to invent when the system is as complicated as many in plasma physics. In contrast, an alternative approach is to try to preserve the {structures} that underpin the conservation laws. It is difficult to explicitly define what structures exactly mean here, but within the scope of this thesis, the symmetry, geometry, and Hamiltonian structures, etc., of the systems all fall into this category. Numerical methods obtained in this manner often enjoy general applicability to a class of systems that share similar structures.

Simplest examples of structure-preserving numerical methods are probably the St\"ormer-Verlet method and the leapfrog method for solving Hamilton's equations, which had long been used before people realized that both methods are {symplectic}\footnote{A method is symplectic if it preserves the symplectic two-form of the Hamiltonian system, which is $\mathrm{d}p\wedge\mathrm{d}q$ for a canonical Hamiltonian system with canonical coordinates $(q,p)$.}. Symplectic integrators have superior long-term energy behaviors for preserving the symplectic structures of the Hamiltonian systems \citep{Hairer2006}, and hence have become the standard methods in simulating many-body systems such as molecular or celestial dynamics.

A popular and powerful machinery for deriving structure-preserving numerical methods is {variational integration} \citep{Marsden2001}. The idea is to discretize the Lagrangians, rather than equations of motion, of the systems, and then obtain numerical schemes from a discrete variational principle. These schemes can naturally inherit many of the structures and the consequent conservation laws of the original systems. For finite-dimensional systems with non-degenerate Lagrangians, it is proven that variational integrators are not only symplectic, but also momentum-preserving due to a discrete Noether's theorem \citep{Marsden2001}. The St\"ormer-Verlet method, among many other symplectic integrators, can be derived variationally.

However, life is not so easy for plasma physicists, as things get much more complicated as soon as electromagnetic fields are involved. To solve for the trajectory of a charged particle in a prescribed electromagnetic field, which is the simplest problem in plasma physics, the standard methods of symplectic or variational integration usually lead to schemes that are implicit and impractical. The most popular explicit method for this problem has been the Boris method \citep{Boris1970} for its excellent long-term accuracy. As \citet{Qin2013} recently showed, the underlying reason is that the Boris method preserves phase space volume, even though it is not symplectic. The Boris method had subsequently been generalized to volume-preserving methods with higher orders \citep{He2015b,He2016} and for relativistic dynamics \citep{Zhang2015}. Also, explicit symplectic \citep{Zhang2016}, explicit $K$-symplectic \citep{He2017}, and Lorentz covariant canonical symplectic methods \citep{Wang2016} for the problem have been developed.

A reduced model in plasma physics that is often used to approximate charged particle dynamics is the guiding center, which captures the particle trajectory averaged over the fast gyro-motion. The Lagrangian for the guiding center model was derived by \citet{Littlejohn1983}, and later discretized by \citet{Qin2008} to obtain variational integrators. These integrators are shown to be superior to conventional methods in several cases \citep{Qin2009,Li2011}. In addition, \citet{Zhang2014} performed canonicalization on the non-canonical guiding center coordinates, and developed symplectic integrators accordingly.

However, \citet{Ellison2015} later found that these variational integrators for guiding center dynamics sometimes exhibit unphysical oscillations known as parasitic modes. The reason is, the Lagrangian for the guiding center model is a phase space Lagrangian that is {degenerate}\footnote{A Lagrangian $L(q,\dot q)$ is degenerate when the Hessian $\partial^2L/\partial\dot q\partial\dot q$ is degenerate, so that the Legendre transformation is not invertible.}. Applying the standard techniques of variational integration leads to multistep schemes that preserve a symplectic structure on a larger dimensional space than the physical dynamics. \citet{Ellison2016} subsequently developed degenerate variational integrators that preserve the degeneracy of the continuous systems, which are one-step methods and do not have the parasitic modes. However, so far these methods can only be applied to a limited class of problems, and how to treat a degenerate (non-canonical) Lagrangian in general is still an open question.

Plasma physics covers the dynamics of more than just charged particles, but also electromagnetic fields, which is governed by Maxwell's equations. In computational electrodynamics, Yee's finite-difference time-domain (FDTD) method \citep{Yee1966} has proven powerful, robust, and maintained remarkable popularity. One feature of Yee's method is that it treats the electromagnetic fields in a mimetic manner, and hence exactly preserves a discrete Gauss' law, which eliminates artificial magnetic monopoles. Also, Yee's method enjoys good long-term energy behavior.

\citet{Stern2008,Stern2015} recently showed that Yee's scheme can be derived variationally, by discretizing the Lagrangian for Maxwell's equations using discrete exterior calculus \citep[DEC,][]{Hirani2003,Desbrun2005,Desbrun2008}, which is a theory of discrete differential forms. The DEC framework guarantees a discrete Stokes' theorem, which covers Gauss' law. Meanwhile, variational discretization implies that Yee's method is multisymplectic. With this mindset, Yee's method can be generalized to unstructured meshes \citep{Chen2017}, or higher orders using finite element exterior calculus \citep[FEEC,][]{Arnold2006,Arnold2010}.

Coupling the dynamics of charged particles and electromagnetic fields yields the Vlasov-Maxwell equations. One popular way to simulate this system is the particle-in-cell (PIC) method \citep{Birdsall2004}, where the plasma is treated as discrete particles moving on a mesh, on which the electromagnetic fields are solved for. Given how successful the Boris method and Yee's scheme have been in their respective applications, not surprisingly, people often use the former to push the particles and the latter as the field solver. Nonetheless, there are various ways to couple these two methods, which involve interpolation of fields and deposition of charge and current. Some of them may violate conservation laws such as charge conservation, thus producing unphysical artifacts.

Recently, a series of structure-preserving PIC methods have been developed. Low's Lagrangian \citep{Low1958} for the Vlasov-Maxwell system is discretized to obtain variational PIC schemes \citep{Squire2012,Xiao2013}. Alternatively, the Morrison-Marsden-Weinstein Bracket \citep{Marsden1982} for the Vlasov-Maxwell system is discretized in space to derive finite-dimensional Poisson brackets, and then integrated in time using symplectic or Poisson integrators \citep{Xiao2015,Qin2016,He2016b,Kraus2017}. In particular, the Hamiltonian splitting method developed by \citet{He2015}, which builds on \citet{Crouseilles2015,Qin2015}, enables explicit and practical time integration. These methods show excellent long-term energy and momentum behaviors \citep{Xiao2017}, and feature exact charge conservation, etc. As presented in \citet{Kraus2017}, structure-preserving collisionless full PIC is in a close to mature status that is ready to be used in production. The remaining challenges include incorporating collisional effects \citep{Hirvijoki2017,Hirvijoki2017a,Kraus2017a} or gyrokinetic approximations \citep{Burby2015,Burby2017} in a structure-preserving manner. 

Nonetheless, kinetic simulations can be too costly in many circumstances, and reduced fluid modelings become useful in these cases. Since ideal (magneto-) fluids have Hamiltonian structures and many consequent conservation laws \citep{Newcomb1962,Morrison1980,Morrison1998}, naturally it makes sense to develop numerical methods that preserve them. This leads to the original motivation for this thesis, to develop structure-preserving numerical methods, particularly variational integrators, for ideal MHD.

The Lagrangian and Hamiltonian formulations of ideal MHD in Lagrangian and Eulerian labelings are physically equivalent \citep{Newcomb1962}, but the choice between the two labelings can have different consequences computationally. In Eulerian labeling, ideal MHD has a non-canonical Hamiltonian structure \citep{Morrison1980}, and the variational principle is constrained, making the discretization much more difficult. Essentially, it is the theory of Euler-Poincar\'e reduction for continuum mechanics \citep{Holm1998}, of which the ideal MHD variational principle is a special case, that needs to be discretized. Following \citet{Pavlov2011}, \citet{Gawlik2011} employed stochastic matrices to discretize the fluid configuration of incompressible ideal MHD. The caveat with this approach is that a non-holonomic constraint needs to be imposed, in order to remove the non-local interactions in the discrete velocity field. This \textit{ad hoc} constraint is absent in the continuous formulation, and possibly jeopardizes the symplecticity of the variational integrators. 

To properly discretize the Euler-Poincar\'e formulation, one needs a discrete vector field, and then Lie bracket and Lie algebra, constructed in analogy to the differential geometrical description of continuum dynamics. These mathematical tools appear to be lacking at the moment. But if successfully developed, the method should be not only useful for ideal MHD simulations, but also generalizable to other systems with similar Euler-Poincar\'e structure, such as extended MHD \citep{KeramidasCharidakos2014,Burby2017a}, hybrid kinetic-MHD \citep{Tronci2010,Holm2012,Burby2017b}, or many geophysical fluid systems \citep{Holm1998}. In principle, applications to kinetic systems should also be possible by treating the plasma as a fluid in the phase space \citep{Cendra1998,Squire2013}. However, the degeneracy in the Lagrangians of these kinetic models may again give rise to problems.

One way to bypass the difficulty with the constrained variational principle of ideal MHD in Eulerian labeling is to discretize the formal Lagrangian instead, which is constructed from the ideal MHD equations using Lagrange multipliers \citep{Kraus2013,Kraus2016,Kraus2017b}. The resulting variational integrators preserve energy and cross helicity exactly. This approach has also been applied to other systems such as guiding center dynamics, the Vlasov-Poisson system, or even non-variational differential equations \citep{Kraus2013,Kraus2015}. Given that the number of variables is doubled in the system, caution is needed when dealing with the discrete Noether theorem to ensure the corresponding conservation law is the desired physical one. This can be somewhat cumbersome when the system becomes more complicated.

In Lagrangian labeling, ideal MHD has a canonical Hamiltonian structure, and the variational principle is unconstrained \citep{Newcomb1962}. Therefore it is relatively straightforward to discretize the Lagrangian and derive variational integrators, which we shall in Chap.\,\ref{ch:integrator}. We adopt DEC for spatial discretization for the sake of a discrete Stokes' theorem. Variational time discretization guarantees the schemes to be symplectic and momentum-preserving. The advection equations are built into the continuous Lagrangian and equation of motion, and this feature is inherited by our discretization. Physically, this means the discrete advected quantities, such as magnetic flux, are exactly advected by the motion of a finite-dimensional mesh. Without errors from solving the advection equations, the method does not suffer from artifacts such as numerical reconnection when current singularities are present, in contrast to Eulerian ones. This feature makes our method optimal for studying the computationally challenging problem of current singularity formation, which we shall in Chaps.\,\ref{ch:2D} and \ref{ch:3D}.

Despite the excellence in preserving the advected quantities, moving-mesh methods are generally vulnerable to the mesh distortion caused by strong shear flow, which largely limits their domains of application. Particle-based Lagrangian methods do not have such a problem, so it is worth considering smoothed-particle (magneto-) hydrodynamics \citep{Bonet1999,Price2004} with a structure-preserving mindset. Following the aforementioned works on PIC methods, \citet{Xiao2016} discretized the Poisson bracket for the two-fluid system by treating the fluids as smoothed particles and the electromagnetic fields on a fixed mesh. However, this approach cannot be readily applied to ideal MHD, where the magnetic field is passively advected by the fluid motion, rather than governed by the full Maxwell's equations.

In this section, we briefly summarize the recent progress accomplished in the development of structure-preserving numerical methods for plasma physics. For more detailed discussions, we refer to the latest review by \citet{Morrison2017}. We do not purport that structure-preserving numerical methods are panaceas that are superior to all others. All numerical methods have strengths and weaknesses, and so do structure-preserving ones. The diversity of numerical methods in the modern era allows users to make choices based on the specific requirements of the problems to study. We do encourage readers to consider structure-preserving numerical methods when dealing with problems that prioritize conservation properties. The problem of current singularity formation, which we shall introduce in Sec.\,\ref{intro:formation} and study in Chaps.\,\ref{ch:2D} and \ref{ch:3D}, is exactly a problem of this kind.

\section{Formation of current singularities}
\label{intro:formation}
Singularities have always been the most intriguing topics in (mathematical) physics. Black holes are singularities in gravitational fields. The existence and smoothness of solutions to the Navier-Stokes equations remains an unsolved Millennium Prize Problem. In plasma physics, singularities are just as interesting, and have profound real world consequences. \citet{Grad1967} first proposed that 3D MHD equilibria with nested toroidal flux surfaces generally do not exist, due to pathologies that arise at rational surfaces. This theory has greatly impacted the studies of not only intrinsically non-axisymmetric toroidal fusion devices such as stellarators, but also nominally axisymmetric ones like tokamaks, since they can be subject to 3D perturbations \citep{Garabedian1998,Boozer2005,Helander2014}. Nonetheless, one possible way to restore nested flux surfaces is to allow for weak solutions with singular current densities on rational surfaces \citep{Cary1985,Hegna1989,Bhattacharjee1995}. 

Singularities in general are difficult to resolve computationally. Although perturbed ideal equilibrium codes are able to capture the singularities in the linear solutions \citep{Nuhrenberg2003,Park2007}, conventional nonlinear ideal MHD equilibrium codes such as VMEC \citep{Hirshman1983} cannot handle them due to the assumption of smooth functions. It was only recently that the existence of 3D equilibria with current singularities at rational surfaces can be demonstrated numerically \citep{Loizu2015,Loizu2015b}, using the SPEC code that implements the theory of multi-region, relaxed MHD \citep[MRxMHD,][]{Hudson2012}. In MRxMHD, magnetic topology is discretely constrained at a finite number of ideal interfaces, where discontinuities in the tangential magnetic field are allowed. In particular, the class of solutions that \citet{Loizu2015,Loizu2015b} obtained yields rotational transform profiles that are discontinuous at the resonant surfaces.

\citet{Rosenbluth1973} first showed how current singularities can emerge dynamically at resonant surfaces, by obtaining an nonlinear perturbed equilibrium solution with a singularity for the $m=1,n=1$ ideal internal kink instability of a cylindrical plasma. The formation of a current singularity also appears in a reduced 2D problem proposed by Taylor and studied by
\citet[the HKT problem,][]{Hahm1985}, which models the effect of boundary forcing on a resonant surface. \citet{Boozer2010} later applied the techniques of \citet{Rosenbluth1973} to this problem, while incorporating the improvements by \citet{Waelbroeck1989}.

For the sake of simplicity, all these works discussed above considered equilibrium solutions when studying current singularity formation. One challenge with this approach is to ensure that the equilibrium solution preserves the magnetic topology of the initial condition, which is supposed to be smooth. After all, current singularity formation is usually considered under the ideal assumption of infinite conductivity. In fact, the solution with current singularity by \citet{Hahm1985} contains residual islands on both sides of the resonant surface \citep{Dewar2013}, while \citet{Loizu2017} recently showed that the solution in \citet{Rosenbluth1973} yields a discontinuity in the rotational transform profile. That is, strictly speaking, neither of these solutions preserves the magnetic topology of the respective initial conditions. This topological constraint is also difficult to enforce computationally. When studying the nonlinear saturation of the internal kink mode, \citet{Park1980} needed to introduce an artificial magnetic field to effectively relieve the resonance, otherwise numerical reconnection would take place.

Nevertheless, in toroidal fusion plasmas, it is widely acknowledged that rational surfaces, where closed field lines exist, are susceptible to current singularity formation in the ideal limit when subject to resonant perturbations. The story is much more complicated in the solar corona, where field lines are anchored into the boundaries and do not close on themselves, as often modeled with the so-called line-tied geometry. \cite{Parker1972} argued that current singularities would still tend to frequently form in the solar corona, and the subsequent magnetic reconnection events can account for substantial heating. He claimed that this mechanism can explain the anomalously high temperature of the solar corona, i.e., the coronal heating conundrum. This conjecture has remained controversial to this day. In this thesis, we refer to the yes-no question of whether current singularities can emerge in 3D line-tied geometry as \textit{the Parker problem}.

\citet{Parker1972} originally considered a uniform axial magnetic field anchored at both ends by perfectly conducting plates representing the photosphere, which are then subject to random motions that carry the footpoints of the field lines with them. Based on perturbative analysis on the magnetostatic equilibrium equation, he argued that smooth equilibria generally do not exist, unless they are axially symmetric. This theory was followed by several concurrences \citep{Yu1973,Parker1983,Tsinganos1984}, as well as objections \citep{Rosner1982,VanBallegooijen1985,Zweibel1987,Antiochos1987}.

In particular, \citet{VanBallegooijen1985} first emphasized the crucial effect of the line-tied geometry in this problem. Using what is essentially the orderings of reduced MHD \citep[RMHD,][]{Strauss1976}, he showed that a current singularity, if it exists, must penetrate into the line-tied boundaries. This conclusion is ironclad when the RMHD approximations are valid, and a similar one can also be drawn in force-free (zero pressure) MHD. \citet{VanBallegooijen1985} then made a further interpretation that current singularities are only possible when the footpoint motions are discontinuous. Its subsequent variants can be found in \citet{VanBallegooijen1988,Longcope1994b,Cowley1997}. However, we remark that this argument, as well as its variants, does not stand if one considers the effect of a finite shear in the footpoint motions, or equivalently, in the magnetic field. Our counter argument is detailed in Appx.\,\ref{ch:CLS}.

Another key ingredient in the Parker problem is the assumption of perfect conductivity, and its implication is, again, the preservation of magnetic topology. Following \citet{Parker1972}, most of the analytical works on this problem are performed on the MHD equilibrium equation in Eulerian labeling. However, due to the underdetermined nature of this equation, it is not guaranteed that its solutions preserve the initial magnetic topology. That is, simply finding singular solutions to this equation does not necessarily prove Parker's conjecture. On the other hand, some smooth 3D equilibrium solutions purported to contradict Parker's theory \citep{Bogoyavlenskij2000,Bogoyavlenskij2000b} actually do not satisfy this topological constraint.

Recently, in \citet{Low2010,Janse2010}, the significance of the topological constraint was underlined again, while the difficulty in enforcing it was also acknowledged. In order to simplify the enforcement, simple compression or expansion of potential magnetic fields is considered and argued to be sufficient for current singularity formation in \citet{Low2006,Low2007,Janse2009}. This theory encountered objections from \citet{Huang2009,Aly2010,Pontin2012}.

\citet{Zweibel1987} first noted that if one adopts Lagrangian labeling, instead of Eulerian labeling that is used in the studies discussed above, the topological constraint is automatically built into the equilibrium equation and its solutions, so that the Parker problem becomes mathematically explicitly posed. Lagrangian labeling was also employed, at least partially, in \citet{VanBallegooijen1988,Craig2005,Low2010a}. In this thesis, we consider Lagrangian labeling as the more natural and preferable description of the Parker problem. 

In Lagrangian labeling, \citet{Zweibel1987} undertook linear analysis on perturbations to Parker's uniform field, as well as a sheared field. In both cases, they found the linear solutions to be smooth in 3D line-tied geometry. The latter case is especially important since it is a generalization of the HKT problem in 2D, of which the linear solution contains a current singularity \citep{Hahm1985}.  We shall discuss this problem extensively in Chaps.\,\ref{ch:2D} (in 2D) and \ref{ch:3D} (in 3D line-tied geometry). 

At that point, the original theory by \citet{Parker1972} had been demolished, as \citet{Parker2000} acknowledged himself. Nonetheless, \citet{Parker1994} argued that the counter examples involve either only weak deformation from a uniform field or a symmetry degeneracy, appearing to suggest that his conclusion still stands if the perturbations are of sufficient amplitude and complexity \citep{Craig2005}. Unfortunately, the conventional approach of perturbative analysis do not apply to such finite-amplitude perturbations \citep{Rosner1982}.

In addition, deformation with sufficient amplitude and complexity may trigger an ideal instability, which invalidates Parker's original assumption of quasistatic evolution. Meanwhile, instabilities such as internal kink \citep{Rosenbluth1973} or coalescence \citep{Finn1977,Longcope1993} are also believed to produce current singularities. Consequently, these instabilities have been generalized to 3D line-tied geometry \citep{Mikic1990,Craig1990,Longcope1994,Longcope1994b,Huang2006}. \citet{Ng1998} further offered a proof that when a smooth line-tied equilibrium is continuously deformed and becomes unstable, there is no smooth equilibrium for the system to relax to, and current singularities must form.

Overall, the necessity to consider nonlinear effects and instabilities seems to have taken the Parker problem out of the reach of purely analytical studies. Over the years, Eulerian methods with increasing sophistication have been used to computationally study this problem \citep{Mikic1989,Longcope1994,Ng1998,Rappazzo2013}. In general, the tendency towards formation of very localized current structures can be observed. However, due to the inevitable errors in solving the induction (frozen-in) equation, artificial reconnection would take place in the presence of nearly singular current densities, such that topologically constrained equilibrium solutions could not be obtained.

Not surprisingly, when it comes to computationally enforcing the topological constraint in the Parker problem, Lagrangian labeling proves superior to Eulerian as well. In Lagrangian labeling, the frozen-in equation is built into the equilibrium equation, and do not need to be solved. Accordingly, discretizing this formulation leads to numerical schemes that avoid errors in solving the frozen-in equation, and hence do not suffer from artificial reconnection in practice. Intuitively speaking, these schemes use the motion of finite-dimensional meshes to simulate the evolution of infinite-dimensional fluid configurations. 

A Lagrangian relaxation scheme of this kind was developed by \citet{Craig1986} using conventional finite difference for spatial discretization, and has since been extensively used to study the Parker problem \citep{Longbottom1998,Craig2005,Wilmot-Smith2009,Wilmot-Smith2009b,Craig2014}. In this method, the inertia in the momentum equation is entirely replaced by frictional damping, which has been argued to cause unphysical artifacts by \cite{Low2013}. Also, \citet{Pontin2009} showed that charge conservation can be violated. Both of these issues have been fixed in \citet{Candelaresi2014}: the former by retaining the inertia during the frictional relaxation, and the latter with mimetic discretization. These relaxation schemes can also be used as equilibrium solvers for general purposes \citep{Craig1990,Smiet2017}.

A similar numerical method is derived in Chap.\,\ref{ch:integrator}, but in a structure-preserving manner. Specifically, it is obtained by discretizing the ideal MHD Lagrangian in Lagrangian labeling \citep{Newcomb1962} using discrete exterior calculus \citep{Hirani2003,Desbrun2005,Desbrun2008}. Compared with existing ones \citep{Craig1986,Candelaresi2014}, our method exactly preserves many conservation laws including charge conservation. Our discrete force is conservative, which means the equilibrium solution minimizes a discrete potential energy. Constructed on unstructured meshes, the method allows resolution to be devoted to where it is most needed, such as the vicinity of a potential current singularity.

Although adopting Lagrangian labeling automatically enforces the topological constraint, computational challenges still remain in studying the Parker problem. For starters, one needs reliable diagnostics to identify a current singularity. In \citet{Longbottom1998}, within the limits imposed by numerical resolution, no convergence is found for a class of equilibrium solutions. With enhanced numerical capabilities, \citet{Candelaresi2015} revisited these solutions and showed that they actually converge at higher resolutions. The point is, since numerical solutions can at best only resolve a finite length scale, it is extremely difficult, if not impossible, to distinguish between a genuine current singularity and a very thin current layer \citep{Ng1998}.

And our strategy is, instead of using {divergence} (of peak current density, more specifically) as sole evidence for current singularity like \citet{Longbottom1998} did, we show {convergence} of numerical solutions to ones with singularities. In Chap.\,\ref{ch:2D}, we show that the topologically constrained equilibrium solutions to the HKT problem \citep{Hahm1985} and the coalescence instability \citep{Longcope1993} in 2D converges, respectively, everywhere except where there are singularities. We consider these results to be the strongest numerical evidence in the extant literature for current singularity formation in 2D.

In 3D line-tied geometry, a series of ideal relaxation simulations found converged smooth equilibrium solutions in the absence of magnetic nulls \citep{Craig2005,Wilmot-Smith2009,Wilmot-Smith2009b,Pontin2012,Craig2014,Candelaresi2015,Pontin2015,Pontin2016}. Admittedly, the
implication is that current singularity formation in 3D line-tied geometry is not as easy as \citet{Parker2012} insists. In the mean time, these results do not rule out the possibility of current singularity formation in other line-tied cases. That is, the fact that certain prescriptions for constructing current singularities in 3D line-tied geometry do not prove successful does not mean others cannot either.

What we believe is a probable formula for current singularity formation in 3D line-tied geometry is to build on 2D cases that are confirmed to produce current singularities, such as the HKT problem discussed in Chap.\,\ref{ch:2D}. In Chap.\,\ref{ch:3D}, we extend this problem to 3D line-tied geometry. The linear solution, which is singular in 2D, is found to be smooth for all system lengths. However, with finite amplitude, the linear solution can become pathological when the system is sufficiently long. The nonlinear solutions turn out to be smooth for short systems, as the peak current density converges with increasing resolution, in contrast to the divergence in the 2D case. Nonetheless, the scaling of peak current density vs.\,system length suggests that the nonlinear solution may become singular at a finite length. 

As the peak current density increases with system length, so does the shear in the motion near the footpoints. In fact, (nearly) singularly sheared motion is an inherent feature of the Parker problem. However, our numerical method, and moving-mesh methods in general,  is vulnerable to mesh distortion due to strongly sheared motion. Consequently, we are unable to obtain solutions with system length near the extrapolated critical value, and hence cannot confirm or rule out the existence of the suspected finite-length singularity. Further progress towards a conclusive resolution of the problem requires improvement of our numerical method in terms of robustness against strongly sheared motion.

As this thesis concludes, the Parker problem of whether current singularities can form in 3D line-tied geometry remains open. One might argue that practically, a genuine current singularity is indistinguishable from a current layer with a finite thickness that is below the relevant dissipative scale \citep{Longcope1994b}. The counter-argument to that is the tendency towards formation of a true current singularity has fundamental consequences for the reconnection and turbulent dynamics of coronal plasmas \citep{Ng1998}. We believe the Parker problem is still of practical relevance and deserves persistent investigation.


\chapter{Variational integration for ideal MHD\label{ch:integrator}}

In this chapter, variational integrators for ideal magnetohydrodynamics (MHD) are derived by discretizing Newcomb's Lagrangian for ideal MHD in Lagrangian labeling using discrete exterior calculus. Besides being symplectic and momentum-preserving, the schemes inherit built-in advection equations from Newcomb's formulation, and therefore avoid solving them and the accompanying error and dissipation. We use the method to study test cases with current singularities present and show that artificial field line reconnection does not take place. The work in this chapter is largely published in \citet{Zhou2014}.

\section{Introduction}
\label{integrator:intro}

Ideal MHD is a fundamental model in fusion, space, and astro plasma physics. It describes an ideal fluid with mass, entropy, and magnetic flux advected by the motion of the fluid elements. It is a Hamiltonian system with zero dissipation \citep{Newcomb1962,Morrison1980}. \citet{Newcomb1962} first presented the Lagrangian formulation for ideal MHD, in both Eulerian and Lagrangian labelings. In this formulation, only the momentum equation follows from Hamilton's action principle, while the advection equations are either applied as constraints of variation in Eulerian labeling, or built into the Lagrangian in Lagrangian labeling. The theory of Euler-Poincar\'e reduction \citep{Holm1998} is a subsequent generalization of Newcomb's theory.

Ideal MHD is also a nonlinear system complicated enough that numerical simulations are usually needed to study it. However, most, if not all, existing numerical methods for ideal MHD suffer from artificial dissipation, which restricts their domains of applicability. Interestingly, numerical dissipation in ideal MHD simulations is characterized by more than just energy dissipation. A numerical scheme for ideal MHD can be energy conserving \citep{Liu2001} yet dissipative due to error introduced by solving the advection equations, which could lead to artifacts violating, for instance, the frozen-in condition. That is, field lines would typically break and reconnect when current singularities are present in ideal MHD simulations \citep{Gardiner2005}. Such a numerical artifact will be referred to as ``numerical reconnection" and focused on in this chapter, as it is directly observable as an indication for the dissipation introduced by solving the advection equations.

Variational integrators \citep{Marsden2001} are derived from discretized Lagrangians as numerical schemes for simulating Hamiltonian systems. These schemes are favorable for managing numerical dissipation because they often inherit many of the conservation laws of the continuous systems. For finite-dimensional systems with non-degenerate Lagrangians, variational integrators are known to be symplectic and momentum-preserving \citep{Marsden2001}. For non-canonical systems like guiding-centers \citep{Qin2008,Qin2009,Li2011,Kraus2013,Ellison2016} and magnetic field lines \citep{Ellison2016}, or infinite-dimensional systems such as electrodynamics \citep{Stern2008,Stern2015}, Vlasov-Maxwell system \citep{Squire2012,Xiao2013}, and incompressible fluids \citep{Pavlov2011}, it becomes more challenging to discretize the Lagrangians while preserving the desired conservation laws. Readers interested in more detailed discussions of these methods are referred to Sec.~\ref{intro:structure}.

Nevertheless, there have been efforts to develop variational integrators for ideal MHD that respect its conservation laws \citep{Gawlik2011,Kraus2013,Kraus2016,Kraus2017b}. In \citet{Kraus2013,Kraus2016,Kraus2017b}, formal Lagrangians are used instead of the physical Lagrangians. In contrast to Newcomb's formulation, the frozen-in equation follows from the Euler-Lagrange equations associated with the formal Lagrangians. When current singularities are present, the method still suffers from numerical reconnection, despite conserving energy exactly \citep{Kraus2013,Kraus2017b}. In \citet{Gawlik2011}, discrete volume-preserving diffeomorphism groups are constructed on isolated cells such that non-local interactions between any two cells are included. As a result, non-holonomic constraints are required to localize the discrete velocity fields \citep{Pavlov2011}. However, it is not clear whether the method remains symplectic with these \textit{ad hoc} constraints. In addition, such a discretization cannot be readily applied to compressible MHD. 

Considering the imperfections of the previous attempts, we choose to develop variational integrators for ideal MHD along a different path. One key guideline is to treat the equation of motion and the advection equations differently, respecting the nature of ideal MHD. Another is to construct the discrete Lagrangian on a discrete manifold that resembles a continuous one more closely than just isolated cells. Therefore, we choose to discretize Newcomb's physical Lagrangian using discrete exterior calculus \citep[DEC,][]{Hirani2003,Desbrun2005,Desbrun2008}. 

In this chapter, we will perform this exercise, and then assess the viability of the resulting variational integrators. The discretization will be carried out on the ideal MHD Lagrangian in Lagrangian labeling. First, the Lagrangian will be spatially discretized using DEC, with built-in advection equations inherited from Newcomb's formulation. By so doing, we will be using the motion of a finite-dimensional mesh to simulate the infinite-dimensional MHD. The spatially discretized Lagrangian will have a many-body form, which we will straightforwardly discretize in time and obtain variational integrators \citep{Marsden2001} to numerically solve for the motion of the mesh. We will show with numerical examples that the schemes can handle current singularities without suffering from numerical reconnection. 

The proposed method brings about the first variational symplectic integrators for ideal compressible MHD. The two steps in discretization account for the two highlights of the integrators, respectively. Variational temporal discretization makes the schemes symplectic and momentum-preserving. By building in the advection equations, we avoid solving them and the accompanying errors. Therefore, the method is very effective in problems where the preservation of the advection equations is of dominant importance, outperforming any existing Eulerian methods in terms of mitigating numerical reconnection. This suggests that it may be best suited for studying current singularity formation in ideal MHD \citep{Parker1994}, as we will do in Chaps.\,\ref{ch:2D} and \ref{ch:3D} .

This chapter is organized as follows. First, Newcomb's Lagrangian formulation for ideal MHD in both Lagrangian and Eulerian labeling is reviewed in Sec.\,\ref{integrator:Lagrangian}. Next, we introduce DEC to spatially discretize Newcomb's Lagrangian, and derive the variational integrators in Sec.\,\ref{integrator:discretization}. Then we implement the method in 2D and show numerical results that artificial reconnection does not take place with current singularities present in Sec.\,\ref{integrator:numerical}. In the end, the strengths and weaknesses of the method will be summarized and discussed in Sec.\,\ref{integrator:summary}.

\section{Lagrangian formulation for ideal MHD}
\label{integrator:Lagrangian}
The equations of ideal MHD include, respectively, the continuity, adiabatic, and induction (frozen-in) equations,
\begin{subequations}\label{advectionE}
\begin{align}
\partial_t\rho&=-\nabla\cdot(\rho\mathbf{v}),\label{massE}\\
\partial_t(p/\rho^\gamma) &=-\mathbf{v}\cdot\nabla (p/\rho^\gamma),\label{adiabaticE}\\
\partial_t\mathbf{B}&=\nabla\times(\mathbf{v}\times\mathbf{B}).\label{fronzeninE}
\end{align}
\end{subequations}
Here $\mathbf{v},\rho,p,\mathbf{B}$, $\gamma$ are the fluid velocity, mass density, pressure, magnetic field, and adiabatic index, respectively. In this thesis, we categorize these equations as the advection equations, since they can all be cast into the following forms of exact advection,
\begin{subequations}\label{advectionET}
\begin{align}
\mathrm{d}(\rho\,\mathrm{d}^3{x})/\mathrm{d}t&=0,\label{massET}\\
\mathrm{d}(p/\rho^\gamma)/\mathrm{d}t&=0,\label{adiabaticET}\\
\mathrm{d}(\mathbf{B}\cdot\mathrm{d}\mathbf{S})/\mathrm{d}t&=0,\label{fronzeninET}
\end{align}
\end{subequations}
 where $\mathrm{d}^3{x}$ and $\mathrm{d}\mathbf{S}$ are arbitrary volume and area elements, respectively. That is, Eqs.\,(\ref{advectionE}), or equivalently Eqs.\,(\ref{advectionET}), portray the exact advection of mass, entropy, and magnetic flux, respectively, in ideal MHD. 
 
The system of ideal MHD is closed by including the momentum equation,
\begin{equation}
\rho\partial_t\mathbf{v}+\rho\mathbf{v}\cdot\nabla\mathbf{v}=(\nabla\times\mathbf{B})\times\mathbf{B}-\nabla p.\label{momentumE}
\end{equation}
In contrast to Eqs.\,(\ref{advectionE}), this equation cannot be expressed in the form of exact advection akin to Eqs.\,(\ref{advectionET}).

In order to develop variational integrators for ideal MHD, we need a Lagrangian first. One can easily write one down, by subtracting the potential (internal and magnetic) energy from the kinetic energy,
\begin{equation}
L[\mathbf{v},\rho,p,\mathbf{B}]=\int\left(\frac{1}{2}\rho{v}^2-\frac{p}{\gamma-1}-\frac{1}{2}{B}^2\right)\mathrm{d}^3x.\label{LagrangianE}
\end{equation}
Here the Lagrangian $L$ is a functional of $\mathbf{v},\rho,p,\mathbf{B}$. However, one then finds that naively applying Hamilton's action principle to this Lagrangian, allowing $\mathbf{v},\rho,p,\mathbf{B}$ to vary freely, does not lead to the expected ideal MHD equations (\ref{advectionE}) and (\ref{momentumE}). 

\citet{Newcomb1962} first presented the proper Lagrangian formulation for ideal MHD, by having the advection equations (\ref{advectionE}) either built into the Lagrangian in Lagrangian labeling, or applied as constraints of variation in Eulerian labeling. In either case, only the momentum equation follows from Hamilton's action principle as the equation of motion. Here, we first review Newcomb's Lagrangian formulation for ideal MHD in Lagrangian labeling, which lays the foundation of the work in this thesis. Then the equivalent Lagrangian formulation in Eulerian labeling is presented as well. Although it does not relate directly to the work in this thesis, it serves the purposes of completing the context and inspiring future work.

The ideal MHD equations (\ref{advectionE}) and (\ref{momentumE}), as well as the Lagrangian (\ref{LagrangianE}), are expressed in Eulerian labeling $(\mathbf{x},t)$, where each fluid element is labeled by its {current} position $\mathbf{x}$ at time $t$. In contrast, in Lagrangian labeling $(\mathbf{x}_0,t)$, a fluid element is labeled by its {original} position $\mathbf{x}_0$ at time $t=0$. The two labelings are connected by the particle relabeling map
 $\mathbf{x}(\mathbf{x}_0,t)$, which stands for the position at time $t$ of a fluid element that is located at $\mathbf{x}_0$ at $t=0$. The mapping $\mathbf{x}(\mathbf{x}_0,t)$ is sometimes referred to as the fluid configuration as well.
Utilizing this mapping, one may integrate the advection equations (\ref{advectionET}) and obtain
\begin{subequations}\label{advection}
\begin{align}
\rho\,\mathrm{d}^3{x}&=\rho_0\,\mathrm{d}^3x_0,\label{continuity}\\
p/\rho^\gamma&=p_0/\rho_0^\gamma,\label{adiabatic}\\
B_i\,\mathrm{d}S_i&=B_{0i}\,\mathrm{d}S_{0i},\label{frozenin}
\end{align}
\end{subequations}
with $\rho_0=\rho(\mathbf{x}_0,0)$, $p_0=p(\mathbf{x}_0,0)$, and $\mathbf{B}_0=\mathbf{B}(\mathbf{x}_0,0)$. Physically, they indicate that the mass enclosed in a moving volume element $\mathrm{d}^3{x}$, the specific entropy [$s=\ln(p/\rho^\gamma)$] at a moving point $\mathbf{x}(\mathbf{x}_0,t)$, and the magnetic flux through a moving area element $\mathrm{d}\mathbf{S}$ do not change as the fluid configuration evolves. Then, $\rho, p, \mathbf{B}$ can be expressed in Lagrangian labeling,
\begin{subequations}\label{advectionJ}
\begin{align}
 \rho&=\rho_0/J,\label{continuityJ}\\
 p&=p_0/J^\gamma,\label{adiabaticJ}\\
 B_i&=x_{ij}B_{0j}/J.\label{frozeninJ}
\end{align}
\end{subequations}
Here $x_{ij}=\partial x_i/\partial x_{0j}$ is the element of the deformation matrix, and the Jacobian $J=\det(x_{ij})$. With these equations and $\mathbf{v}(\mathbf{x},t)=\dot{\mathbf{x}}(\mathbf{x}_0,t)$ substituted into Eq.\,(\ref{LagrangianE}), we can obtain the Lagrangian in Lagrangian labeling,
\begin{align}
{L}[\mathbf{x}]=\int\left[ \frac{1}{2}\rho_0\dot{{x}}^2-\frac{p_0}{(\gamma-1)J^{\gamma-1}}-\frac{x_{ij}x_{ik}{B}_{0j}B_{0k}}{2J}\right]\mathrm{d}^3x_0.\label{Lagrangian3}
\end{align}
This Lagrangian has the form of a standard field Lagrangian, as it is a functional of a field $\mathbf{x}(\mathbf{x}_0,t)$ with dependence on its derivatives, $\dot{\mathbf{x}}$ and $x_{ij}$.
The Euler-Lagrange equation that follows from Hamilton's action principle is the momentum equation for ideal MHD in Lagrangian labeling,
\begin{align}
\rho_0\ddot{x}_i-B_{0j}\frac{\partial}{\partial x_{0j}}\left(\frac{x_{ik}B_{0k}}{J}\right)+\frac{\partial J}{\partial x_{ij}}\frac{\partial }{\partial x_{0j}}\left(\frac{p_0}{J^\gamma}+\frac{x_{kl}x_{km}B_{0l}B_{0m}}{2J^2}\right)=0,\label{momentum3}
\end{align}
which is the one and only ideal MHD equation in Lagrangian labeling. This formulation can easily be projected into 2D and 1D. It also has a few interesting variations. For instance, with the internal energy term dropped and an extra volume-preserving constraint $J=1$ added via a Lagrange multiplier, we have a Lagrangian for incompressible MHD. If one further sets $x_3=x_{03}$ and $B_{03}=1$, a Lagrangian for reduced MHD \citep{Strauss1976} is obtained.

What is learned from deriving the variational principle in Lagrangian labeling is then used by \citet{Newcomb1962} to obtain the equivalent formulation in Eulerian labeling. That is, the advection equations (\ref{advectionE}) are not equations of motion, but prescribed information that constrains the dynamics of the system. In other words, the variations of $\rho$, $p$, and $\mathbf{B}$ are not arbitrary, but subject to the constraints that the advection of mass, entropy, and magnetic flux cannot be violated.

Suppose the system depends on a virtual time-like parameter $\epsilon$, and take the derivative of Eqs.\,(\ref{advection}) with regard to $\epsilon$, one has
\begin{subequations}\label{advectionVT}
\begin{align}
\mathrm{d}(\rho\,\mathrm{d}^3{x})/\mathrm{d}\epsilon=0&,\label{massVT}\\
 \mathrm{d}(p/\rho^\gamma)/\mathrm{d}\epsilon=0&,\label{adiabaticVT}\\
\mathrm{d}(\mathbf{B}\cdot\mathrm{d}\mathbf{S})/\mathrm{d}\epsilon=0&.\label{fronzeninVT}
\end{align}
\end{subequations}
Similar to the equivalency between Eqs.\,(\ref{advectionE}) and Eqs.\,(\ref{advectionET}), these equations can be cast into the following forms,
\begin{subequations}\label{advectionV}
\begin{align}
\delta\rho&=-\nabla\cdot(\rho\bm{\xi}),\label{massV}\\
\delta (p/\rho^\gamma) &=-\bm{\xi}\cdot\nabla (p/\rho^\gamma),\label{adiabaticV}\\
\delta\mathbf{B}&=\nabla\times(\bm{\xi}\times\mathbf{B}).\label{fronzeninV}
\end{align}
\end{subequations}
Here the virtual derivative $\partial_\epsilon$ is replaced by the more commonly used notation of variation, $\delta$, with the displacement $\bm{\xi}(\mathbf{x},t)=\delta\mathbf{x}(\mathbf{x}_0,t)$. In the mean time, taking the total time derivative of the displacement yields $\partial_t\bm{\xi}+\mathbf{v}\cdot\nabla\bm{\xi}=\delta\dot{\mathbf{x}}$, while taking the variation of the velocity $\mathbf{v}(\mathbf{x},t)=\dot{\mathbf{x}}(\mathbf{x}_0,t)$ leads to $\delta\mathbf{v}+\bm{\xi}\cdot\nabla\mathbf{v}=\delta\dot{\mathbf{x}}$. So we end up with
\begin{align}
\delta\mathbf{v}=\partial_t\bm{\xi}+\mathbf{v}\cdot\nabla\bm{\xi}-\bm{\xi}\cdot\nabla\mathbf{v}.\label{velocityV}
\end{align}
\citet{Newcomb1962} then applied Eqs.\,(\ref{advectionV}) and (\ref{velocityV}) as constraints when applying Hamilton's action principle on the Lagrangian (\ref{LagrangianE}), and obtained the momentum equation (\ref{momentumE}). This is the Lagrangian formulation for ideal MHD in Eulerian labeling. This theory can also be used for stability analysis of ideal MHD equilibria \citep{Zhou2013}.

A key distinction between the formulations in two labelings is the number of time-dependent variables. In Eulerian labeling there are $\mathbf{v},\rho,p,\mathbf{B}$, whereas in Lagrangian labeling the only one is the fluid configuration $\mathbf{x}(\mathbf{x}_0,t)$, while $\rho_0,p_0,\mathbf{B}_0$ are parameters (specifically, initial conditions) that do not depend on time. The reduction of the number of time-dependent variables in Lagrangian labeling is a result of building in the advection equations, in contrast to Eulerian labeling where the advection equations are applied as constraints of variation. In the next section, we will discretize the Lagrangian in Lagrangian labeling in order to avoid the difficulty in discretizing the constrained variational principle.


\section{Discretization}
\label{integrator:discretization}
Newcomb's formulation was later revisited from the perspective of geometric mechanics, and generalized into the theory of Euler-Poincar\'e reduction \citep{Holm1998}. From such a perspective, $\rho,s,\mathbf{B}$ are categorized as advected quantities and treated equivalently, although the advection equations look somewhat different because the mass density is a 3-form, the specific entropy is a 0-form, while the magnetic field is a 2-form. 
Therefore, it seems natural to respect their identities as differential forms when discretizing Newcomb's formulation, and DEC \citep{Hirani2003,Desbrun2005,Desbrun2008} offers an appropriate framework for that. DEC has also been successfully applied to geometrically discretizing the Lagrangians for electrodynamics \citep{Stern2008,Stern2015} and Vlasov-Maxwell systems \citep{Squire2012}.

DEC is a theory of differential forms on a discrete manifold, such as a simplicial complex \citep{Munkres1984}, i.e., a collection of simplices. In 3D, it is a tetrahedral mesh $K^3$, with the tetrahedra and their faces, edges, and vertices as 3-, 2-, 1-, and 0-simplices, respectively. A discrete $k$-form $\alpha^k$ assigns a real number to each $k$-simplex $\sigma^k$, denoted by $\langle\alpha^k,\sigma^k\rangle$, which can be interpreted as the discrete analog of $\int_{\sigma^k}\alpha^k$. Operations such as exterior derivatives, wedge products and hodge stars can be defined in a way that parallels their continuous definitions. For a complete treatment of DEC, see \citet{Hirani2003,Desbrun2005,Desbrun2008}. In this chapter, we will only discuss those parts of the theory that are crucial to our work.

The ideal MHD Lagrangian in Lagrangian labeling (\ref{Lagrangian3}) is not easy to discretize directly. Therefore we choose to first discretize the Lagrangian in Eulerian labeling (\ref{LagrangianE}), and then use discrete advection equations to pass into Lagrangian labeling. The same approach was adopted by Newcomb in the continuous case. In Eulerian labeling $(\sigma^k,t)$, we have a static tetrahedral mesh $K^3=\{\sigma^k\}$. The variables $\mathbf{B}$, $\rho$ and $p$ are discretized into discrete 2-form and 3-forms, respectively, while $\mathbf{v}$ is discretized as a map from the vertices $\sigma^0$ to $\mathbb{R}^3$. Physically, $\mathbf{v}$ is the Eulerian velocity at the vertices. 

The first term in the Lagrangian is the kinetic energy, and discretizing it involves discretizing the operation of multiplying a 3-form $\rho\,\mathrm{d}^3x$ by a 0-form ${v}^2$, evaluated as $\langle{v}^2,\sigma^0\rangle=||\mathbf{v}(\sigma^0,t)||^2$. The multiplication is discretized as follows,
\begin{equation}
\int\rho{v}^2\,\mathrm{d}^3x\rightarrow\sum_{\sigma^3}\langle\rho,\sigma^3\rangle\frac{1}{4}\sum_{\sigma^0\prec\sigma^3}\langle{v}^2,\sigma^0\rangle.\label{kineticE}
\end{equation}
The second summation is essentially averaging $\langle{v}^2,\sigma^0\rangle$ stored at the 4 vertices $\sigma^0$ of a tetrahedron $\sigma^3$. It is then multiplied with $\langle\rho,\sigma^3\rangle$ stored in this tetrahedron, and then summed over every tetrahedra in $K^3$. Note that barycentric subdivision \citep{Hirani2003} is implied with this discretization. We choose barycentric subdivision here because it makes averaging easier than circumcentric subdivision \citep{Hirani2003}.

The second term in the Lagrangian is the internal energy, and its discretization is straightforward by discretizing a 3-form $p\,\mathrm{d}^3x$,
\begin{equation}
\int p\,\mathrm{d}^3x\rightarrow\sum_{\sigma^3}\langle p,\sigma^3\rangle.\label{internalE}
\end{equation}

The last term is the magnetic energy. Mathematically, it involves the norm of a 2-form, $\mathbf{B}\cdot\mathrm{d}\mathbf{S}$. With DEC, this norm is discretized as \citep{Desbrun2005,Desbrun2008,Stern2008}
\begin{equation}
\int {B}^2\,\mathrm{d}^3x\rightarrow\sum_{\sigma^3}\sum_{\sigma^2\prec\sigma^3}\frac{|* \sigma^2|}{| \sigma^2|}\langle B,\sigma^2\rangle^2,\label{magneticE}
\end{equation}
where $| \sigma^2|$ is the volume (area) of $\sigma^2$, and $|* \sigma^2|$ is the volume of its dual cell, namely the distance from $\sigma^2$ to the circumcenter of the tetrahedron that it is a face of. Note that this norm is defined with circumcentric subdivision, because to our knowledge there is not a good discretization of such a norm with barycentric subdivision. There is no conflict that we know of between using circumcentric subdivision here and barycentric subdivision in the kinetic energy term. 

After applying the substitutions in Eqs.\,(\ref{kineticE})\,-\,(\ref{magneticE}) to Eq.\,(\ref{LagrangianE}), we obtain a discrete Lagrangian in Eulerian labeling,
\begin{align}
L(\mathbf{v},\rho,p,\mathbf{B})=\sum_{\sigma^3}\left[\frac{\langle\rho,\sigma^3\rangle}{8}\sum_{\sigma^0\prec\sigma^3}\langle{v}^2,\sigma^0\rangle-\frac{\langle p,\sigma^3\rangle}{\gamma-1}-\sum_{\sigma^2\prec\sigma^3}\frac{|*\sigma^2|}{2| \sigma^2|}\langle B,\sigma^2\rangle^2\right].\label{LagrangianED}
\end{align}
We believe there should be a discrete constrained variational principle associated with this Lagrangian, which could lead to a variational integrator in Eulerian labeling. However, due to our current lack of understanding of discrete vector fields and Lie derivatives, we do not know how to properly discretize the variational constraints yet. 

Instead, we relabel this Lagrangian into Lagrangian labeling, where we have a moving mesh with each simplex $\sigma^k$ labeled by its origin $\sigma^k_0$. $\rho,p,\mathbf{B}$ are relabeled using the following discrete advection equations,
\begin{subequations}\label{advectionD}
\begin{align}
\langle\rho,\sigma^3\rangle&=\langle\rho_0,\sigma_0^3\rangle,\label{continuityD}\\
\frac{\langle p,\sigma^3\rangle|\sigma^3|^{\gamma-1}}{\langle\rho,\sigma^3\rangle^\gamma}&=\frac{\langle p_0,\sigma_0^3\rangle|\sigma_0^3|^{\gamma-1}}{\langle\rho_0,\sigma_0^3\rangle^\gamma},\label{adiabaticD}\\
\langle B,\sigma^2\rangle&=\langle B_0,\sigma_0^2\rangle.\label{frozeninD}
\end{align}
\end{subequations}
These equations can be interpreted as discrete analogs of Eqs.\,(\ref{advection}), with $\sigma^3$, barycenter of $\sigma^3$, and $\sigma^2$ regarded as discrete analogs of volume element, point, and area element, respectively. Note that if the discrete magnetic field is initially divergence-free ($\mathrm{d}B=0$), it will be guaranteed to remain so by Eq.\,(\ref{frozeninD}). Details on the discrete exterior derivative $\mathrm{d}$ can be found in \citet{Hirani2003,Desbrun2005,Desbrun2008}.  The velocity at the vertices can be relabeled by $\mathbf{v}(\sigma^0,t)=\dot{\mathbf{x}}(\sigma_0^0,t)$, where the discrete configuration $\mathbf{x}(\sigma_0^0,t)$ stands for the position of the vertex labeled by $\sigma_0^0$. Then we can express the discrete Lagrangian in Lagrangian labeling, 
\begin{align}
L(\mathbf{x},\dot{\mathbf{x}})=\sum_{\sigma_0^3}\left[\frac{\langle\rho_0,\sigma_0^3\rangle}{8}\sum_{\sigma_0^0\prec\sigma_0^3}\dot{x}^2-\frac{\langle p_0,\sigma_0^3\rangle}{(\gamma-1)J^{\gamma-1}}-\sum_{\sigma_0^2\prec\sigma_0^3}\frac{|*\sigma^2|}{2| \sigma^2|}\langle B_0,\sigma_0^2\rangle^2\right],\label{Lagrangian3D}
\end{align}
where $J=|\sigma^3|/|\sigma_0^3|$ is the discrete Jacobian. Note that $|\sigma^2|,|* \sigma^2|$ and $J$ can all be expressed solely in terms of $\mathbf{x}(\sigma_0^0,t)$, which hence becomes the only variable. There is a subtlety here that we need to comment on, associated with the magnetic energy term discretized with circumcentric subdivision. As the mesh evolves, it may lose the property of being well-centered. That is, the circumcenters may move out of the tetrahedra, and $|* \sigma^2|$ will therefore become negative. However, when that happens in practice, the discretization (\ref{magneticE}) still seems to be functional, and so does our integrator.

The Lagrangian \eqref{Lagrangian3D} is a geometric spatial discretization of Eq.\,(\ref{Lagrangian3}). Furthermore, by regarding its last two terms as a discrete potential energy $W[\mathbf{x}(\sigma_0^0,t)]$, and rearranging the first term to be summing over vertices, one can rewrite the Lagrangian as
\begin{equation}
L(\mathbf{x},\dot{\mathbf{x}})=\sum_{\sigma_0^0}\frac{1}{2}M(\sigma_0^0)\dot{x}^2-W(\mathbf{x}),\label{Lagrangian3D2}
\end{equation}
where $M(\sigma_0^0)=\sum_{\sigma_0^3\succ\sigma_0^0}\langle\rho_0,\sigma_0^3\rangle/4$ is the effective mass of vertex $\sigma_0^0$. This Lagrangian has the form of a conservative many-body Lagrangian, and the Euler-Lagrange equation that follows from it has the simple form of a Newton's equation,
\begin{equation}
M(\sigma_0^0)\ddot{\mathbf{x}}=-\partial W/\partial\mathbf{x}=\mathbf{F}(\sigma_0^0).\label{momentum3D}
\end{equation}
Keep in mind that this is a spatial discretization of the ideal MHD momentum equation in Lagrangian labeling (\ref{momentum3}).

So far, by spatial discretization, we have used a moving mesh to simulate the evolution of the fluid configuration. The spatially discretized system still has built-in advection equations and is Hamiltonian, with a conserved energy $E=\sum_{\sigma_0^0}M\dot{x}^2/2+W$. Moreover, the system is momentum conserving, in the sense that it can only gain momentum from external sources, either via forcing like gravity, or through boundaries. In our formulation, boundary conditions are applied as holonomic constraints. Examples include periodic and no-slip boundaries, or rigid walls. The system cannot gain momentum from periodic boundaries. From rigid walls it can, but only in the normal direction, not the tangential directions. As for no-slip boundaries, momentum can be gained in both normal and tangential directions.

Next we shall discretize the system in time in order to solve for the motion of the mesh. The built-in advection equations will be inherited after any temporal discretization. However, energy and momentum behavior is highly dependent on choice of temporal discretization. One way to guarantee favorable energy and momentum behavior is to employ variational integrators \citep{Marsden2001,Kraus2013}. The idea is to temporally discretize the Lagrangian (\ref{Lagrangian3D2}) and obtain the update scheme from the discrete Euler-Lagrange equation, rather than discretizing the equation of motion (\ref{momentum3D}) directly. For example, with trapezoidal discretization, the update equations (in one-step form) are
\begin{subequations}\label{update}
\begin{align}
\mathbf{x}^{n+1}&=\mathbf{x}^{n}+\tau\mathbf{p}^{n}/M+\tau^2\mathbf{F}^{n}/(2M),\\
\mathbf{p}^{n+1}&=\mathbf{p}^{n}+\tau(\mathbf{F}^{n}+\mathbf{F}^{n+1})/2,
\end{align}
\end{subequations}
where $n$ and $\tau$ are the number and size of the time step, respectively, and $\mathbf{p}=M\dot{\mathbf{x}}$ is the momentum. This is the well-known (velocity) Verlet method, which is explicit and second-order accurate. In our numerical implementation, we use such a scheme as it is fast and reasonably stable. There are also other choices, such as midpoint discretization \citep{Marsden2001,Kraus2013}.

According to \citet{Marsden2001,Kraus2013}, such schemes preserve the canonical symplectic structure on $T^*G^{N}$, the cotangent bundle of the discrete configuration space $G^{N}$, i.e., the phase space of the spatially discretized system. As $N$ becomes large, $G^{N}$ becomes ``close" to the continuous configuration space $\mathrm{Diff}(G)$, namely the diffeomorphism group \citep{Holm1998} on the domain $G$, and $T^*G^{N}$ becomes ``close" to the continuous phase space $T^*\mathrm{Diff}(G)$. Thus, we are preserving the canonical symplectic structure on a space that approximates the true fluid phase space. This is not the same as preserving the symplectic structure of the continuous system. Yet with the symplectic structure on $T^*G^{N}$ preserved, the error of energy $E$ will be bounded in our simulations \citep{Marsden2001,Kraus2013}. Besides, a discrete Noether's theorem \citep{Marsden2001} states that the schemes are momentum-preserving, which means that momentum gain can only come from external sources.

Being symplectic and momentum-preserving is a major advantage of our ideal MHD integrators. However, we will not show numerical results on this feature in this chapter, for the following two reasons. First, such properties of variational integrators have been thoroughly discussed in \citet{Marsden2001}. In addition, energy conservation does not necessarily mean the system is free of dissipation, as resistive MHD also has energy conservation. An ideal MHD algorithm can have exact energy conservation but still suffer from numerical reconnection \citep{Kraus2013,Kraus2017b}. 

Instead, a priority of ideal MHD discretizations should be to treat the advection equations in a dissipation-free manner. After all, it is the advection of mass, entropy, and magnetic flux that defines ideal MHD. And that is exactly the second highlight of our method, which comes along with spatial discretization where discrete advection equations (\ref{advectionD}) are built into the spatially discretized Lagrangian (\ref{Lagrangian3D2}). The point is, we avoid error and dissipation that come with solving advection equations, now that we do not need to solve them. Such built-in advection equations are what make our schemes excel as ideal MHD integrators. In the next section, we will show results from two numerical tests demonstrating that our method does not suffer from numerical reconnection when current singularities are present, thanks to the built-in frozen-in equation.

\section{Numerical tests}
\label{integrator:numerical}
In the previous section, all the discretization is carried out in the context of 3D compressible MHD. However, just like the continuous formulation, our integrators can easily be projected to lower dimensions. Or, with an extra discrete volume-preserving constraint, we can obtain integrators for incompressible MHD. Constructed with DEC, our machinery applies to any coordinates, while the specific schemes derived for implementation depend on the choice of coordinates. In Appx.\,\ref{ch:scheme}, the schemes in cartesian coordinates are derived for 1D, 2D, and 3D compressible MHD.

\begin{figure}[h]
\centering
\includegraphics[scale=0.75]{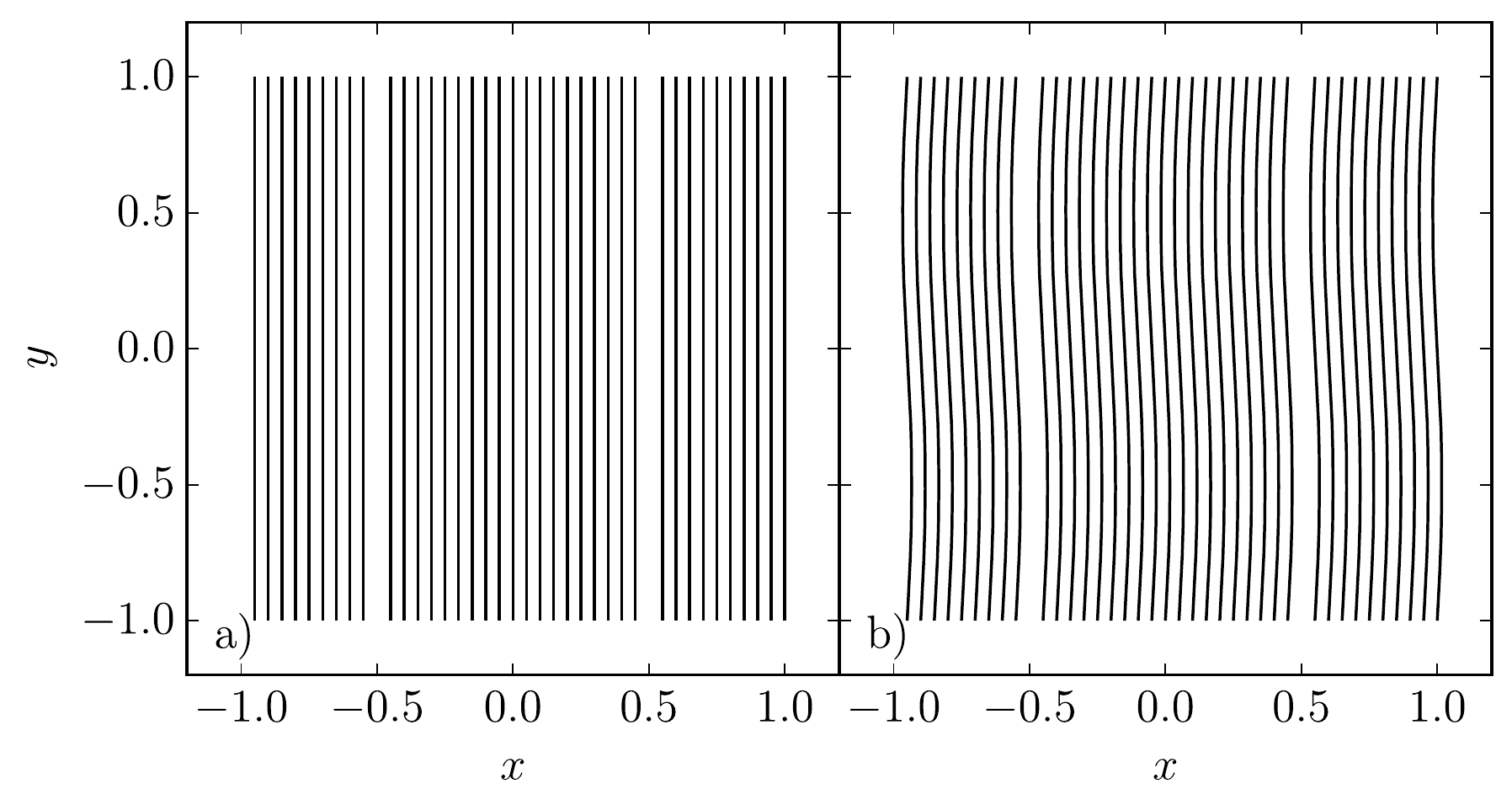}
\caption{\label{currentsheet}Field line configurations at $t=0$ (a) and $t=100$ (b). Current singularities are initially located at $x=\pm0.5$.}
\end{figure}

In this section, we show results from two numerical tests with our scheme implemented for 2D compressible MHD. The first test studies an equilibrium with two singular current sheets perturbed by a standing shear Alfv\'en wave. We borrow the setup from \citet{Gardiner2005}, which is also used in \citet{Gawlik2011,Kraus2013,Kraus2017b}. The domain is $[-1,1]^2$ with a resolution of $128^2$ and periodic boundaries. The initial equilibrium is set up with $\rho_0=1$, $p_0=0.1$, $\gamma=5/3$, $B_{0y}=1$ for $0.5<|x_0|\le1$, and $B_{0y}=-1$ for $|x_0| \le 0.5$, and perturbed with $\dot x=0.1\sin\pi y_0$. This equilibrium is stable in ideal MHD context, but unstable to tearing modes when finite resistivity exists. In \citet{Gardiner2005,Kraus2013,Kraus2017b}, magnetic islands are observed to develop along the singular current sheets at $|x_0|=0.5$. In \citet{Gawlik2011}, no numerical reconnection is shown for the duration of the run time, which is short (till $t=4$). Our simulation is run for much longer time (till $t=100$), and shows no numerical reconnection. Fig.\,\ref{currentsheet} shows that the magnetic field lines at $t=100$ preserve the topology of those at $t=0$.

The first test shows that our method can handle prescribed current singularities, which suggests that it might be best suited to studying current singularity formation in ideal MHD \citep{Parker1994}. The next test is borrowed from \citet{Craig2005b} to demonstrate that the method can handle spontaneously formed current singularities, and distinguish them from intense but ultimately smooth current layers. This problem considers a 1D perfectly conducting plasma magnetized by a initial field $B_{y0}=x_0$, and there is no guide field or pressure. The domain is $[-a,a]$ with ideal walls at $x_0=\pm a$. The system is not in equilibrium and will collapse towards one with a quadratic fluid mapping $x = x_0|x_0|/a$.
The Jacobian $J = 2|x_0|/a$ is zero at the neutral line, where the equilibrium field $B_y = B_{0y}/J = \text{sgn}(x)a/2$ yields a current singularity. Here $\text{sgn}(x)$ stands for the sign function. However, if there is a constant guide field $B_{z0}=\kappa$ (or equivalently pressure, but the algebra is easier with guide field) present, the equilibrium solution reads:
\begin{align}
x = \frac{1}{2C}\left\{x_0\sqrt{\kappa^2+x_0^2}+\kappa^2\left[\ln\left(x_0+\sqrt{\kappa^2+x_0^2}\right)-\ln\kappa\right]\right\},\label{1Dcollapse}
\end{align}
with the constant $C$ given by $x(a)=a$. The equilibrium field $B_y= C|{x_0}|/\sqrt{\kappa^2+x_0^2}$ turns out smooth (with $x_0$ being a smooth function of $x$) due to the presence of finite compressibility. When $\kappa$ is small, the current is concentrated in a very thin layer with a finite width.

We use the 2D implementation to study this 1D problem by setting, in $y$, the resolution to be 1, the boundary conditions to be rigid walls, and the system length to be 0.5. Thanks to the symmetry of the problem, we can simulate half the domain in $x$, namely $[0,a]$. We choose $a=0.5$ with a spatial resolution in $x$ of 128. The vertices are distributed non-uniformly to better resolve the vicinity of $x=0$. To obtain the equilibrium solutions, a friction term $-\nu\rho_0\dot{\mathbf{x}}$ is added to the RHS of the momentum equation (\ref{momentum3}) in order to dissipate the kinetic energy. Accordingly, the spatially discretized momentum equation \eqref{momentum3D} becomes 
\begin{equation}
M(\sigma_0^0)\ddot{\mathbf{x}}=-\partial W/\partial\mathbf{x}-\nu M(\sigma_0^0)\dot{\mathbf{x}}=\mathbf{F}(\sigma_0^0).\label{momentumF}
\end{equation}
Our temporal discretization of this friction term follows from \citet{Kane2000}. That is, the update equations \eqref{update} still apply, except that $\mathbf{F}^{n}=-\partial W^n/\partial\mathbf{x}^n-\nu {\mathbf{p}}^n$ now depends on not only $\mathbf{x}^{n}$ but also $\mathbf{p}^{n}$. This approach of frictional relaxation will be used throughout the rest of the thesis whenever equilibrium solutions are obtained.

\begin{figure}[h]
\centering
\includegraphics[scale=0.75]{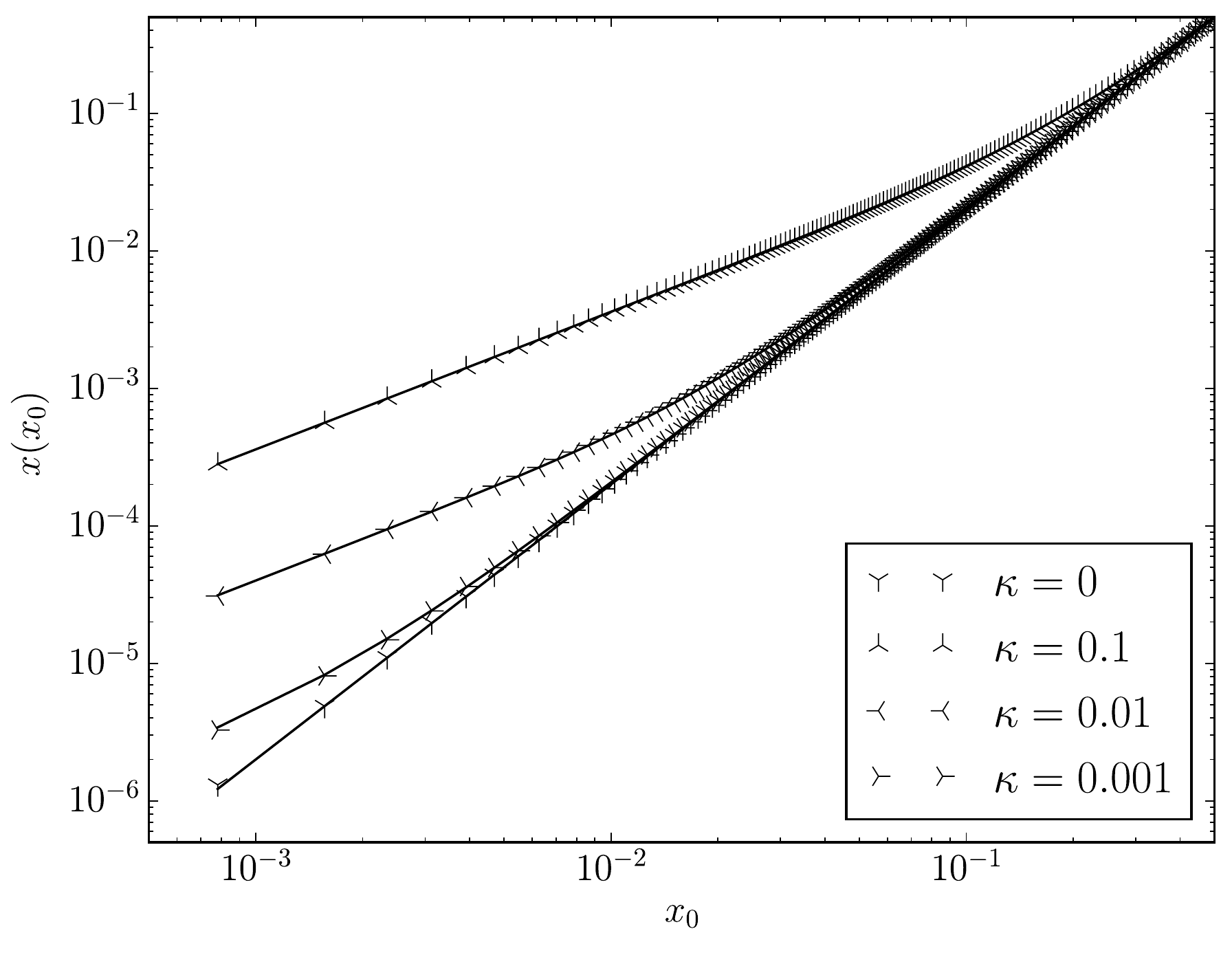}
\caption{\label{collapse}Equilibrium solution $x(x_0)$ with different values of $\kappa$ for the 1D collapse problem. The numerical solutions (markers) agree with the respective analytical solutions (solid lines).}
\end{figure}

In Fig.\,\ref{collapse}, equilibrium solutions $x(x_0)$ for $\kappa=0,0.1,0.01,0.001$ are shown to agree with the respective analytical ones, demonstrating that the scheme can indeed distinguish the formation of a genuine current singularity from that of intense but ultimately smooth current layers.

\section{Summary and discussion}
\label{integrator:summary}
In this chapter, variational integrators for ideal MHD with built-in advection equations are derived by discretizing Newcomb's Lagrangian for ideal MHD in Lagrangian labeling using DEC. The integrators possess two significant advantages. First, they are symplectic and momentum-preserving, which results from the variational temporal discretization. Second, with the advection equations built-in, we avoid solving them and the accompanying error and dissipation. The latter is especially important since it allows our method to accomplish what previous methods cannot, such as handling current singularities without encountering numerical reconnection. In addition, the method is physically transparent. The moving mesh practically simulates the motion of the fluid elements. Arguably, it captures the physics of ideal MHD more closely than other methods.

It is worthwhile to compare our method with other Lagrangian schemes. As a moving-mesh method, ours differ from particle-based Lagrangian methods, where solving the frozen-in equation can also be avoided by advecting the so-called Euler potentials with the particles \citep{Rosswog2007}. However, due to the errors introduced by interpolating and reconstructing the magnetic fields, these particle-based methods can still suffer from numerical reconnection. Simulating MHD with a moving mesh is not a new idea \citep{Pakmor2011}, but to our knowledge we are the first to discretize the ideal MHD Lagrangian and obtain variational integrators on it. 

Among existing methods, the one that bears the most similarity to ours is developed by discretizing Eq.\,(\ref{momentum3}) with conventional finite difference \citep{Craig1986}. One caveat of this approach is that the discrete force may not be conservative. Moreover, \citet{Pontin2009} has recently shown that its current density output can violate charge conservation ($\nabla\cdot\mathbf{j}= 0$, where $\mathbf{j}$ is the current density), and mimetic discretization has been applied to fix it \citep{Candelaresi2014}. In contrast, the force in our method is derived from a discrete potential and hence guaranteed conservative. Many of the conservation laws of ideal MHD, including charge conservation, are naturally inherited by our method, thanks to the adoption of DEC.

While numerical results suggest that the method we propose here is promising, we should emphasize that it is not a panacea for all ideal MHD simulations, at least not in its current shape. One vulnerability of the method we have recognized is that when strong shear flow is present, the simplices can get extremely deformed, and the mesh will be torn up. Such mesh distortion is a well-known problem for most numerical methods constructed on a moving mesh \citep{Springel2010}. Re-meshing is a popular strategy for handling it, but it appears difficult to apply it to our method in a consistent variational manner. It would be worthwhile to consider smoothed particle (magneto-) hydrodynamics with a structure-preserving mindset as well.

Another issue is that presently the method can at best go only as far as ideal MHD. However, ideal MHD itself is a model with limited applicability. For example, ideal MHD fails when shocks develop. Shocks are not adiabatic and therefore some kind of dissipation is needed to capture it. A possible way to introduce viscosity or resistivity to the schemes is via certain discrete random-walk processes. It would be challenging to avoid ruining the conservation of energy and momentum, or the divergence-freeness of the magnetic field, etc.

There is another possibility to resolve all these issues. That is, to develop an Eulerian variational integrator using the discrete Lagrangian in Eulerian labeling (\ref{LagrangianED}), as discussed in Sec.\,\ref{integrator:discretization}. In that case, the mesh will be fixed, therefore mesh distortion will no longer be a problem. On the other hand, resistivity and viscosity can be added to the scheme via discrete Laplacians, which has been done successfully in \citet{Mullen2009}. Previous works along this line of thought \citep{Gawlik2011,Kraus2013,Kraus2017b} are not entirely satisfactory. The key obstacle is the lack of a viable discrete vector field (and hence Lie bracket, algebra, and derivatives), but recent developments in finite element exterior calculus (FEEC) \citep{Arnold2006,Arnold2010} may have improved our chances. Besides, adopting FEEC instead of DEC in our Lagrangian scheme may improve its accuracy as it allows higher order spatial discretization than DEC. FEEC can also potentially offer a better discretization of the magnetic energy, and hence the Ampere's force.

Despite these issues, the strengths of the proposed method still makes it favorable for studying certain classes of ideal MHD problems. Generally speaking, it is most suitable for problems that are shock-free, quasi-static, and highly prioritize preserving the advection equations. An immediately applicable problem of this kind is current singularity formation \citep{Parker1994}. The results from the test cases shown in Sec.\,\ref{integrator:numerical} suggest that it is promising. In Chaps.\,\ref{ch:2D} and \ref{ch:3D}, results from applying our variational integrators to studying current singularity formation will be presented.

When studying current singularity formation, it is customary to focus on the equilibria rather than the dynamics. However, being symplectic and momentum-preserving, the variational integrators for ideal MHD are also capable of performing certain dynamical studies with exceptional long-term fidelity. A potential application that could exploit this feature is to study nonlinear (Alfv\'en) wave heating, another candidate mechanism for coronal heating. The study would allow non-resistive effects to be distinguished from others, being in the perfectly conducting limit with bounded energy error. However, this topic will not be investigated in this thesis.

In this chapter we present our method entirely in the context of ideal MHD. Yet it can be straightforwardly generalized to other systems with similar Euler-Poincar\'e structures, such as nematic liquid crystals, elastodynamics, or many geophysical fluids \citep{Holm1998}. The consequent methods will likely share the strengths and weaknesses of the ideal MHD ones. Accordingly, we advise those who are interested in pursuing in this direction to identify suitable problems of interest first before undertaking numerical developments.

\chapter{Current singularity formation in 2D\label{ch:2D}}

In Chap.\,\ref{ch:integrator}, variational integrators for ideal MHD in Lagrangian labeling are derived. With the  frozen-in equation built-in, the method does not suffer from numerical reconnection, and thus is optimal for studying current singularity formation. In this chapter, we investigate this problem in 2D. We first study the Hahm-Kulsrud-Taylor (HKT) problem, which considers the response of a 2D plasma magnetized by a sheared field under sinusoidal boundary forcing. We obtain an equilibrium solution that preserves the magnetic topology of the initial field exactly, with a fluid mapping that is non-differentiable. Unlike previous studies that use diverging peak current densities as sole evidences for current singularities, we show that our numerical solutions converge to one with a current singularity. These results are benchmarked with a constrained Grad-Shafranov solver. The same signature of current singularity can be found in other cases with more complex magnetic topologies, such as the coalescence instability of magnetic islands. A modification to the HKT problem with a current singularity embedded in the initial condition is also discussed. The work in this chapter is largely published in \citet{Zhou2016}.

\section{Introduction}
\label{2D:intro}
Current singularity formation has long been an issue of interest in plasma physics.
In toroidal fusion plasmas, closed field lines exist at rational surfaces. It is believed that current singularities tend to occur when these surfaces are subject to resonant perturbations \citep{Grad1967,Rosenbluth1973,Hahm1985,Cary1985,Hegna1989,Bhattacharjee1995,Boozer2010,Dewar2013,Helander2014,Loizu2015}, which jeopardizes the existence of 3D equilibria with nested flux surfaces. In the solar corona, field lines are tied into the boundaries and do not close on themselves. Yet \citet{Parker1972} argued that current singularities would still tend to form frequently, and the subsequent magnetic reconnection events can lead to substantial heating. This theory has stayed controversial to this day \citep{Rosner1982,Parker1983,Parker1994,Tsinganos1984,VanBallegooijen1985,VanBallegooijen1988,Zweibel1985,Zweibel1987,Antiochos1987,Longcope1994b,Ng1998,Longbottom1998,Bogoyavlenskij2000,Craig2005,Low2006,Low2010,Wilmot-Smith2009,Wilmot-Smith2009b,Janse2010,Rappazzo2013,Craig2014,Pontin2015,Candelaresi2015,Pontin2016}. A brief history of studying current singularity formation is reviewed in Sec.\,\ref{intro:formation}.

Albeit inherently a dynamical problem, current singularity formation is usually treated by examining magnetostatic equilibria for the sake of simplicity. The justification is, if the final equilibrium that an initially smooth magnetic field relaxes to contains current singularities, they must have formed during the relaxation. Here the plasma is supposed to be perfectly conducting, so the equilibrium needs to preserve the magnetic topology of the initial field. This topological constraint is mathematically challenging to explicitly describe and attach to the magnetostatic equilibrium equation,
\begin{equation}
(\nabla\times\mathbf{B})\times\mathbf{B}=\nabla p,\label{JcrossB}
\end{equation}
and its solutions \citep{Janse2010,Low2010}. Eq.\,(\ref{JcrossB}) is simply the momentum equation in Eulerian labeling (\ref{momentumE}) with $\mathbf{v}=0$. A given set of boundary conditions usually allows for more than one solution to this equation, and additional information is needed to identify a specific one. Often it is prescribed to the equilibrium, such as the pressure and guide field profiles in the Grad-Shafranov equation \citep{Grad1967}. To distinguish a topologically constrained equilibrium solution from others, the additional information needed is the very constraint to preserve the initial magnetic topology. To enforce this constraint is a major challenge for studying current singularity formation, either analytically or numerically. 

It turns out this difficulty can be overcome by adopting Lagrangian labeling instead of the commonly used Eulerian labeling. As reviewed in Sec.\,\ref{integrator:Lagrangian}, in Lagrangian labeling, the motion of the fluid elements is traced in terms of a continuous mapping from the initial position $\mathbf{x}_0$ to the current position $\mathbf{x}(\mathbf{x}_0,t)$. In this formulation, the advection (continuity, adiabatic, and frozen-in) equations (\ref{advection}) are built into the ideal MHD Lagrangian (\ref{Lagrangian3}), the momentum equation (\ref{momentum3}), and the equilibrium equation in Lagrangian labeling, 
\begin{align}
-B_{0j}\frac{\partial}{\partial x_{0j}}\left(\frac{x_{ik}B_{0k}}{J}\right)
+\frac{\partial J}{\partial x_{ij}}\frac{\partial }{\partial x_{0j}}\left(\frac{p_0}{J^\gamma}+\frac{x_{kl}x_{km}B_{0l}B_{0m}}{2J^2}\right)=0,\label{equilibrium3}
\end{align}
which is Eq.\,(\ref{momentum3}) without time dependence. Its solutions will satisfy not only Eq.\,(\ref{JcrossB}) but automatically the topological constraint in studying current singularity formation, since the initial field configuration $\mathbf{B}_0$ is built into it.
In contrast, not all solutions to Eq.\,(\ref{JcrossB}) can necessarily be mapped from given initial conditions. Therefore, the equilibrium equation in Lagrangian labeling offers a more natural and mathematically explicit description for the problem of current singularity formation, which simply becomes whether there exist singular solutions to such equation, given smooth initial and boundary conditions. If the initial field $\mathbf{B}_0$ is smooth, any singularity in the equilibrium field $\mathbf{B}$ should trace back to that in the fluid mapping $\mathbf{x}(\mathbf{x}_0)$. 

Analytically, \citet{Zweibel1987} first used the advantageous Lagrangian labeling to study current singularity formation. Numerically, a Lagrangian relaxation scheme has been developed using conventional finite difference \citep{Craig1986}, and extensively used to study current singularity formation  \citep{Longbottom1998,Craig2005,Craig2005b,Pontin2005,Wilmot-Smith2009,Wilmot-Smith2009b,Craig2014,Craig2014b,Pontin2015}. 

In Chap.\,\ref{ch:integrator}, variational integrators for ideal MHD in Lagrangian labeling are derived in a geometric and field-theoretic manner such that many of the conservation laws of ideal MHD are naturally inherited. As discussed in Sec.\,\ref{integrator:summary}, these schemes are superior to that in \citet{Craig1986} in terms of guaranteeing charge conservation and conservativeness of the force, flexibility due to unstructured meshing, etc. 

In this chapter, we present the results of applying this optimal method to studying current singularity formation in 2D. It is widely believed that current singularity can form in 2D, as supported by numerous studies \citep{Syrovatskii1971,Hahm1985,Low1988,Wang1992,Longcope1993,Wang1994,Ma1995,Scheper1998,Craig2005b}. However, the numerical results presented in these studies are not sufficiently conclusive. One of the reasons is the lack of a convincing method to numerically identify current singularities. We present one here, and hence offer stronger than previous numerical evidences for current singularity formation in 2D. The work in this chapter serves as a building-block for further examining Parker's controversial conjecture of current singularity formation in 3D line-tied geometry, since the methodology we employ in 2D is still applicable there, as will be shown in Chap.\,\ref{ch:3D}.

We begin with a fundamental prototype problem first proposed by Taylor and studied by \citet[][the HKT problem from here on]{Hahm1985}, where a 2D plasma in a sheared magnetic field is subject to sinusoidal boundary forcing. It was originally designed to study forced magnetic reconnection induced by resonant perturbation on a rational surface. In the context of studying current singularity formation, we refer to finding a topologically constrained equilibrium solution to it as the ideal HKT problem. The linear solution by \citet{Zweibel1987} to this problem  contains a current singularity but also a discontinuous displacement that is unphysical. It has remained unclear whether the nonlinear solution to it is ultimately singular or smooth.

In Sec.\,\ref{2D:HKT}, we study how the nonlinear numerical solution to the ideal HKT problem converges with increasing spatial resolution, and find that the fluid mapping along the neutral line is non-differentiable. Unlike previous studies that depend heavily on the current density diagnostic that is more vulnerable to numerical inaccuracies \citep{Longbottom1998,Craig2005,Craig2005b,Pontin2005,Craig2014b,Craig2014,Candelaresi2015}, we identify a current singularity from the quadratic fluid mapping normal to the neutral line.
Prompted by these results, we employ a Grad-Shafranov solver where the equilibrium guide field is not prescribed \textit{a priori} but constrained by flux conservation \citep{Huang2009} to independently verify the accuracy of our Lagrangian method.

Then, in Sec.\,\ref{2D:modified}, we take a detour to look into a modification of the ideal HKT problem due to \citet{Dewar2013,Loizu2015,Loizu2015b}, where a finite jump is introduced to the initial field at the neutral line. The linear displacement becomes continuous, yet can still be pathological when the amplitude of the perturbation is considered finite and sufficiently large. The nonlinear solution of the fluid mapping becomes continuous and differentiable, and the numerical solutions converge, even when the linear solution is pathological with finite amplitude.

Finally, in Sec.\,\ref{2D:coalescence}, we investigate the coalescence instability of magnetic islands \citep{Finn1977,Longcope1993}, a problem with more complex magnetic topology than the HKT problem. We identify the exact same signature of current singularity in the topologically constrained equilibrium solution. That is, the current singularity that forms in the coalescence instability is locally indistinguishable from the one in the ideal HKT problem. This demonstrates the generality of the recipe for current singularity formation in 2D that we find. Further discussions follow in Sec.\,\ref{2D:discussion}.
 
\section{The Hahm-Kulsrud-Taylor problem}
\label{2D:HKT}
\begin{figure}[h]
\centering
\includegraphics[scale=0.75]{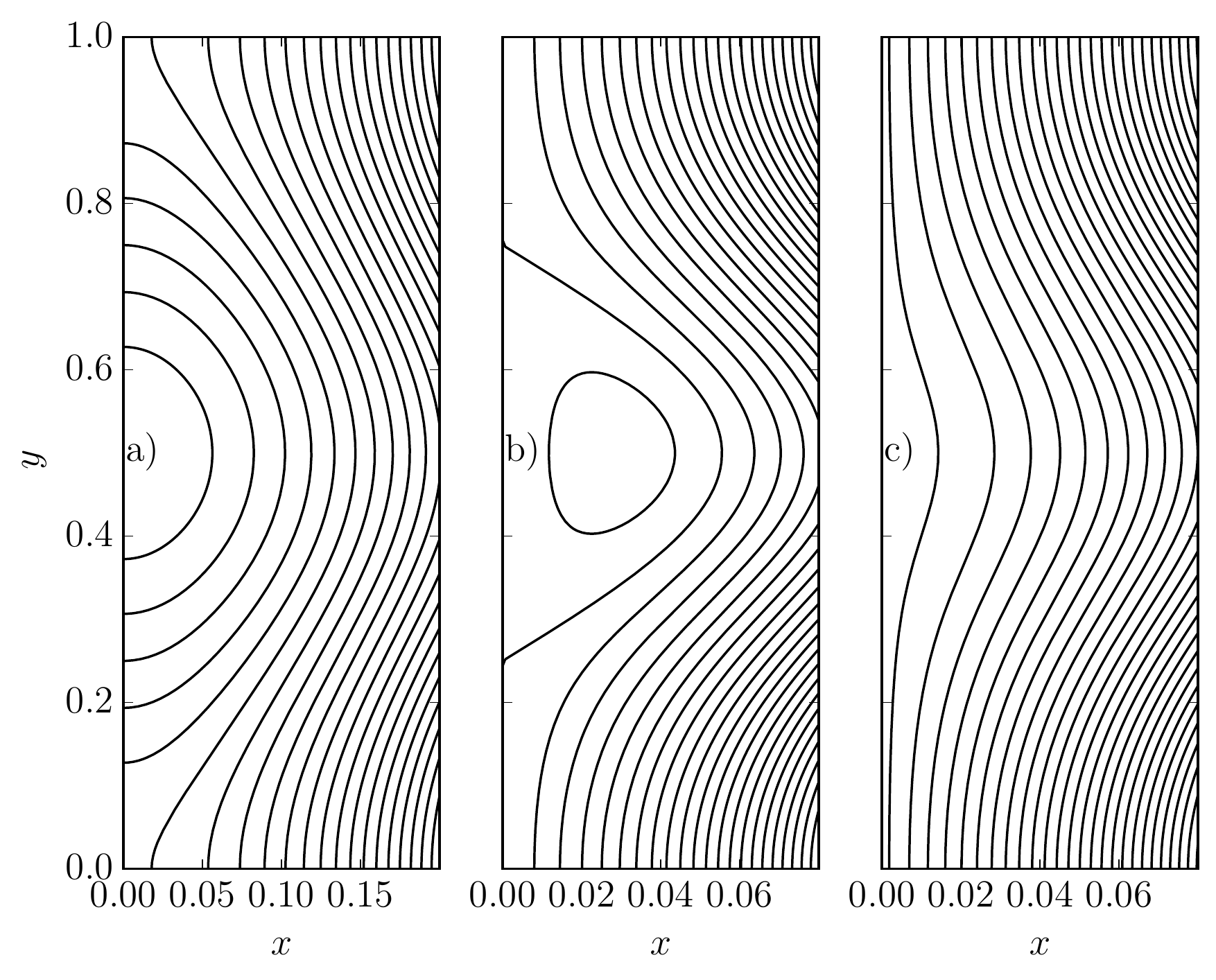}
\caption{\label{residual} Field line configurations of the exact but not topologically constrained solutions to the HKT problem: (\ref{island}) (a), (\ref{HK}) (b), and (\ref{Dewar}) (c). Parameters used are $a=0.5$, $k=2\pi$, $\delta=0.1$, and $\alpha=ka\delta/\sinh ka$. Note the different scales in $x$.}
\end{figure}

The HKT problem \citep{Hahm1985} originally considers a 2D incompressible plasma magnetized by a sheared equilibrium field $B_{0y} =  x_0$. The perfectly conducting boundaries at $x_0=\pm a$ are deformed sinusoidally into the shapes that $x(\pm a, y_0)=\pm \left[a-\delta\cos ky(\pm a, y_0)\right]$, assuming $k>0$. Two branches of solutions were originally obtained by solving the Eulerian equilibrium equation (\ref{JcrossB}) perturbatively, but later found to be nonlinearly exact. One branch is
\begin{equation}
B_y =  x +ka\delta\sinh kx\cos ky/\cosh ka,\label{island}
\end{equation}
which contains magnetic islands with width of $O(\delta^{1/2})$ along the neutral line $x=0$, as shown in Fig.\,\ref{residual}(a).
The other branch reads
\begin{equation}
B_y =  x +\text{sgn}(x)ka\delta\cosh kx\cos ky/\sinh ka,\label{HK}
\end{equation}
where the sign function $\text{sgn}(x)$ introduces a jump in $B_y$, namely a current singularity at the neutral line. However, it can be shopdfwn that this solution introduces residual islands with width of $O(\delta)$ on both sides of the neutral line \citep{Boozer2010,Dewar2013}, as plotted in Fig.\,\ref{residual}(b). \citet{Dewar2013} further generalized solution (\ref{HK}) into
\begin{equation}
B_y =  x +\text{sgn}(x)\left(\alpha+ka\delta\cosh kx\cos ky/\sinh ka\right).\label{Dewar}
\end{equation}
The introduction of the positive constant $\alpha$ can eliminate these residual islands when $\alpha\geq ka\delta/\sinh ka$, as plotted in Fig.\,\ref{residual}(c). However, the magnetic flux contained in $[0,a]$, $a(\alpha +a/2)$, is then more than that of the initial field, $a^2/2$. In other words, there are more field lines in solution (\ref{Dewar}) than the initial condition.

Therefore, none of the above solutions preserve the magnetic topology of the initial field, which means they are not topologically constrained equilibria. This highlights the difficulty in enforcing the topological constraint when one studies current singularity formation with Eq.\,(\ref{JcrossB}).

So we switch to the advantageous Lagrangian labeling to study the ideal HKT problem. For linear solutions, linearizing Eq.\,(\ref{equilibrium3}) is equivalent to linearizing Eq.\,(\ref{JcrossB}) while applying the advection constraints (\ref{advectionV}). Therefore the linearized equilibrium equation in Lagrangian labeling is simply $\mathbf{F}(\bm{\xi})=0$, where $\bm{\xi}$ is the displacement and $\mathbf{F}$ is the standard ideal MHD force operator \citep{Schnack2009linearized}. The subsequent linear solution to the ideal HKT problem obtained by \citet{Zweibel1987} takes the incompressible form, $\bm{\xi}=\nabla\chi\times\hat z$, where the stream function $\chi$ has Fourier dependence $\chi(x_0,y_0)=\bar\chi(x_0)\sin(ky_0)$.
Substituting $B_{0y}=x_0$ into the linearized equilibrium equation $\mathbf{F}(\bm{\xi})=0$ leads to one simple ordinary differential equation,
\begin{align}\label{HKE}
(\partial_{x_0}^2-k^2)(x_0\bar\chi)=0.
\end{align}
There are two branches of solutions to this equation, $\bar\chi\sim x_0^{-1}\sinh kx_0$ and $\bar\chi\sim x_0^{-1}\cosh kx_0$. The latter is discarded as it diverges at $x_0=0$.
Then the boundary condition $\bar\chi(\pm a)=\mp \delta/k$ determines that there only exists a weak solution that is discontinuous at $x_0=0$,
\begin{align}\label{HKS}
\bar\chi=-\frac{\delta a\sinh kx_0}{k|x_0|\sinh ka}.
\end{align}
This linear solution agrees with Eq.\,(\ref{HK}) linearly and hence contains a current singularity at the neutral line. However, the discontinuity in $\xi_x=\partial\chi/\partial{y_0}$ is not physically permissible, which results in the residual islands shown in Fig.\,\ref{residual}(b) [also see Fig.\,\ref{alpha}(a) for how such discontinuity causes field lines to intersect]. The failure at the neutral line is expected from the linear solution since the linear assumption breaks down there. Similar discontinuity in the displacement also appears in a linear analysis of the internal kink instability \citep{Rosenbluth1973}.

It is worth noting that instead of enforcing incompressibility ($J=1$), \citet{Zweibel1987} used a guide field $B_{0z}=\sqrt{1-x_0^2}$ so that the unperturbed equilibrium is force-free. Their solution (\ref{linear}) turns out to be linearly incompressible. Even near the neutral line, the plasma should still be rather incompressible because the guide field dominates there. Therefore, the physics of the ideal HKT problem is not affected by such alteration in setup, which we shall adopt in our numerical studies. 

\begin{figure}[h]
\centering
\includegraphics[scale=0.75]{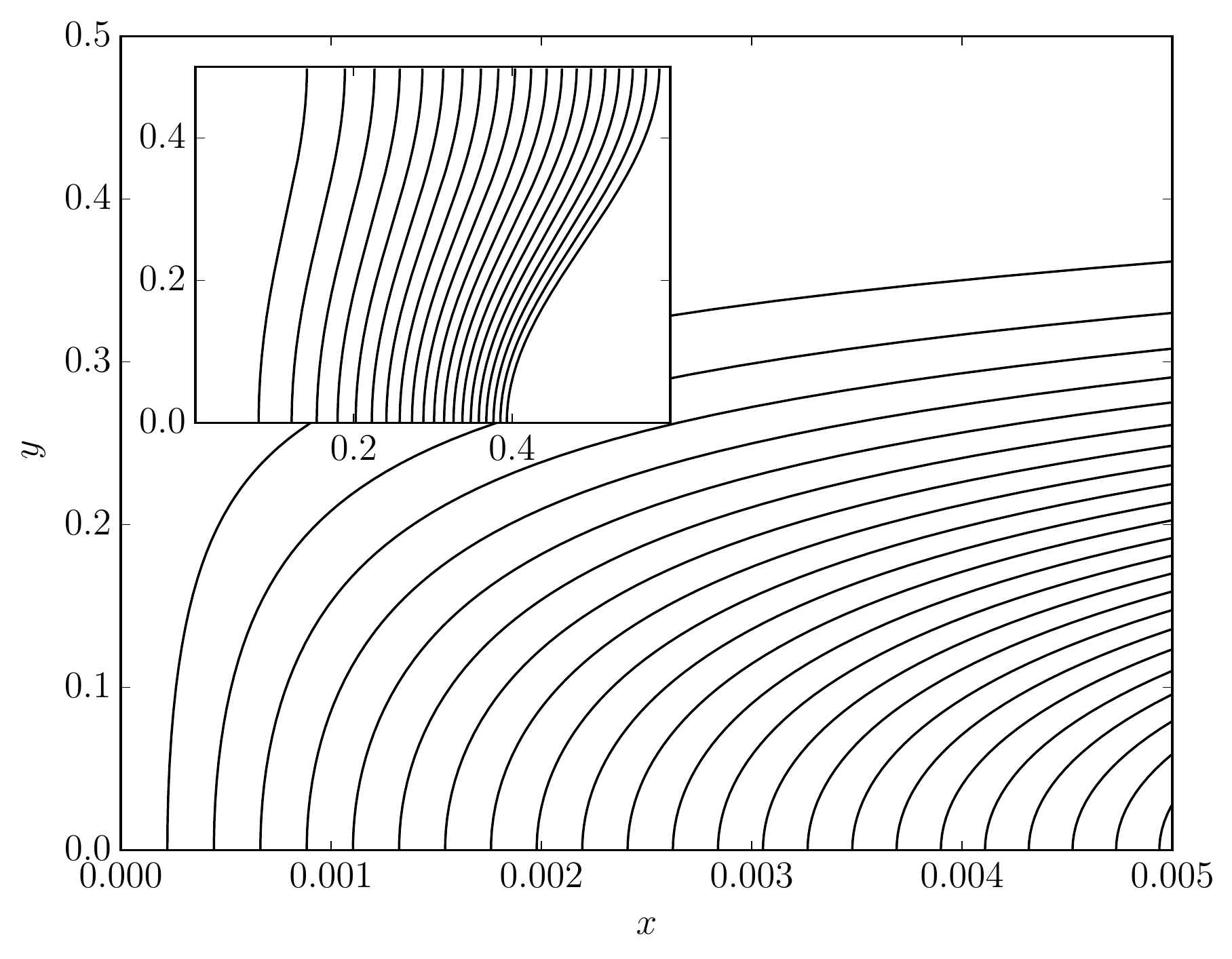}
\caption{\label{magnetic} Equilibrium field line configuration in the vicinity of the neutral line, and the entire domain (inset). The field lines appear equally spaced along $y=0$ near the neutral line.}
\end{figure}

We use the variational integrator for ideal MHD derived in Chap.\,\ref{ch:integrator}, and introduce friction to dynamically equilibrate the system and obtain a solution to Eq.\,(\ref{equilibrium3}). The details of the 2D scheme in Cartesian coordinates can be found in Appx.\,\ref{scheme:2D}. 
For the ideal HKT problem, we use a structured triangular mesh. Thanks to the symmetry in this problem, we can simulate only a quarter of the domain, $[0,a]\times[0,\pi/k]$. The boundary at $x_0=a$ is constrained such that $x=a-\delta\cos ky$. The vertices on all the boundaries are allowed to move tangentially along but not normally to them. These boundary conditions are exactly consistent with the original HKT setup. The parameters we choose are $\rho_0=1$, $a=0.5$, $k=2\pi$, and $\delta=0.1$. We apply a large perturbation so that the nonlinear effect is more significant and easier to resolve. The vertices are distributed uniformly in $y$ but non-uniformly in $x$ in order to devote more resolution to the region near the neutral line. The system starts from a smoothly perturbed configuration $\mathbf{x}$ consistent with the boundary conditions and relaxes to equilibrium. The specific choice of the initial fluid configuration does not affect the equilibrium solution. In Fig.\,\ref{magnetic} we plot the field line configuration of the equilibrium. 

An observation from Fig.\,\ref{magnetic} is that $B_y(x,0)$ approaches a finite constant near the neutral line. To better illustrate the origin of such tangential discontinuity, we recall the simple yet instructive 1D collapse problem \citep{Craig2005b} discussed in Sec.\,\ref{integrator:numerical}. With no guide field or pressure present, its exact nonlinear solution yields a current singularity: the same sheared initial field $B_{0y}= x_0$, relaxes to an equilibrium with a quadratic fluid mapping $x=x_0|x_0|/a$, which accounts for a discontinuous equilibrium field $B_y=B_{0y}/J= \text{sgn}(x)a/2$.
As we shall show next, the current singularity in the ideal HKT problem develops from the same ingredients, sheared initial field and quadratic fluid mapping.

\begin{figure}[h]
\centering
\includegraphics[scale=0.75]{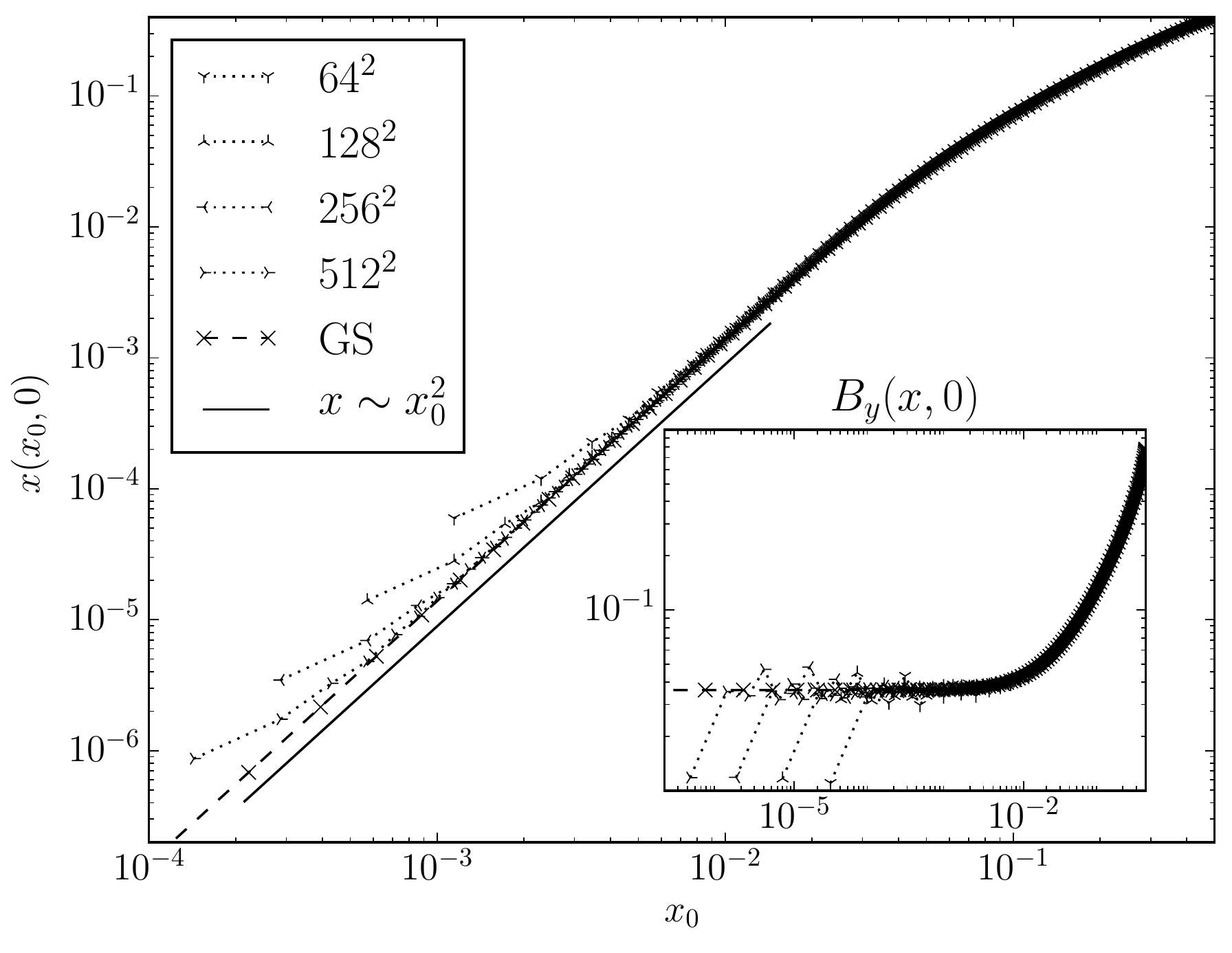}
\caption{\label{mappingx} Numerical solutions of $x(x_0,0)$ and $B_y(x,0)$ (inset) for different resolutions (dotted lines). The converged parts agree with the results obtained with a constrained Grad-Shafranov solver (dashed lines). Near the neutral line, $x(x_0,0)$ shows a quadratic power law, while $B_y(x,0)$ approaches a finite constant. The solutions do not converge for the few vertices closest to the neutral line.}
\end{figure}

We check how the equilibrium solutions converge with increasing spatial resolutions, from $64^2$ to $128^2$, $256^2$, and $512^2$. 
In Fig.\,\ref{mappingx}, we plot the equilibrium fluid mapping normal to the neutral line at $y_0=0$, namely $x(x_0,0)$. For the part near $x_0=0$, where the solutions converge, quadratic power law $x\sim  x_0^2$ can be identified. As discussed in the 1D collapse problem, together with a sheared field $B_{0y}\sim x_0$, such a mapping leads to a magnetic field $B_y=B_{0y}/(\partial x/\partial x_0)\sim \text{sgn}(x)$ [note that $J=(\partial x/\partial x_0)(\partial y/\partial y_0)$ at ${y_0=0}$] that is discontinuous at $x_0=0$, as plotted in the inset of Fig.\,\ref{mappingx}.

Despite the remarkable resemblance on the mechanism of current singularity formation, there is a key distinction between the 1D collapse and the ideal HKT problem. For the former, the solution is singular if and only if the plasma is infinitely compressible. In that case, $J=2|x_0|/a$ is zero at the neutral line, yet the equilibrium fluid mapping is continuous and differentiable. If there is guide field or pressure, no matter how small, to supply finite compressibility that prevents the Jacobian from reaching zero, the equilibrium solution (\ref{1Dcollapse}) would be smooth with no current singularity \citep{Craig2005b}. In the ideal HKT problem, the plasma is (close to) incompressible. This is confirmed by our numerical solutions which show $J\approx1+O(\delta^2)$. As a result, the equilibrium fluid mapping turns out to be non-differentiable.

\begin{figure}[h]
\centering
\includegraphics[scale=0.75]{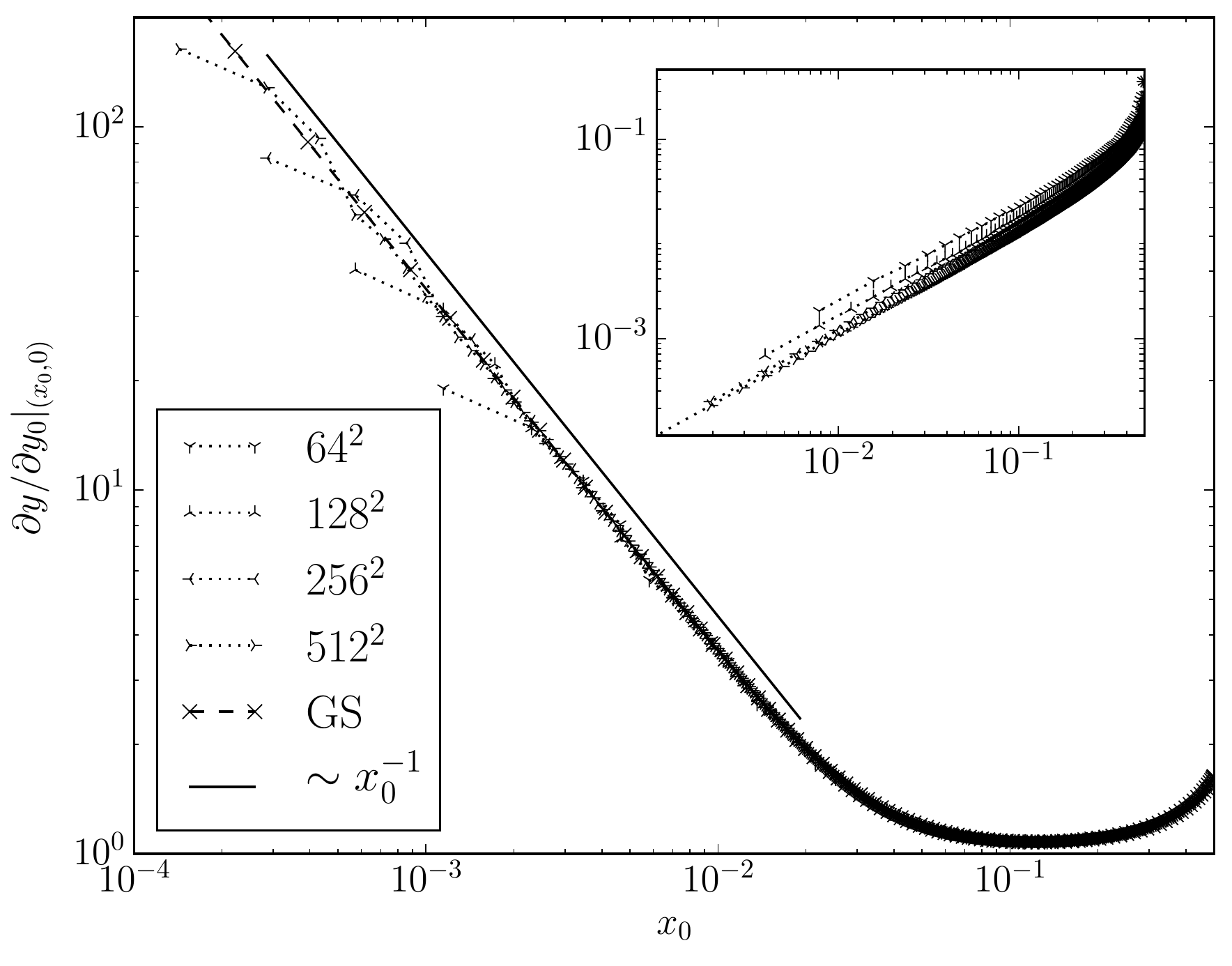}
\caption{\label{mappingy} Numerical solutions of $\partial y/\partial y_0|_{(x_0,0)}$ for different resolutions (dotted lines). The converged parts agree with the results obtained with a constrained Grad-Shafranov solver (dashed lines). Near the neutral line, $\partial y/\partial y_0|_{(x_0,0)}$ shows $x_0^{-1}$ power law. The solutions do not converge for the few vertices closest to the neutral line. In the inset of, the final versus initial distance to $(0,0.5)$ for the vertices on the neutral line, i.\,e.\, $0.5-y(0,y_0)$ vs.\,$0.5-y_0$ for different resolutions are shown to not converge.}
\end{figure}

At $y_0=0$, the converged power law $x\sim  x_0^2$ suggests that $\partial x/\partial x_0\sim  x_0$ would vanish as $x_0$ approaches $0$. To ensure incompressibility, there should be $\partial y/\partial y_0\sim  x_0^{-1}$, which would diverge at $x_0=0$. This is shown in Fig.\,\ref{mappingy}. Physically, this means the fluid element at the origin (0,0) is infinitely compressed in the normal direction ($x$), while infinitely stretched in the tangential direction ($y$) of the neutral line.

However, it is difficult to numerically resolve a diverging $x_0^{-1}$ power law at $x_0=0$. Hence, the numerical solutions $x(x_0,0)$ and $\partial y/\partial y_0|_{(x_0,0)}$ both deviate from the converged power law for the few vertices closest to the neutral line. This deviation consistently reduces but does not vanish with increasing resolutions. The inset of Fig.\,\ref{mappingy} shows that the vertices on the neutral line become more packed at $(0,0.5)$ as the resolution increases, suggesting that the solutions do not converge on the neutral line.

These numerical results are benchmarked with the solutions obtained by Yi-Min Huang using a constrained Grad-Shafranov (GS) solver. In this solver, the equilibrium guide field is determined self-consistently with a constraint to preserve its flux at each flux surface \citep{Huang2009}, unlike conventional ones where it is prescribed as a flux function. Without this feature the solver would not be capable for studying the ideal HKT problem. As shown in Figs.\,\ref{mappingx} and \ref{mappingy}, the GS results are in excellent agreement with the converged part of those obtained with the Lagrangian scheme. Since the fluid mapping is inferred rather than directly solved for, the GS solver is able to achieve better agreement with the $x_0^{-1}$ power law shown in Fig.\,\ref{mappingy}. However, it should be pointed out that the applicability of the GS solver is limited to 2D problems with nested flux surfaces, while the Lagrangian scheme can be readily generalized to problems with complex magnetic topologies or to 3D. In Sec.\,\ref{2D:coalescence}, we will use the Lagrangian scheme to study the coalescence instability of magnetic islands \citep{Finn1977,Longcope1993}, a problem with more complex magnetic topology such that the GS solver is not applicable to it. The same signature of current singularity as shown in Figs.\,\ref{mappingx} and \ref{mappingy} will be identified. Before that, we investigate a modification to the ideal HKT problem in the next section, where the initial magnetic field yields a finite jump at the neutral line.

pdf\section{The modified Hahm-Kulsrud-Taylor problem}
\label{2D:modified}
It is worthwhile to compare our result with the recent work of \citet{Loizu2015,Loizu2015b}, which also studies the ideal HKT problem, but in the context of finding well-defined ideal MHD equilibria with nested flux surfaces. For the original HKT setup, they could not find such an equilibrium. Then they suggest a modification to the problem, which in our terminology is equivalent to making the initial magnetic field discontinuous with $B_{0y}=x_0+\text{sgn}(x_0)\alpha$, where $\alpha$ is a positive constant. For clarification, we refer to such a problem as the modified HKT problem.

\begin{figure}[h]
\centering
\includegraphics[scale=0.75]{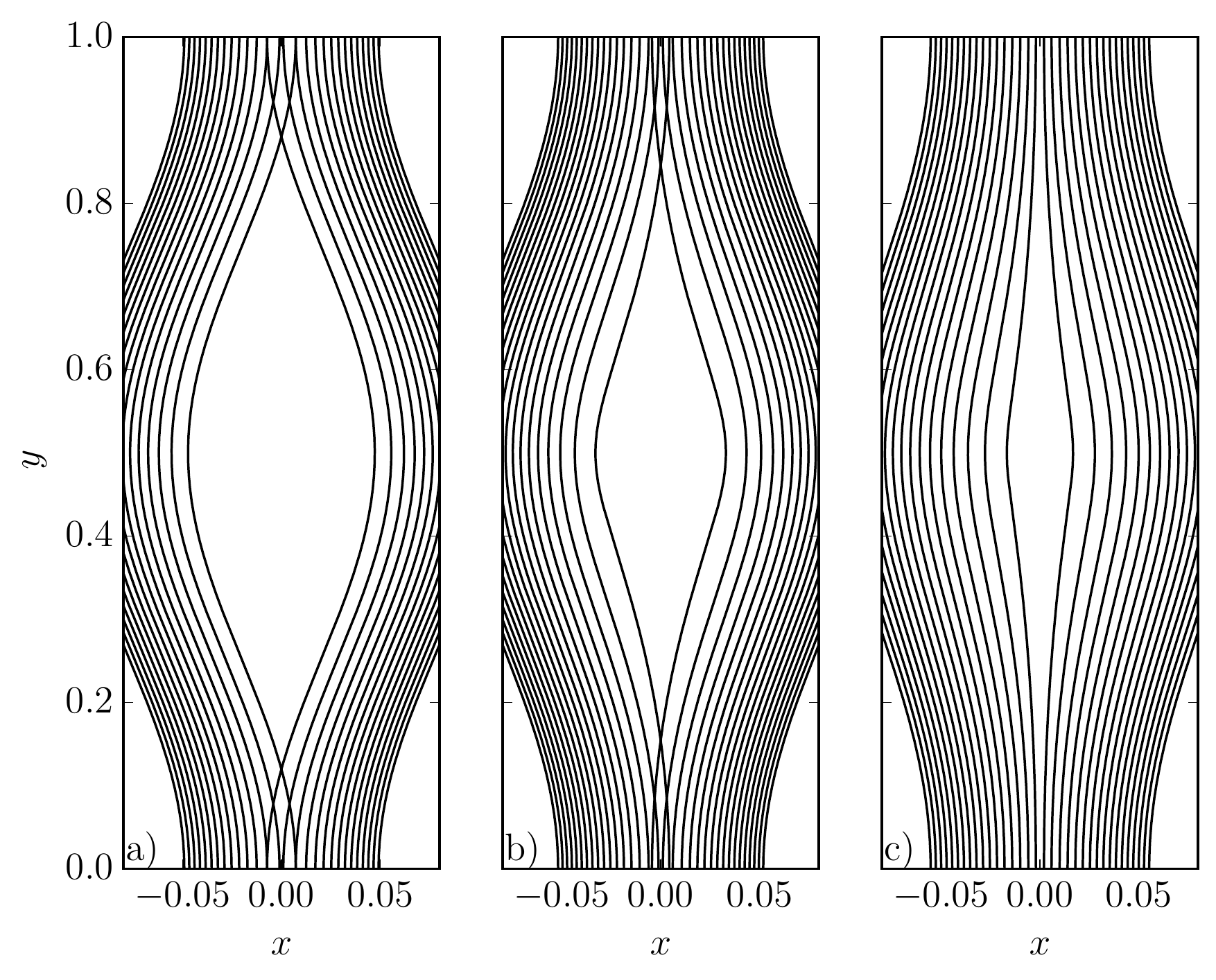}
\caption{\label{alpha} Field line configurations of the linear solution (\ref{linearA}) for different values of $\alpha$: 0 (a), $ka\delta/(4\sinh ka)$ (b), and $ka\delta/\sinh ka$ (c). Parameters used are $a=0.5$, $k=2\pi$, and $\delta=0.1$. Only the field lines for $x_0\in[-0.08,0.08]$ are plotted. The intersections of field lines in (a) and (b) are not physically permissible in ideal MHD. The solution shown in (c) corresponds to the nonlinear solution \eqref{Dewar} as shown in Fig.\,\ref{residual}(c)}
\end{figure}

The linear solution to the modified HKT problem still has the incompressible form of $\bm{\xi}=\nabla\chi\times\hat z$, with $\chi(x_0,y_0)=\bar\chi(x_0)\sin(ky_0)$. However, with $B_{0y}=x_0+\text{sgn}(x_0)\alpha$, the linearized equilibrium equation (\ref{HKE}) becomes
\begin{align}\label{MHKE}
(\partial_{x_0}^2-k^2)[B_{0y}(x_0)\bar\chi]=0.
\end{align}
The two branches of solutions are $\bar\chi\sim (|x_0|+\alpha)^{-1}\sinh kx_0$ and $\bar\chi\sim (|x_0|+\alpha)^{-1}\cosh kx_0$. The boundary condition $\bar\chi(\pm a)=\mp\delta/k$ suggests that the solution is odd in $x_0$, while the latter branch is even. Therefore, similar to Eq.\,(\ref{HKS}), the linear solution only contains the former branch,
\begin{align}\label{linearA}
\bar\chi(x_0)=-\frac{\delta(a+\alpha)\sinh kx_0}{k\left(|x_0|+\alpha\right)\sinh ka}.
\end{align}
This linear solution is continuous for $\alpha>0$, in contrast to the discontinuous solution for $\alpha=0$ as in the ideal HKT problem. However, when the amplitude $\delta$ is finite, the linear solution can still be pathological. That is, when $\alpha$ is not sufficiently large, the field lines can still intersect, as plotted in Fig.\ref{alpha}(b), which is not physically permissible. To eliminate these unphysical intersections, the fluid mapping $x$ mush be not only continuous, but also monotonic in $x_{0}$, $\partial x/\partial x_0\ge0$. Since $x=x_0+\xi_x$, we need $\partial\xi_x/\partial x_0\ge-1$, which means $|k\bar\chi'|\le1$ \citep{Loizu2015b}. Then follows the criterion on $\alpha$ for the linear solution to be well-defined, $\alpha\ge ka\delta/(\sinh ka - k\delta)$. One solution of this kind is shown in Fig.\ref{alpha}(c).

Recall that for solution (\ref{Dewar}) \citep{Dewar2013}, the condition to eliminate the residual islands is $\alpha\ge ka\delta/\sinh ka$, which agrees linearly (when $\delta\ll a$) with the above criterion for the linear solution above to be well-defined. In fact, solution (\ref{Dewar}) agrees linearly with the linear solution (\ref{linearA}) for $B_{0y}=x_0+\text{sgn}(x_0)\alpha$. Interestingly, in \citet{Loizu2015,Loizu2015b}, the numerical solutions can only be obtained for large $\alpha$. Therefore, we speculate that the solutions in \citet{Loizu2015,Loizu2015b} are physically equivalent to solution (\ref{Dewar}). They are both solutions to the modified HKT problem in the  large-$\alpha$ regime, where the finite-amplitude pathology in the linear displacement is absent.

\begin{figure}[h]
\centering
\includegraphics[scale=.75]{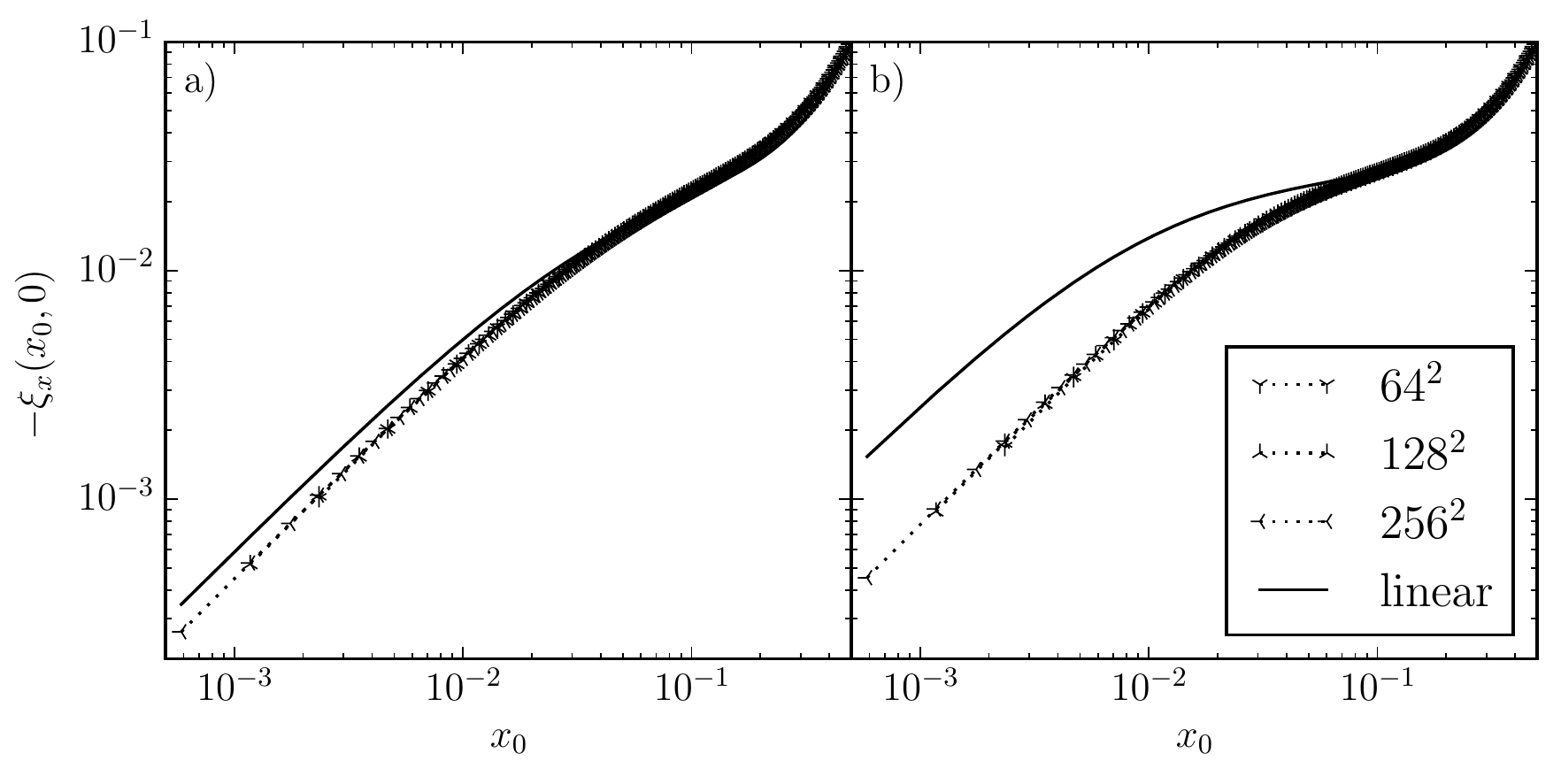}
\caption{\label{mappingxi}For both $\alpha=0.05$ (a) and $\alpha=0.01$ (b), numerical solutions of $-\xi_{x}(x_0,0)$ for different resolutions (dotted lines) converge. The linear solution (\ref{linearA}) (solid line) captures the behavior of the nonlinear solution rather well for (a), but not so for (b).}
\end{figure}

We use the variational integrator to obtain the nonlinear solution to the modified HKT problem. The parameters used are again $a=0.5$, $k=2\pi$, and $\delta=0.1$. 
The solution for $\alpha=0.05>ka\delta/\sinh ka\approx0.027$, when the linear solution is well-defined, is plotted in Fig.\,\ref{mappingxi}(a). The numerical solutions converge, and 
the linear solution offers a good approximation of the true solution. This case corresponds to solution (\ref{Dewar}) and those in \citet{Loizu2015,Loizu2015b}. In contrast, the solution for $\alpha=0.01<ka\delta/\sinh ka$, when the linear solution is not well-defined, is presented in Fig.\,\ref{mappingxi}(b). The numerical solutions still converge, but the linear solution is no longer a good approximation of the true solution. This case is not captured by solution (\ref{Dewar}) and those in \citet{Loizu2015,Loizu2015b}. 

One may compare the converged numerical solutions to the modified HKT problem here with the solutions to the ideal HKT problem that do not converge at the neutral line (Figs.\,\ref{mappingx} and \ref{mappingy}). With $B_{0y}\sim\text{sgn}(x_0)\alpha$ at $x_0=0$, both $\partial x/\partial x_0$ and $\partial y/\partial y_0$ are well-behaved there, and thus convergence is easily achieved. For the ideal HKT problem ($\alpha=0$), the diverging $\partial y/\partial y_0$ is improbable to resolve, resulting in the non-convergence at the neutral line.

This calls for clarification on the contextual difference between the ideal HKT problem (Sec.\,\ref{2D:HKT}) and the modified HKT problem (this section). For the former, we begin with a smooth initial condition ($\alpha=0$), rather than one with discontinuity, in order to observe the emergence of a current singularity. On the contrary, the modified HKT problem, with an initially discontinuous magnetic field ($\alpha>0$), is no longer in the context of current singularity formation. Nonetheless, it is still a problem of interest in the context of finding well-defined ideal MHD equilibria.

pdf\section{The coalescence instability}
\label{2D:coalescence}
The ideal HKT problem is originally designed to study the effect of resonant perturbation on a rational surface in slab geometry. Similarly, it has been shown that the $m=1$ internal kink instability can induce a current singularity at the resonant surface \citep{Rosenbluth1973}. An obvious distinction between the two cases is the drive: the former is subject to external boundary forcing, while the latter is due to an internal instability. For instability driven systems, the linear equilibrium equation $\mathbf{F}(\bm{\xi})=0$ usually does not have non-trivial solutions (except when there exist marginally stable eigenmodes). It is natural to wonder whether such distinction has any effect on the formation of current singularities. 

However, we have not implemented the variational integrator in cylindrical geometry, so we choose to study the coalescence instability instead. Although its magnetic topology is more complex, it can be studied in either slab \citep{Finn1977} or doubly periodic geometry \citep{Longcope1993}. In this section, we only present the results obtained with the doubly periodic configuration, while the conclusions we make are also supported by the results from the slab configuration.

\begin{figure}[h]
\centering
\includegraphics[scale=.75]{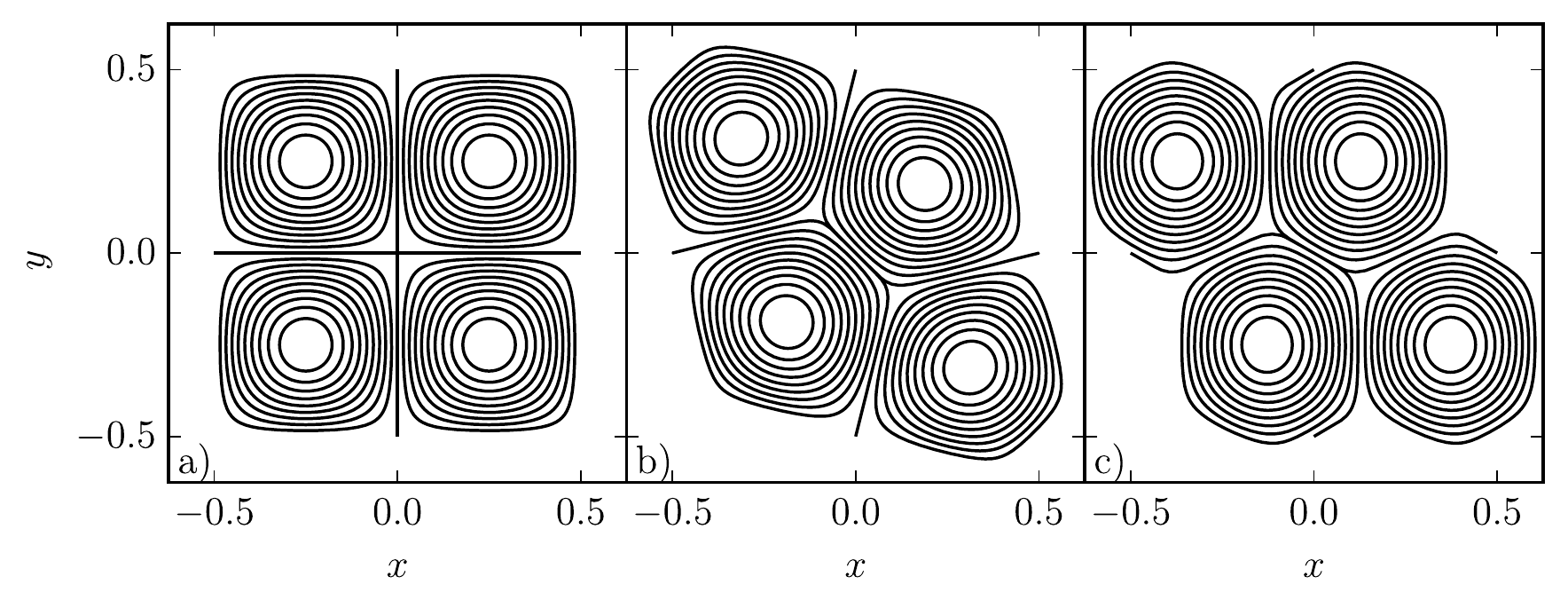}
\caption{\label{CI}Field line configurations, i.e., contours of the flux function $A_0$, of the unstable initial equilibrium (a), the metastable pentagonal equilibrium (b), and the stable hexagonal equilibrium (c).}
\end{figure}

The initial force-free equilibrium is set up in $[-\pi/k,\pi/k]^2$ with $\mathbf{B}_0=\nabla A_0\times\hat z+B_{0z}\hat z$, with the flux function $A_0=\bar A\sin kx_0\sin ky_0$ and $B_{0z}=\sqrt{\bar{B}^2+2k^2A_0^2}$, and $\bar A$ and $\bar{B}$ are constants. As shown in Fig.\,\ref{CI}(a), the equilibrium consists of rectangular arrays of alternately twisted flux tubes. \citet{Longcope1993} showed that the equilibrium is linearly unstable to the attraction between flux tubes with the same helicities. They also presented a numerical simulation of such an instability using an Eulerian method, where the system is observed to approach an intermediate pentagonal equilibrium, then the flux tubes merge due to numerical reconnection.

The parameters used in our simulation are $\rho_0=1$, $k=2\pi$, $\bar A=0.1$, and $\bar{B}=4$. We use a symmetric triangular mesh, with the vertices distributed non-uniformly in both $x$ and $y$, so as to devote more resolution to the separatrices (the X-lines). The initial perturbation is chosen to be $\dot{\mathbf{x}}=-0.001(\sin ky_0,\sin kx_0)$, which is symmetric to the diagonals $y=\pm x$. 
Due to friction, the system relaxes to the pentagonal equilibrium shown in Fig.\,\ref{CI}(b). Interestingly, this equilibrium is still unstable when subject to further perturbations that are not symmetric to the diagonals. In that case, the system ends up in the hexagonal equilibrium shown in Fig.\,\ref{CI}(c), as predicted by \citet{Longcope1993}. This equilibrium is stable.

Nevertheless, we choose to analyze the pentagonal equilibrium for the convenience due to its nice symmetry to the diagonals. The X-point at $(0,0)$ is compressed along $y=x$ and stretched along $y=-x$. Subsequently, we rotate to the normal and tangential coordinates, $x_n=(y-x)/\sqrt{2}$ and $x_t=(y+x)/\sqrt{2}$, respectively. Then follows the normal and tangential components of $\mathbf{B}$, $B_n=(B_y-B_x)/\sqrt{2}$ and $B_t=(B_y+B_x)/\sqrt{2}$.

\begin{figure}
\centering
\includegraphics[scale=.75]{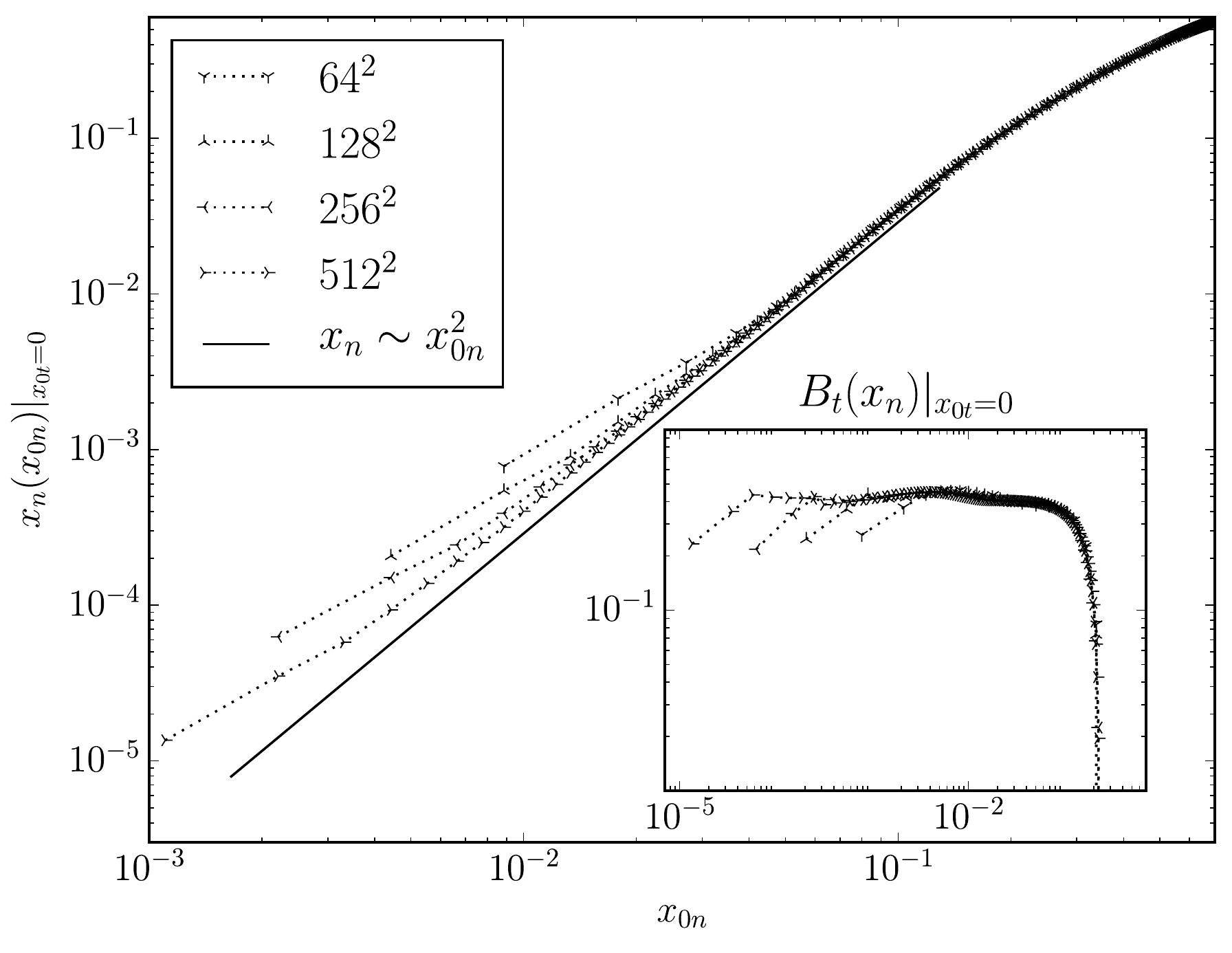}
\caption{\label{mappingc}Numerical solutions of $x_n(x_{0n})|_{x_{0t}=0}$, and $B_t(x_{n})|_{x_{0t}=0}$ (inset) for different resolutions (dotted lines). Near the neutral line, $x_n(x_{0n})|_{x_{0t}=0}$ shows a quadratic power law, while $B_t(x_{n})|_{x_{0t}=0}$ approaches a finite constant. The solutions do not converge for the few vertices closest to the neutral line. Note the similarity between this figure and Fig.\,\ref{mappingx}.}
\end{figure}

Similar to what is done in the ideal HKT problem, we examine how the normal component of the fluid mapping converges. In Fig.\,\ref{mappingc}, the mapping along the diagonal $y=x$, namely $x_n(x_{0n})|_{x_{0t}=0}$, is plotted for increasing resolutions from $64^2$, to $128^2$, $256^2$, and $512^2$. The same quadratic power law as shown in Fig.\,\ref{mappingx}, $x_{n}\sim x_{0n}^2$, can be observed near $x_{0n}=0$, where the solutions converge. Since the initial flux function is approximately hyperbolic near the X-point $(0,0)$, $A_0\sim x_0y_0 \sim x_{0t}^2-x_{0n}^2$, the initial tangential field is close to constantly sheared, $B_{0t}\sim x_{0n}$. Therefore, the equilibrium tangential field $B_{t}\sim B_{0t}/(\partial x_{n}/\partial x_{0n})\sim \text{sgn}(x_{n})$ turns out to be discontinuous, as shown in the inset of Fig.\,\ref{mappingc}. Note that $x_t(x_{0n})|_{x_{0t}=0}=0$ for the pentagonal equilibrium.

The remarkable resemblance between Figs.\,\ref{mappingx} and \ref{mappingc} shows that the current singularities in the ideal HKT problem and the coalescence instability are locally indistinguishable, despite the difference in the drive. Therefore, the recipe for current singularity formation in 2D we find, (constantly) sheared initial field and quadratic fluid mapping, is quite general. From here on, we refer to this mechanism as ``squashing". Whether squashing still functions in 3D line-tied geometry is discussed in Chap.\,\ref{ch:3D}.

\section{Discussion}
\label{2D:discussion}

A straightforward conclusion we can draw from the numerical solutions to both the ideal HKT problem and the coalescence instability is that there exists no smooth equilibrium fluid mapping. However, this does not necessarily leads to the conclusion that there is a genuine current singularity. In the context of studying current singularity formation, one needs to take the further step of differentiating the mapping and confirm the existence of a possible current singularity. This is exactly what we have done in this chapter. 

In previous studies that use similar Lagrangian relaxation methods \citep{Longbottom1998,Craig2005,Craig2005b,Pontin2005,Craig2014,Craig2014b,Candelaresi2015}, current singularities are identified by examining whether the peak current density diverges with increasing spatial resolutions. However, involving second-derivatives, the output of current density is generally less reliable than that of the fluid mapping, especially near the current singularity where the mesh can be highly distorted. Since any singularity in the current density should trace back to that in the more fundamental fluid mapping, we choose to identify current singularities by examining the latter. In this chapter, the current singularities we find originate from the quadratic fluid mapping normal to them. 

More importantly, even if the current density output is reliable, it is not convincing enough to use diverging peak current densities as sole evidences for current singularities. For numerical solutions to be credible, they must converge, with the exception that divergence may be allowed where the solutions are singular. In other words, to numerically demonstrate singularities in solutions, one needs to show not only divergence, but convergence above all. In this section, we show that  the numerical solutions to the ideal HKT problem and the coalescence instability converge to ones with singularities, respectively, except where there are singularities. In this sense, we consider our numerical evidence for current singularity formation in 2D to be the strongest in the extant literature.

Parker's original model considered a uniform field in 3D line-tied geometry that is subject to footpoint motions \citep{Parker1972}. \citet{Zweibel1987} studied the ideal HKT problem as a variation of it, as the sheared field can be realized from a uniform field with sheared footpoint motion. It has also been shown that the initial configuration for coalescence instability can be realized from Parker's uniform field via smooth footpoint motion \citep{Longcope1994}. Therefore, these problems are more closely related to Parker's model than other variations that involve magnetic nulls \citep{Pontin2005,Craig2014,Craig2014b}. Now that we have confirmed that current singularities can form in the 2D ideal HKT problem and coalescence instability, naturally the next step is to find out whether they survive in 3D line-tied geometry. In fact, in \citet{Zweibel1987} and \citet{Longcope1994}, it is conjectured that current singularities would not form in the 3D line-tied ideal HKT problem and coalescence instability, respectively. In the next chapter, we will use the same investigative mindset developed in this chapter, to perform convergence studies on the solutions to these problems and look for the potential current singularities.

\chapter{Current singularity formation in 3D line-tied geometry\label{ch:3D}}
We revisit Parker's conjecture of current singularity formation in 3D line-tied plasmas
using the numerical method developed in Chap.\,\ref{ch:integrator}, variational integration for ideal MHD in Lagrangian labeling. With the frozen-in equation built-in, the method is free of artificial reconnection, hence arguably an optimal tool for studying current singularity formation. In Chap.\,\ref{ch:2D}, the formation of current singularity in the Hahm-Kulsrud-Taylor problem in 2D has been confirmed using this method. In this chapter, we extend this problem to 3D line-tied geometry. The linear solution, which is singular in 2D, is found to be smooth for arbitrary system length. However, with finite amplitude, the linear solution can become pathological when the system is sufficiently long. The nonlinear solutions turn out to be smooth for short systems. Nonetheless, the scaling of converged peak current density vs.~system length suggests that the nonlinear solution may become singular at a finite length. With the results in hand, we can neither confirm nor rule out this possibility conclusively, since we cannot obtain solutions with system length near the extrapolated critical value. The work in this chapter is largely published in \citet{Zhou2017}.

\section{Introduction}
\label{3D:intro}
A long-standing problem in solar physics is why the solar corona, a nearly perfectly conducting plasma where the Lundquist number $S$ can be as high as $10^{14}$, has an anomalously high temperature that conventional Ohmic heating cannot explain. Decades ago, \citet{Parker1972} proposed that convective motions in the photosphere will tend to induce current singularities in the corona, and the subsequent magnetic reconnection events can account for substantial heating. This conjecture has remained controversial to this day \citep{Rosner1982,Parker1983,Parker1994,Tsinganos1984,VanBallegooijen1985,VanBallegooijen1988,Zweibel1985,Zweibel1987,Antiochos1987,Longcope1994b,Ng1998,Longbottom1998,Bogoyavlenskij2000,Craig2005,Low2006,Low2010,Wilmot-Smith2009,Wilmot-Smith2009b,Janse2010,Rappazzo2013,Craig2014,Pontin2015,Candelaresi2015,Pontin2016}.

This controversy fits into the larger context of current singularity formation, which is also a problem of interest in toroidal fusion plasmas \citep{Grad1967,Rosenbluth1973,Hahm1985,Loizu2015b}. However, the solar corona, where magnetic field lines are anchored in the photosphere, is often modeled with the so-called line-tied geometry. This is a crucial difference from toroidal fusion plasmas where closed field lines can exist. For clarification, in this thesis, we refer to the problem of whether current singularities can emerge in 3D line-tied geometry as \textit{the Parker problem}. Readers interested in the historical achievements in studying current singularity formation are referred to Sec.\,\ref{intro:formation}.

Although the Parker problem is inherently dynamical, it is usually treated by examining magnetostatic equilibria for simplicity, as \citet{Parker1972} did in the first place. As discussed in Sec.\,\ref{2D:intro}, with the assumption of perfect conductivity, the equilibrium needs to preserve the magnetic topology of the initial field. Analytically, this constraint is difficult to explicitly attach to the magnetostatic equilibrium equation \eqref{JcrossB}. Numerically, most standard methods for ideal MHD are susceptible to artificial field line reconnection in the presence of (nearly) singular current densities. Either way, to enforce this topological constraint is a major challenge for studying the Parker problem. 

Fortunately, one can overcome this difficulty by adopting Lagrangian labeling, where the frozen-in equation is built into the equilibrium equation, instead of the commonly used Eulerian labeling. \citet{Zweibel1987} first noticed that this makes the mathematical formulation of the Parker problem explicit and well-posed. Moreover, not solving the frozen-in equation numerically avoids the accompanying error and resultant artificial reconnection. A Lagrangian relaxation scheme with this feature has been developed using conventional finite difference \citep{Craig1986}, and extensively used to study the Parker problem  \citep{Longbottom1998,Craig2005,Wilmot-Smith2009,Wilmot-Smith2009b,Craig2014}. 
\citet{Pontin2009} has later found that its current density output can violate charge conservation, and mimetic discretization has been applied to fix it \citep{Candelaresi2014}. 

In Chap.\,\ref{ch:integrator}, a variational integrator for ideal MHD in Lagrangian labeling has been developed by using discrete exterior calculus \citep{Desbrun2005}. Derived in a geometric and field-theoretic manner, it naturally preserves many of the conservation laws of ideal MHD, including charge conservation. Constructed on unstructured meshes, the method allows resolution to be devoted to where it is most needed, such as the vicinity of a potential current singularity. It is arguably an optimal tool for studying current singularity formation.

In Chap.\,\ref{ch:2D}, this method is used to study the Hahm-Kulsrud-Taylor (HKT) problem \citep{Hahm1985}, a fundamental prototype problem for current singularity formation in 2D, where a plasma in a sheared magnetic field is subject to boundary forcing. The formation of current singularity is conclusively confirmed via convergence study, and its signature is also identified in other 2D cases with more complex topology, such as the coalescence instability of magnetic islands discussed in Sec.\,\ref{2D:coalescence}.

In this chapter, we extend the HKT problem to 3D line-tied geometry. \citet{Zweibel1987} showed that the linear solution, which is singular in 2D, should become smooth. This prediction is confirmed by our numerical results. However, we also find that given finite amplitude, the linear solution can be pathological when the system is sufficiently long. We speculate that this finite-amplitude pathology may trigger a finite-length singularity in the nonlinear solution.

We perform convergence study on the nonlinear solutions for varying system length $L$.
For short systems, the nonlinear solutions converge to smooth ones. The peak current density approximately scales with $(L_n-L)^{-1}$, suggesting that the solution may become singular above a finite length $L_n$. However, the solutions for longer systems inherently involve strongly sheared motions, which often lead to mesh distortion in our numerical method. As a result, we cannot obtain solutions for systems with lengths close to $L_n$, and hence cannot conclude whether such a finite-length singularity does exist. Nonetheless, our results are suggestive that current singularity may well survive in this line-tied system, in accordance with the arguments in \citet{Ng1998}.

This chapter is organized as follows. In Sec.\,\ref{3D:Parker} we formulate the Parker problem in Lagrangian labeling, specify the setup in line-tied geometry, and introduce the conventions of reduced MHD. In Secs.\,\ref{3D:linear} and \ref{3D:nonlinear}, we present our linear and nonlinear results, respectively, in 3D line-tied geometry. Discussions follow in Sec.\,\ref{3D:discussion}.

\section{The Parker problem}
\label{3D:Parker}
\citet{Parker1972} originally considered a perfectly conducting plasma magnetized by a uniform field $\mathbf{B}=\hat z$ threaded between two planes at $z=0,L$, which are often referred to as the footpoints. The footpoints are then subject to random motions such that the magnetic field becomes nonuniform. He argued that in general, there exists no smooth equilibrium for the system to relax to, and therefore current singularities must form. This conjecture is based on perturbative analysis of the magnetostatic equilibrium equation \eqref{JcrossB}.
Many of the subsequent works on the Parker problem are performed on this equation as well \citep{Rosner1982,Parker1983,Tsinganos1984,VanBallegooijen1985,Antiochos1987,Bogoyavlenskij2000,Low2006,Low2010,Janse2010}.

As discussed in Sec.\,\ref{2D:intro}, a caveat of this approach is that Eq.\,(\ref{JcrossB}) is usually underdetermined. That is, a given set of boundary conditions may allow for more than one solution to this equation, and additional information is needed to identify a specific one. For the Parker problem, the information is the very constraint to preserve the initial magnetic topology. The implication is, identifying singular solutions to Eq.\,(\ref{JcrossB}) does not necessarily prove Parker's conjecture, since these solutions may not be topologically constrained.

However, this topological constraint is mathematically challenging to explicitly attach to Eq.\,(\ref{JcrossB}) and its solutions \citep{Janse2010,Low2010}. Nonetheless, it can be naturally enforced if one adopts Lagrangian labeling for ideal MHD, instead of Eulerian labeling that is used in Eq.\,(\ref{JcrossB}), as first noticed by \citet{Zweibel1987}.
In Lagrangian labeling, the advection (continuity, adiabatic, and frozen-in) equations \eqref{advection} are built into the ideal MHD Lagrangian \eqref{Lagrangian3}, the Euler-Lagrange (momentum) equation \eqref{momentum3}, and subsequently the equilibrium equation \eqref{equilibrium3}. That is, the solutions to Eq.\,\eqref{equilibrium3} will satisfy not only Eq.\,(\ref{JcrossB}) but automatically the topological constraint implied in the Parker problem, since the initial field configuration $\mathbf{B}_0$ is built into the equation. In contrast, not all solutions to Eq.\,(\ref{JcrossB}) can necessarily be mapped from given initial conditions. 

Thus, the equilibrium equation in Lagrangian labeling offers a more natural and mathematically explicit description for the Parker problem, which simply becomes whether there exist singular solutions to Eq.\,\eqref{equilibrium3}, given certain smooth initial and boundary conditions. If the initial field $\mathbf{B}_0$ is smooth, any singularity in the equilibrium field $\mathbf{B}$ should trace back to that in the fluid mapping $\mathbf{x}(\mathbf{x}_0)$. 

To specify the smooth initial and boundary conditions, Parker's original model can be characterized with a uniform initial field $\mathbf{B}_0=\hat z$ and prescribed smooth footpoint motion $\mathbf{x}_{\perp}(\mathbf{x}_{0\perp})$ at ${z_0=0,L}$ while $z|_{z_0=0,L}=z_0$. The subscript $\perp$ denotes the in-plane components ($x,y$). Certain classes of footpoint motion were considered in \citet{VanBallegooijen1988,Mikic1989,Longbottom1998,Ng1998,Craig2005}, which are referred to as braiding experiments in the recent review by \citet{Wilmot-Smith2015}.

An alternative is to consider a nonuniform $\mathbf{B}_0$, referred to as initially braided field in \citet{Wilmot-Smith2015}, with no-slip footpoints ($\mathbf{x}=\mathbf{x}_{0}$ at $z_0=0,L$). Note that to remain relevant to Parker's original model, the nonuniform $\mathbf{B}_0$ must be realizable from Parker's uniform field via smooth footpoint motion. Examples include the coalescence instability \citep{Longcope1994,Longcope1994b} and the threaded X-point \citep{Craig2014}, but exclude those with magnetic nulls \citep{Pontin2005,Craig2014b}.

We adopt the latter approach for its two advantages. One is reduced computational complexity. More importantly, these initially braided fields are usually extended from 2D cases that are susceptible to current singularity formation \citep{Longcope1993,Craig2005b}. Unlike Parker's original setup, this allows one to focus on the effect of 3D line-tied geometry on current singularity formation. In Secs.\,\ref{3D:linear} and \ref{3D:nonlinear}, we will extend the HKT problem in 2D \citep{Hahm1985}, where current singularity formation is confirmed in Chap.\,\ref{ch:2D}, to 3D line-tied geometry. 

Reduced MHD \citep[RMHD,][]{Strauss1976} is a reduction of MHD in the strong guide field limit that is often used to model the solar corona. \citet{VanBallegooijen1985} first used what essentially is RMHD to study the Parker problem. 
In Eulerian labeling, RMHD approximations include uniform guide field ($B_{z}=1$), removal of $z$ dynamics (${v}_z=0$), and incompressibility ($\nabla\cdot\mathbf{v}=0$). The equilibrium equation becomes
\begin{equation}
\mathbf{B}\cdot\nabla j_z=0,\label{reduced}
\end{equation}
which is obtained from the $z$ component of the curl of Eq.\,(\ref{JcrossB}). Here $\mathbf j = \nabla\times\mathbf B$ is the current density.

Physically, Eq.\,(\ref{reduced}) means that $j_z$ is constant along a field line. In RMHD, every field line is threaded through all $z$. Therefore, the implication for the Parker problem is, if an equilibrium solution contains a current singularity, it must penetrate into the line-tied boundaries. Note that this is a very strong condition that applies to all solutions of Eq.\,(\ref{reduced}), topologically constrained or not.
 
 Translated into Lagrangian labeling, RMHD approximations become $B_{0z}=1$, $z=z_0$, and $J=1$. Following Eq.\,(\ref{frozeninJ}), the in-plane field then reads
  \begin{equation}\label{Bperp}
\mathbf{B}_{\perp}=\frac{\partial\mathbf{x}_{\perp}}{\partial\mathbf{x}_{0\perp}}\cdot\mathbf{B}_{0\perp}+\frac{\partial\mathbf{x}_{\perp}}{\partial z_{0}}.
\end{equation}
The first term on the RHS results from the in-plane motion, while the second term is the projection of the tilted guide field that shows up only in 3D. 

At the line-tied boundaries ($z_0=0,L$), where $\mathbf{x}_{\perp}=\mathbf{x}_{0\perp}$, the $z$ component of the (Eulerian) curl of Eq.\,(\ref{Bperp}) reads
\begin{equation}\label{jz}
j_{z}\hat z=j_{0z}\hat z+\nabla_{\perp}\times\frac{\partial\mathbf{x}_{\perp}}{\partial z_{0}}.
\end{equation}
Here $j_{0z}$ is the initial condition that has to be smooth. That is, for $j_z$ to be (nearly) singular at the footpoints, $(\partial\mathbf{x}_{\perp}/\partial z_{0})|_{z_0=0,L}$ must be (nearly) singularly sheared (note that this is compatible with the line-tied boundary condition). Therefore, we assert that strongly sheared motion is an inherent feature of the Parker problem. 

In this chapter we use the variational integrator derived in Chap.\,\ref{ch:integrator} to study the Parker problem. As previously discussed, an Achilles' heel of our numerical method, and others that solve Eq.\,(\ref{momentum3}) with moving meshes, is exactly its vulnerability to mesh distortion due to strongly sheared motion. Unfortunately, this poses a formidable challenge for our numerical endeavor at the very outset.

Throughout the rest of the chapter, RMHD approximations are adopted unless otherwise noted. We comment that the boundary layers close to the footpoints that are identified in full MHD analysis \citep{Zweibel1985,Zweibel1987,Scheper1999} are precluded in RHMD, which makes current singularity formation even more difficult. Nonetheless, we expect that if a current singularity can emerge in RMHD, it will likely survive in full MHD.

\section{Linear results}
\label{3D:linear}
In Chap.\,\ref{ch:2D}, we investigate the HKT problem, which was originally proposed by Taylor and studied by \citet{Hahm1985} in the context of studying forced magnetic reconnection induced by resonant perturbation on a rational surface. 
It is a fundamental prototype problem considering how a 2D incompressible plasma magnetized by a sheared equilibrium field $B_{0y} = x_0$ responds
to external forcing. Specifically, the perfectly conducting boundaries at $x_0=\pm a$ are deformed sinusoidally into the shapes that $x(\pm a, y_0)=\pm [a-\delta\cos ky(\pm a, y_0)]$. 

\begin{figure}[h]
\centering
\includegraphics[scale=0.75]{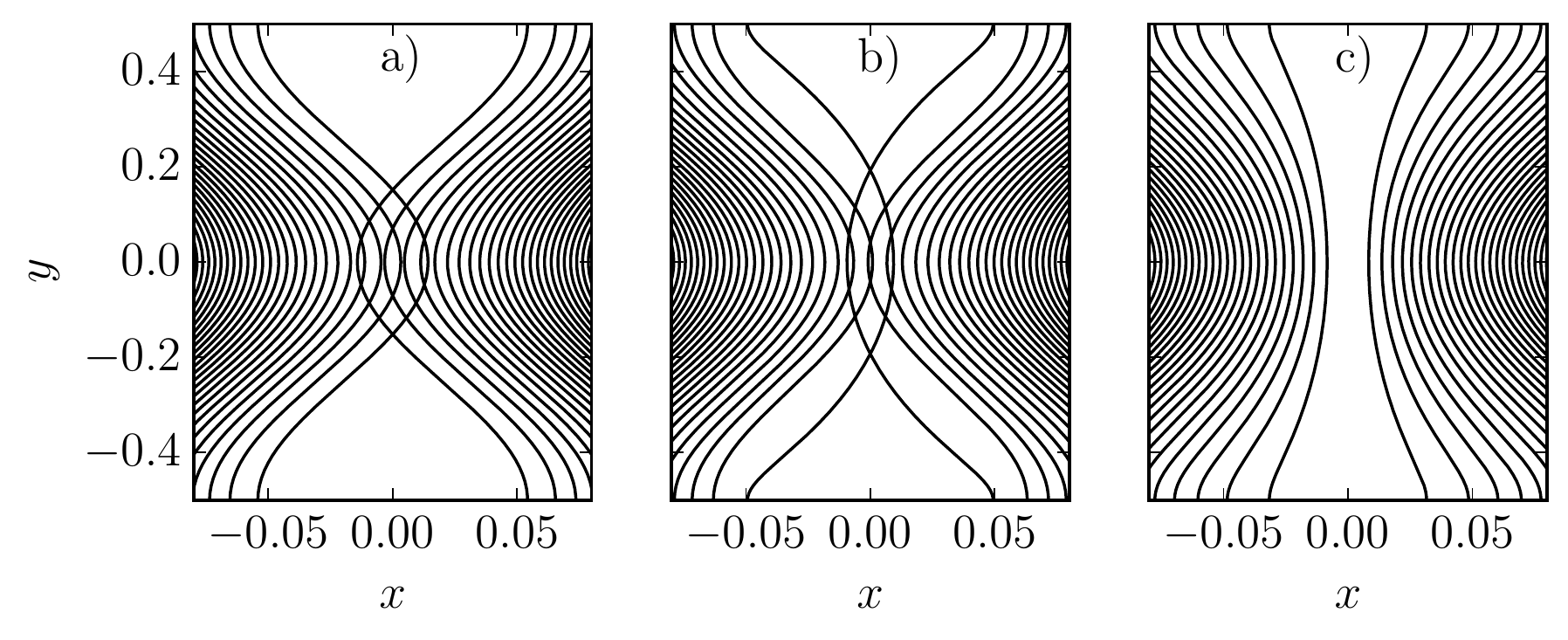}
\caption{\label{pathology}Contours of flux function subject to perturbation with $\delta=0.05$: 2D solution (\ref{HKS}) (a), and 3D solutions at the mid-plane for $L=64$ (b) and $L=16$ (c). The intersection of contours in (a) and (b) are pathological.}
\end{figure}

\citet{Zweibel1987} first connected this problem to the Parker problem, since the sheared initial field is easily realizable from Parker's uniform field via sheared footpoint motion. 
Their linear solution (\ref{HKS}) yields a perturbed magnetic field \eqref{HK}, which contains a current singularity at the neutral line $x_0=0$. This singularity results from the singularity in $\partial\xi_x/\partial{x_0}$. Nonetheless, such a normal discontinuity in the displacement is not physically permissible (see Figs.\,\ref{alpha} and \ref{pathology}, and relevant discussions). The failure at the neutral line is expected from the linear solution since the linear assumption breaks down there. 

In Chap.\,\ref{ch:2D}, we confirm that the nonlinear solution to this problem is singular, using the numerical method described in Chap.\,\ref{ch:integrator}. It is found that the equilibrium fluid mapping normal to the neutral line at $y_0=0$, namely $x(x_0,0)$, converges to a quadratic power law $x\sim  x_0^2$. Due to incompressibility, $(\partial y/\partial y_0)|_{y_0=0}\sim  x_0^{-1}$ diverges at $x_0=0$. With the sheared initial field $B_{0y} = x_0$ substituted into Eq.\,(\ref{Bperp}), such a mapping leads to an equilibrium field $B_y\sim \text{sgn}(x)$ that is discontinuous at the neutral line.

Physically, this means the fluid element at the origin $(0,0)$ is infinitely compressed normally towards, while infinitely stretched tangentially along the neutral line. The exact same signature of current singularity is also identified in other 2D cases with more complex topology, such as the coalescence instability of magnetic islands discussed in Sec.\,\ref{2D:coalescence}. It appears to be a general recipe for current singularity formation in 2D, which we shall refer to as ``squashing" in this chapter. 

In Appx.\,\ref{ch:periodic}, we show that the singularity caused by squashing in the 2D HKT problem survives in 3D periodic geometry. The question then becomes whether squashing works in 3D line-tied geometry. We can learn from Eq.\,(\ref{Bperp}) that squashing is 2D in-plane motion that only contributes to the first term on the RHS. At the footpoints, where in-plane motion is absent and Eq.\,(\ref{jz}) holds, squashing does not work anymore. Consequently, we expect 3D line-tied geometry to have a smoothing effect on current singularities that exist in 2D. Still, we need to find out whether it eliminates these singularities entirely.

For the HKT problem in 2D, the singularity in the linear solution appears to be very suggestive for that in the nonlinear solution. Naturally, when extending the problem to 3D line-tied geometry, we consider the linear solution first.

In 3D line-tied geometry, we modulate the linear displacement on the perfectly conducting boundaries at at $x_0 = \pm a$ into the form of $\xi_x(\pm a,y_0,x_0)=\mp\delta\cos ky_0 \sin(\pi z_0/L)$. The perturbations vanish at the footpoints ($z_0 = 0, L$), consistent with the line-tied (no-slip) boundary condition. Accounting for the initial field $B_{0y}=x_0$, adopting RMHD approximations ($B_{0z}=1$ and $\bm\xi = \nabla \chi\times \hat z $) and Fourier dependence $\chi(x_0,y_0,z_0)=\bar\chi(x_0,z_0)\exp{iky_0}$, the linear equilibrium equation $\mathbf{F}(\bm\xi )=0$ becomes  
\begin{align}\label{linear}
(ikx_0 + \partial_{z_0})(\partial_{x_0}^2-k^2)(ikx_0 + \partial_{z_0})\bar\chi = 0.
\end{align}
When $\partial_{z_0}=0$, the 2D equation \eqref{HKE} and solution \eqref{HKS} can be recovered. 

\begin{figure}[h]
\centering
\includegraphics[scale=0.75]{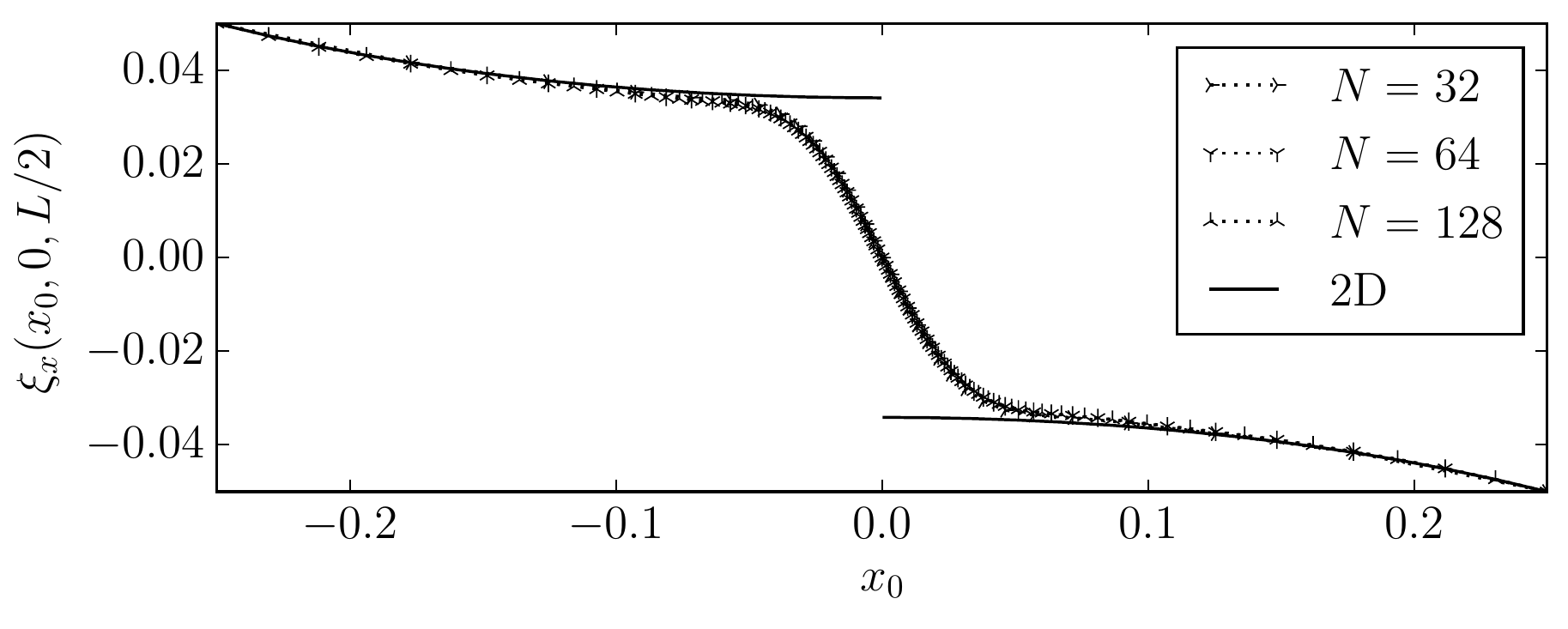}
\caption{\label{convergence}Numerical solutions of $\xi_x(x_0,0,L/2)$ for $L=32$ with $N=32$, $64$, and $128$ converge to a smooth one. The solid line shows the 2D solution(\ref{HKS}) with discontinuity.}
\end{figure}

Eq.\,(\ref{linear}) is solved numerically using second-order finite difference, with boundary conditions $\bar\chi|_{x_0=\pm a}=\pm i(\delta/k)\sin(\pi z_0/L)$ and $\bar\chi|_{z_0=0,L}=0$. The parameters used are $a=0.25$, $k=2\pi$, $\delta=0.05$, with varying $L$ and resolution $N\times NL/8$. For a given $L$, the numerical solutions are found to converge to a smooth one. In Fig.\,\ref{convergence}, $\xi_x(x_0,0,L/2)$ obtained with different resolutions for $L=32$ are shown to converge.

\begin{figure}[h]
\centering
\includegraphics[scale=0.75]{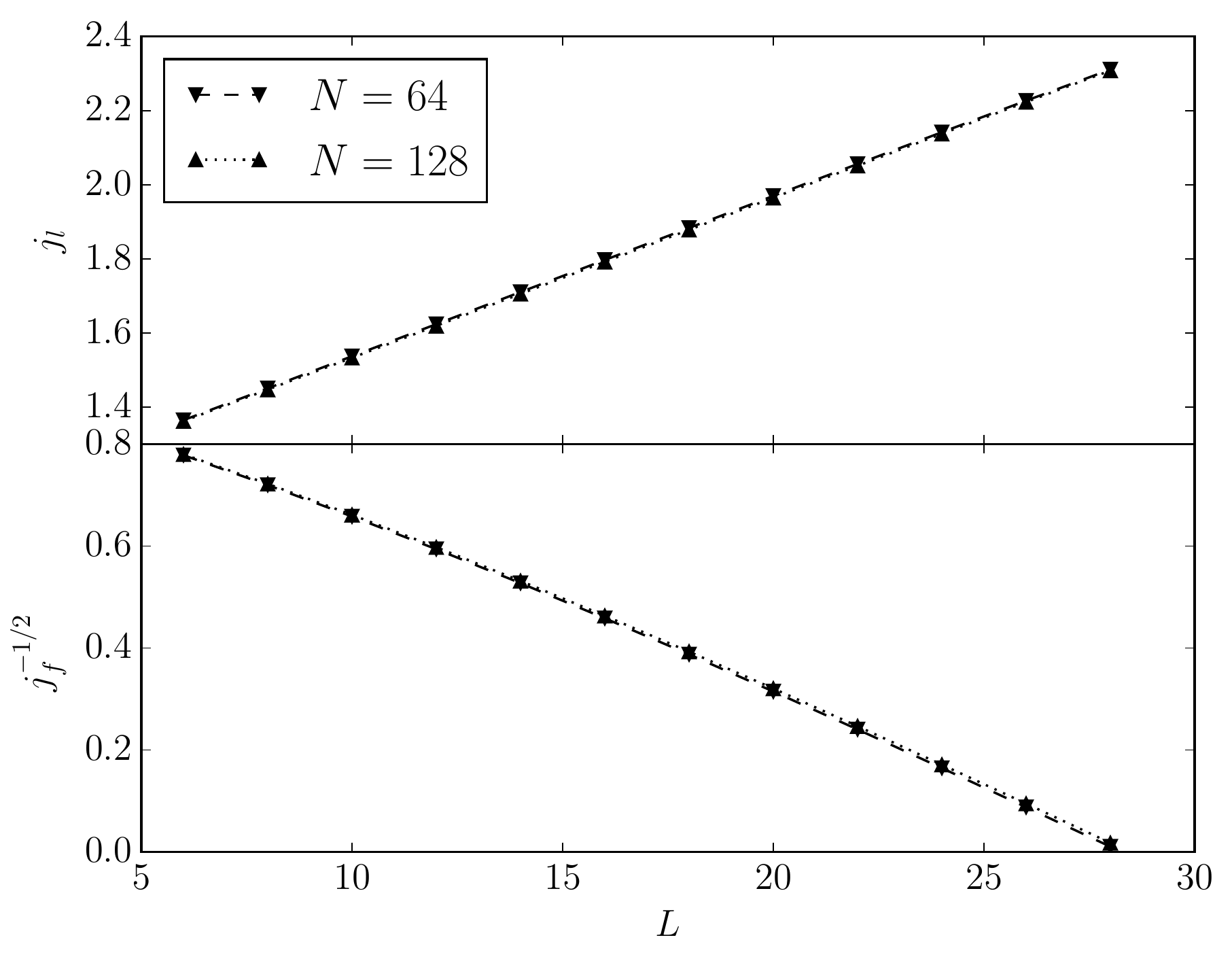}
\caption{\label{deltaj}Maximum of the linearly calculated current density $j_l$ (top), and inverse square root of the maximum of the finite-amplitude current density $j_{f}$ (bottom), vs. system length $L$ for $N=64$ and $128$. $j_l$ increases linearly with $L$, while $j_{f}\sim(L_f-L)^{-2}$ can roughly be observed.}
\end{figure}

We also find that with increasing $L$, $\xi_x(x_0,0,L/2)$ approaches the 2D solution with discontinuity (the solid line in Fig.\,\ref{convergence}) as its gradient at $x_0 = 0$ steepens. Accordingly, the maximum of the linearly calculated current density $j_{0z}+\delta j_z$, where $j_{0z}=1$ and $\delta j_z$ is the perturbed current density, is shown to increase linearly with $L$ in Fig.\,\ref{deltaj} (labeled $j_l$). This suggests that the linear solution is smooth for arbitrary $L$, only becoming singular when $L=\infty$. These results are consistent with the 3D linear analysis by \citet{Zweibel1987}.

So far, all the calculations have been strictly linear, assuming the amplitude of the perturbation $\delta$ to be infinitesimal. The linear solutions, be they $\xi_x$ or $\delta j_z$, are proportional to $\delta$. The magnitude of $\delta$ has no physical impact in this context. The finite amplitude of the perturbation must be accounted for in the fully nonlinear study. But before that, we further exploit the linear solutions by considering the effect of finite amplitude on them.

It is mentioned above that for the HKT problem in 2D, the discontinuity in the linear solution (\ref{HKS}) is not physically permissible. Here we show that given finite amplitude, the linear solution in 3D exhibits similar pathology when the system is sufficiently long. Similar finite-amplitude pathology is also discussed in Sec.\,\ref{2D:modified}.

In Fig.\,\ref{pathology}(a), it is shown that the flux surfaces (constant surfaces of flux function $\psi_0=x_0^2/2$) overlap when they are subject to the perturbed fluid mapping $\mathbf x =\mathbf x_0+\bm\xi$, with $\bm\xi$ given by the 2D linear solution (\ref{HKS}). The cause for this unphysical overlap is that the mapping $x(x_0,0)$ is non-monotonic: $\partial x/\partial x_0=1+\partial \xi_x/\partial x_0<0$ at $(0,0)$, because $ \xi_x( x_0,0)\sim-\text{sgn}(x_0)$. In this case, $\mathbf x$ becomes pathological since its Jacobian $J$ is no longer everywhere positive.

However, in order for the perturbed mapping $x(x_0)$ to be non-monotonic, it is not necessary that $\partial \xi_x/\partial x_0$ is singular as in the 2D solution. The requirement is $\partial \xi_x/\partial x_0<-1$, i.e., the gradient of the displacement is steep enough. For the linear solutions $\xi_x(x_0,0,L/2)$ that are obtained above, this can be achieved when the system is sufficiently long.

As shown in Fig.\,\ref{pathology}(b), the flux surfaces can indeed overlap at the mid-plane when subject to the perturbed fluid mapping given by the 3D linear solution for a long system. When the system length is relatively short, the pathology is absent and the flux surfaces do not overlap, as is shown in Fig.\,\ref{pathology}(c).

Another way to interpret this pathology is to calculate ``nonlinearly'' the magnetic field that the fluid mapping $\mathbf x =\mathbf x_0+\bm\xi$ maps into, using Eq.\,(\ref{frozenin}), and the current density $j_z$ thereafter, which we refer to as the finite-amplitude current density in this chapter. We perform such calculation using the linear solutions obtained (with $\delta=0.05$) above, and notice that the finite-amplitude current density peaks at ${(0,0,L/2)}$. As shown in Fig.\,\ref{deltaj}, the maximum $j_{f}\sim(L_f-L)^{-2}$, which diverges at a critical length $L_f$. When $L>L_f$, the perturbed fluid mapping becomes pathological. The value $L_f\approx28.96$ (using solutions with $N=128$ for fitting) depends on the specific parameters we obtain the linear solutions with, $\delta$ in particular.
It is worthwhile to emphasize that in 2D, $\partial \xi_x/\partial x_0$ is singular, so the pathology exists for infinitesimal amplitude. In 3D line-tied geometry, $\partial \xi_x/\partial x_0$ is smooth, and finite amplitude is required to trigger the pathology at a critical length.



Recall that $\partial x/\partial x_0=0$, the trigger for the pathology, is also a signature of current singularity that is identified in the 2D HKT problem. Interestingly, \citet{Loizu2015b} have also linked similar finite-amplitude pathology of the linear solution to the existence of current singularity in 3D equilibria. We therefore suspect that the nonlinear solution to the line-tied HKT problem may be singular above a finite length, which is presumably comparable to the critical length $L_f$ for the finite-amplitude pathology of the linear solution. Also, we expect the maximum current density of the nonlinear solution to be bounded between $j_l$ and $j_f$. We investigate whether our nonlinear results support these speculations in Sec.\,\ref{3D:nonlinear}.

\section{Nonlinear results}\label{3D:nonlinear}
We solve the line-tied HKT problem numerically using the method described in Chap.\,\ref{ch:integrator}, in a domain of $[-a,a]\times[-\pi/k,\pi/k]\times[0,L]$. The update scheme in 3D is specified in Appx.\,\ref{scheme:3D}. The perfectly conducting walls
at $x_0=\pm a$ re deformed into to the shapes that $x(\pm a, y_0,z_0) = \pm [a - \delta \cos ky(\pm a, y_0,z_0)\sin(\pi z_0/L)]$. The boundary conditions in $y$ and $z$ are periodic and no-slip, respectively. We use the same parameters as used in the linear study, namely $a=0.25$, $k=2\pi$, $\delta=0.05$, with varying $L$.

In addition, we adopt RMHD approximations described in Sec.\,\ref{3D:Parker}, by setting $B_{0z}=1$ and $z=z_0$. For the sake of numerical practicality, we use moderate pressure to approximate incompressibility, instead of enforcing the constraint $J=1$. After all, incompressibility itself is an approximation. Specifically, we initialize with $p_0=0.1-x_0^2/2$ to balance the sheared field $B_{0y}=x_0$, and choose $\gamma=5/3$. In our numerical solutions, we find $|J-1|\lesssim 0.02$, which decreases as resolution
increases. This suggests that our solutions are
very close to incompressible.

A consequence of approximating $J=1$ is that Eq.\,(\ref{reduced}) does not hold anymore, but instead we have $\mathbf{B}\cdot\nabla j_z=\mathbf{j}\cdot\nabla B_z$ with $B_z=1/J$. Since the system is symmetric under rotation by $\pi$ with respect to the $z$ axis ($x,y=0$, the field line of interest in this problem), one finds that $\mathbf{B}_{\perp}=\mathbf{j}_{\perp}=0$, and therefore $j_z(z)=j_z(0)/J(z)$, along the $z$ axis. So in our solutions $j_z(0,0,z)$ should still be approximately constant.

\begin{figure}[h]
\centering
\includegraphics[scale=0.75]{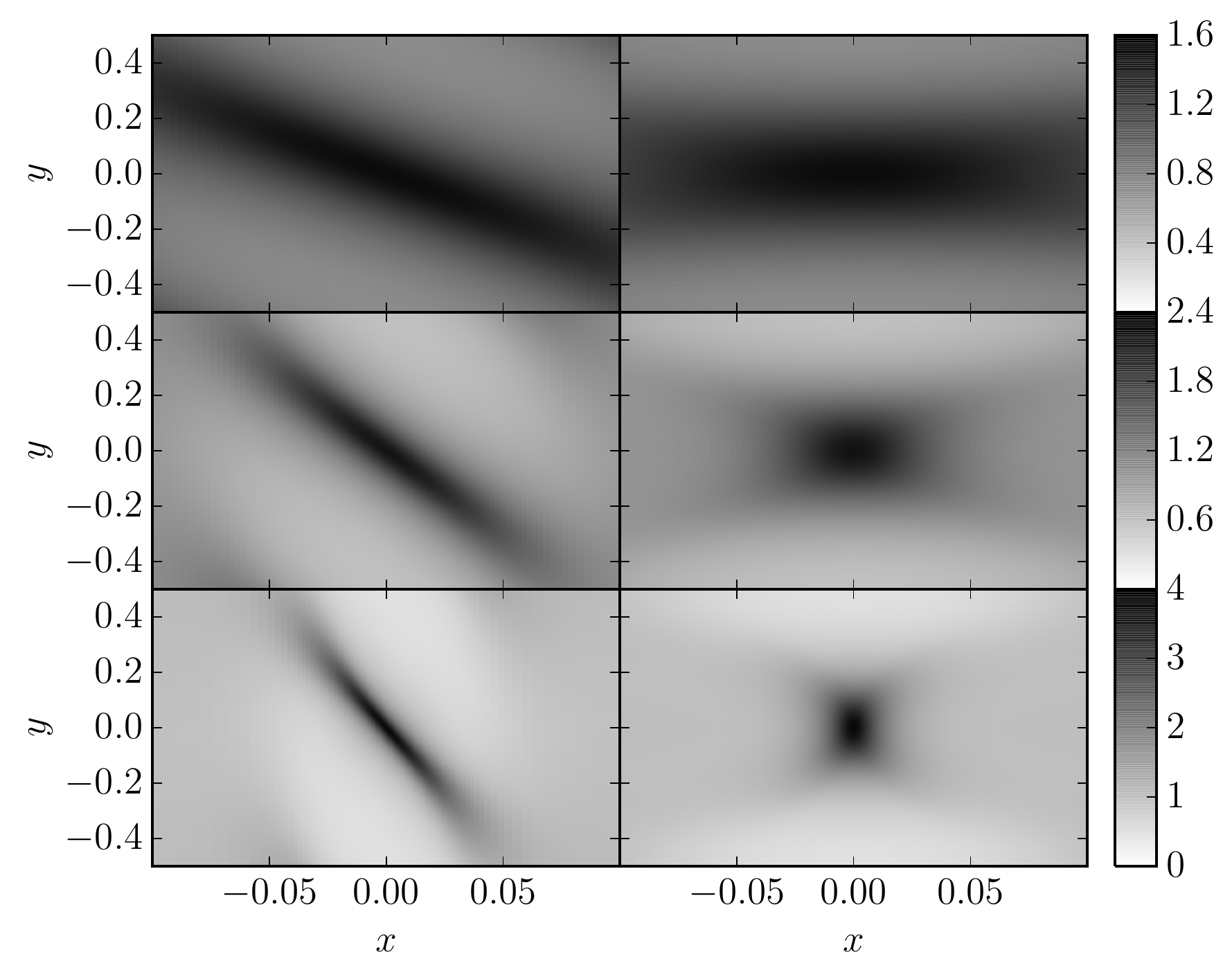}
\caption{\label{current}Distribution of current density $j_z(x,y)$ obtained with $N=160$ for $L=6$, $12$, and $18$ (from top to bottom, respectively) at $z=0$ (left) and $z=L/2$ (right).}
\end{figure}

We use a tetrahedral mesh where the vertices are arranged in a structured manner with resolution $N\times2N\times NL/4$. The grid number in $z$ varies with $L$ so that the grid size does not. The vertices are non-uniformly distributed in $x$ and $y$ to devote more resolution near the $z$ axis. We use a same profile of mesh packing for a given $L$, but adjust it accordingly when $L$ varies.
Similar to the 2D case in Chap.\,\ref{2D:HKT}, the system starts from a smoothly perturbed configuration consistent with the boundary conditions and relaxes to equilibrium due to friction. In Fig.\,\ref{current}, the equilibrium current density distributions obtained with $N=160$ for $L=6$, $12$, and $18$ are shown. Despite that the distributions become significantly more concentrated to the $z$ axis with increasing $L$, all these solutions turn out to be smooth and well-resolved, as our convergence study shows. 

\begin{figure}[h]
\centering
\includegraphics[scale=0.75]{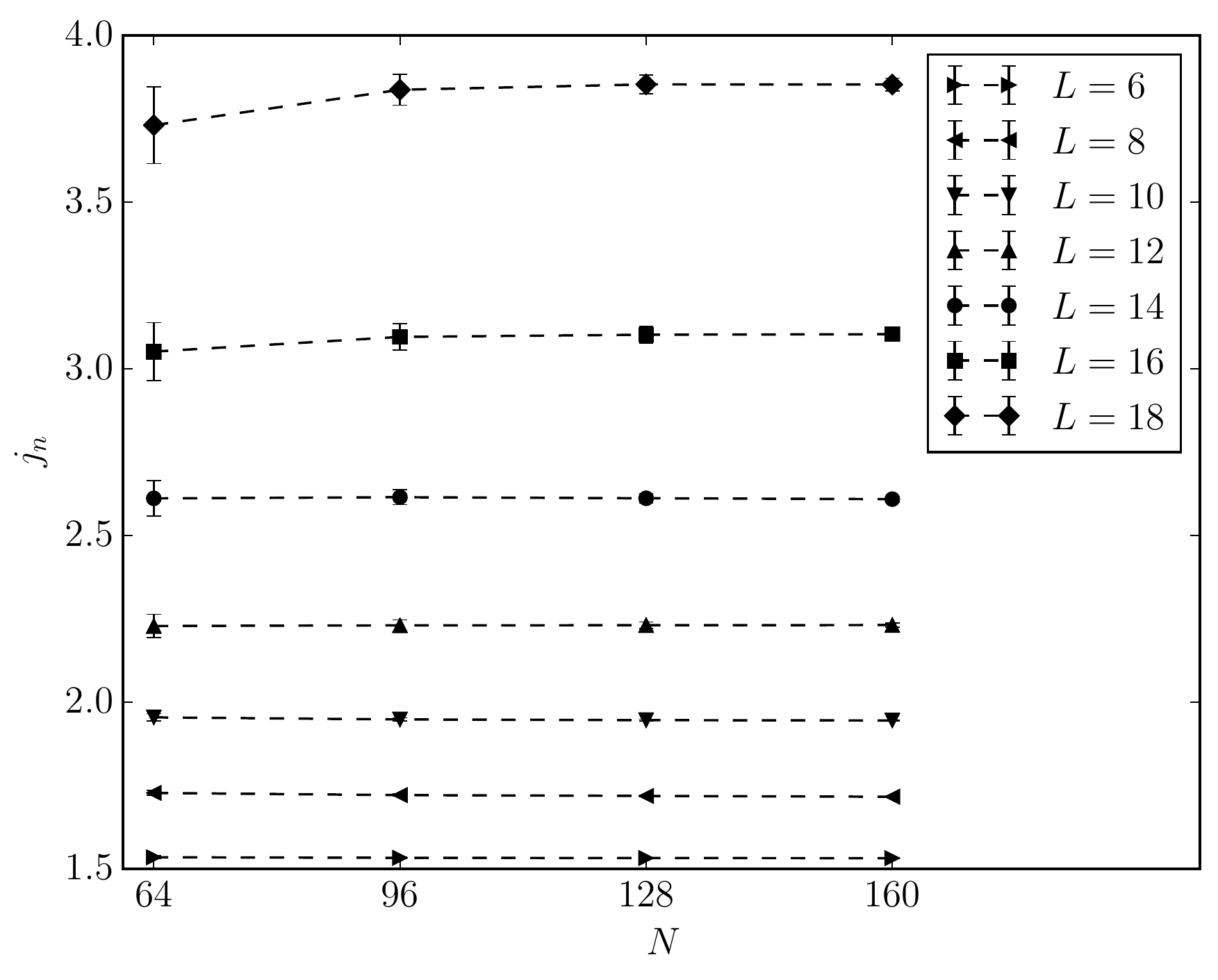}
\caption{\label{resolution}The means ($j_n$) and the standard deviations (error bars) of the current density $j_z(0,0,z)$ for varying system length $L$ are shown to converge with increasing resolution $N$.}
\end{figure}

In Fig.\,\ref{resolution}, the means of $j_z(0,0,z)$ (labeled $j_n$) for varying $L$ are shown to converge with increasing resolution $N$. In addition, the standard deviations of $j_z(0,0,z)$, as shown by the error bars, decrease with increasing $N$. That is, $j_z(0,0,z)$ is indeed approximately constant, demonstrating that our almost incompressible solutions capture the features of the exact RMHD solutions very well. We therefore conclude that the nonlinear solutions for these relatively short systems are smooth.

Another observation from Fig.\,\ref{resolution} is that the standard deviation of $j_z(0,0,z)$ increases with $L$. The reason is, for longer systems, the footpoints are more difficult to resolve than the mid-plane. At the footpoints, there is no in-plane motion, which means the mesh there stays as initially prescribed. Meanwhile, as $L$ increases, the mid-plane bears more resemblance with the 2D case, where the squashing effect spontaneously packs the mesh near the $z$ axis. In fact, in the simulations, we need to pack the mesh more aggressively near the footpoints than the mid-plane, particularly for longer systems, in order to compensate for the self-packing near the mid-plane. To sum up, longer systems are simply much more challenging to resolve computationally than shorter ones.



\begin{figure}[h]
\centering
\includegraphics[scale=0.75]{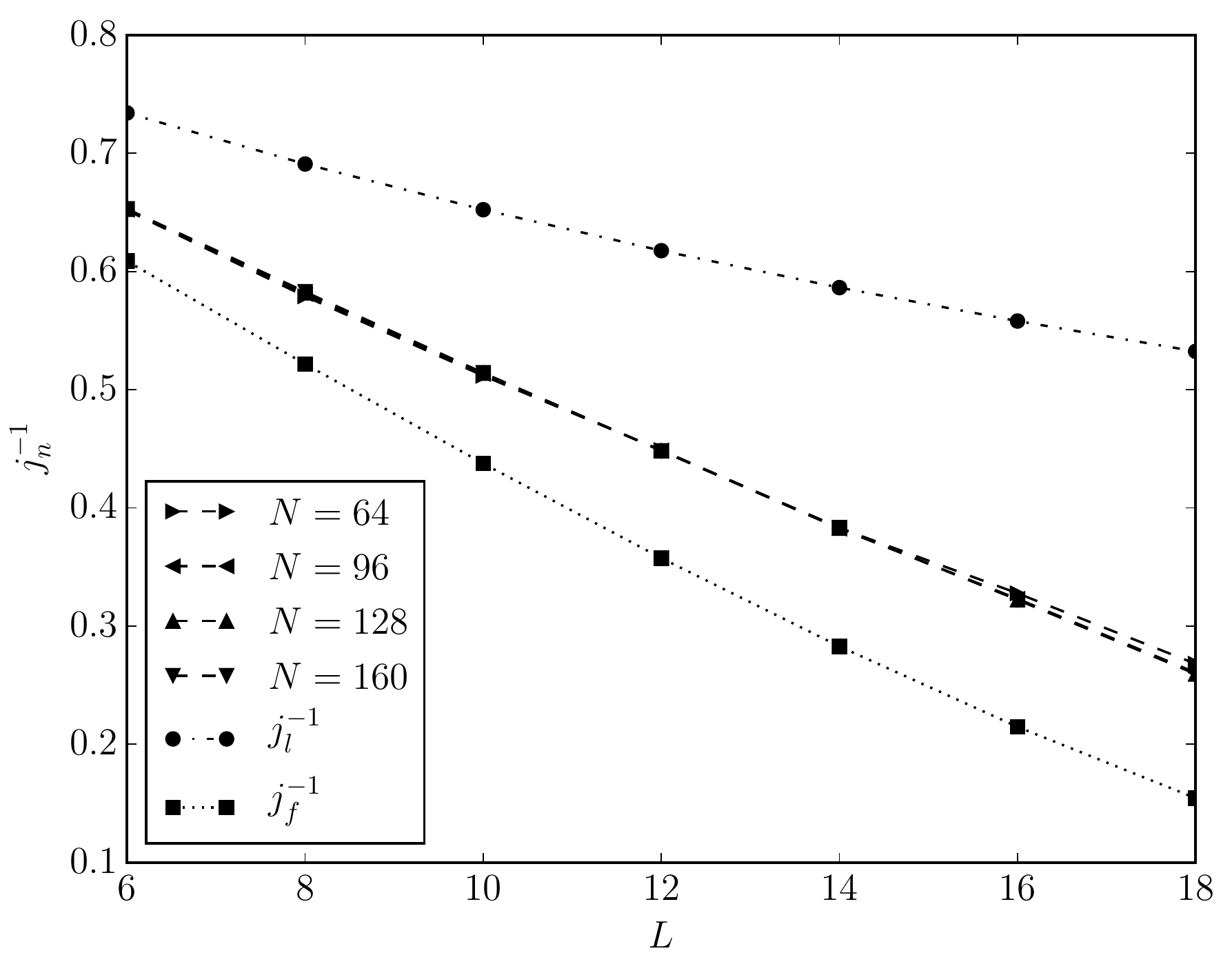}
\caption{\label{length}Inverse of the mean ($j_n$) of the current density $j_z(0,0,z)$ (dashed lines) with varying resolution $N$ vs.\,system length $L$. $j_n\sim (L_n-L)^{-1}$ can roughly be observed. $j_l$ (dash-dot line) and $j_f$ (dotted line) are also shown for comparison (solutions with $N=128$ from Fig.\,\ref{deltaj} are used).}
\end{figure}

Still, one wonders whether the nonlinear solution is smooth for arbitrary $L$. Fig.\,\ref{length} shows that $j_n^{-1}$ decreases roughly linearly with $L$. That is, $j_n\sim (L_n-L)^{-1}$, which diverges at a critical length $L_n$. This suggests that the nonlinear solution may become singular at a finite length. Using the solutions with $N=160$ for fitting, we obtain $L_n\approx25.81$, which is comparable to the critical length $L_f$ for the finite-amplitude pathology discussed in Sec.\,\ref{3D:linear}. Fig.\,\ref{length} also shows that $j_n$ is indeed bounded between $j_l$ and $j_f$, as expected.

In order to validate such a diverging scaling law and confirm the existence of the finite-length singularity, we should examine the solutions for systems with lengths close to or above the critical value $L_n$. Unfortunately, we are not able to obtain (converged) equilibrium solutions for systems with $L=20$ or higher: as $j_z(0,0,z)$ increases with $L$, the motion near the footpoints becomes more strongly sheared, eventually leading to mesh distortion, as discussed in Sec.\,\ref{3D:Parker}. As $L$ increases, $j_z(0,0,z)$ may indeed diverge at a finite length, or convert to a different scaling law that stays well-defined for arbitrary $L$. With the results in hand, we cannot confirm or rule out either possibility conclusively.

\section{Discussion}\label{3D:discussion}
One conclusion we can indeed draw from our results is that 3D line-tied geometry does have a smoothing effect on the current singularity in the 2D HKT problem. In 2D, both the linear and nonlinear solutions are singular. In 3D line-tied geometry, the linear solution is smooth for arbitrary system length; the nonlinear solution is smooth when the system is short. 

Whether the nonlinear solution becomes singular at a finite system length remains yet to be confirmed. Our numerical results show that the maximum current density scales with $(L_n-L)^{-1}$, which implies finite-length singularity. However, since we cannot obtain numerical solutions to validate such a scaling law near the critical value $L_n$, these results can only be considered suggestive, but not conclusive. Nonetheless, we remark that this scaling law is already stronger than the exponential scaling predicted by \citet{Longcope1994b}.

In this chapter, we have prescribed what we believe is an effective formula for realizing possible current singularities in 3D line-tied geometry. The idea is to extend a 2D case with singularity to 3D line-tied geometry, and then make the system really long. In particular, the HKT problem is arguably a simplest prototype, for it captures how a sheared field responds to squashing, both ingredients ubiquitous in nature. Also, the finite-amplitude pathology in its linear solution may be suggestive for the possible finite-length singularity in the nonlinear solution.

The results of the HKT problem can also be suggestive for other cases with more complex magnetic topologies, such as the internal kink instability \citep{Rosenbluth1973,Huang2006} and the coalescence instability \citep{Longcope1993,Longcope1994,Longcope1994b}, the 2D case of which is discussed in Sec.\,\ref{2D:coalescence}. The obvious distinction between the HKT problem and these cases is the former is externally forced, while the latter are instability driven. A subtlety is, for the instability driven cases, the linear equilibrium equation $\mathbf{F}(\bm{\xi})=0$ usually has no nontrivial solutions. In these cases, (fastest-growing) eigenmodes are usually considered to as linear solutions. In addition, eigenmodes do not have intrinsic amplitudes, unlike in the HKT problem where the linear equilibrium solution can reasonably be given the finite amplitude of the boundary forcing. Consequently, the linear solutions in the instability driven cases can be less suggestive for the nonlinear ones. 

Nonetheless, in Sec.\,\ref{2D:coalescence}, it is shown that the current singularity that emerges in the 2D coalescence instability is locally indistinguishable from the one in the 2D HKT problem. Moreover, critical lengths still exist in 3D line-tied geometry for the instability driven cases. That is, these systems are unstable only with lengths above certain finite values \citep{Longcope1994,Longcope1994b,Huang2006}. In fact, \citet{Ng1998} argued that current singularities must emerge when these line-tied systems become unstable. 

Still, what prevents us from obtaining more conclusive results is the limitation of our numerical method, namely its vulnerability to mesh distortion caused by strongly sheared motion. There are a few remedies that are worth investigating. One option is to enforce incompressibility ($J=1$), since the signature of mesh distortion is $J$ becoming negative. However, naively enforcing this constraint means implicitly solving a global nonlinear equation at every time step, which is not practical. What might be a better option is to solve for pressure from a Poisson-like equation, yet that could still be expensive on an unstructured mesh. More importantly, when the motion becomes too strongly sheared for the mesh to resolve, enforcing incompressibility may just not be enough.

An alternative is to employ adaptive mesh refinement. Intuitively, that means to divide a simplex into smaller ones once its deformation reaches a certain threshold. This approach will not work for problems with strong background shear flows where the number of simplices can grow exponentially, but may suffice for the Parker problem that is quasi-static. In addition, one may consider more delicate discretization of Eq.\,(\ref{momentum3}) to make the mesh itself more robust against shear flow.

Finally, we emphasize that the Parker problem is still open and of practical relevance, by echoing the latest review by \citet{Zweibel2016}: \textit{``It is important to determine whether the equilibrium of line-tied magnetic fields has true current singularities or merely very large and intermittent currents, to characterize the statistical properties of the sheets and to determine how the equilibrium level and spatial and temporal intermittency of energy release depend on $S$."}

\appendix 
\chapter{On van Ballegooijen's argument\label{ch:CLS}}
In Sec.\,\ref{intro:formation}, the argument by \citet{VanBallegooijen1985} that in line-tied systems, current singularities are only possible when the footpoint motions are discontinuous,  is briefly mentioned. In this appendix, we counter this argument by showing one of its variants due to \citet{Cowley1997} [see also \citet{VanBallegooijen1988,Longcope1994b}] does not stand, if one properly considers the effect of a finite shear in the footpoint motions, or equivalently, the magnetic field.

\begin{figure}[h]
\centering
\includegraphics[scale=0.75]{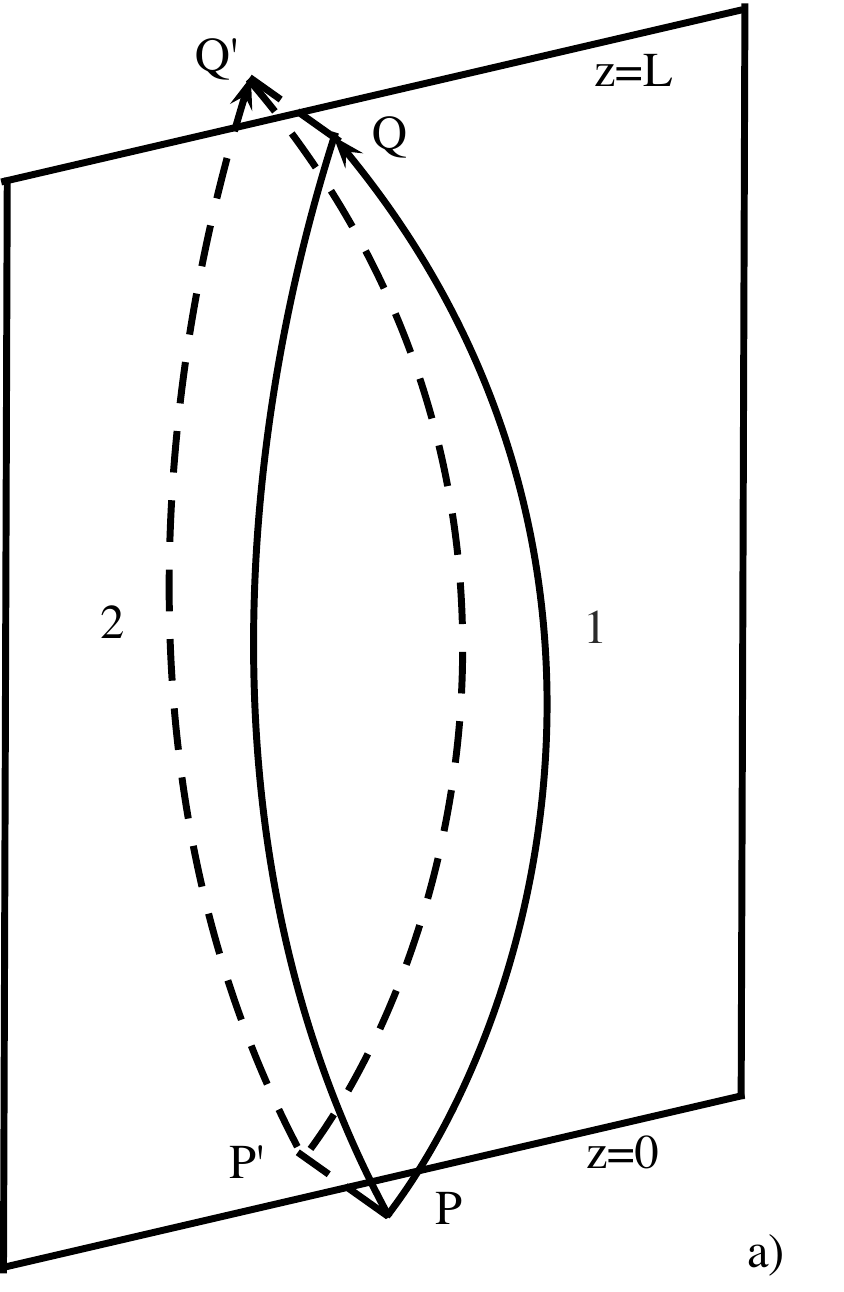}
\includegraphics[scale=0.75]{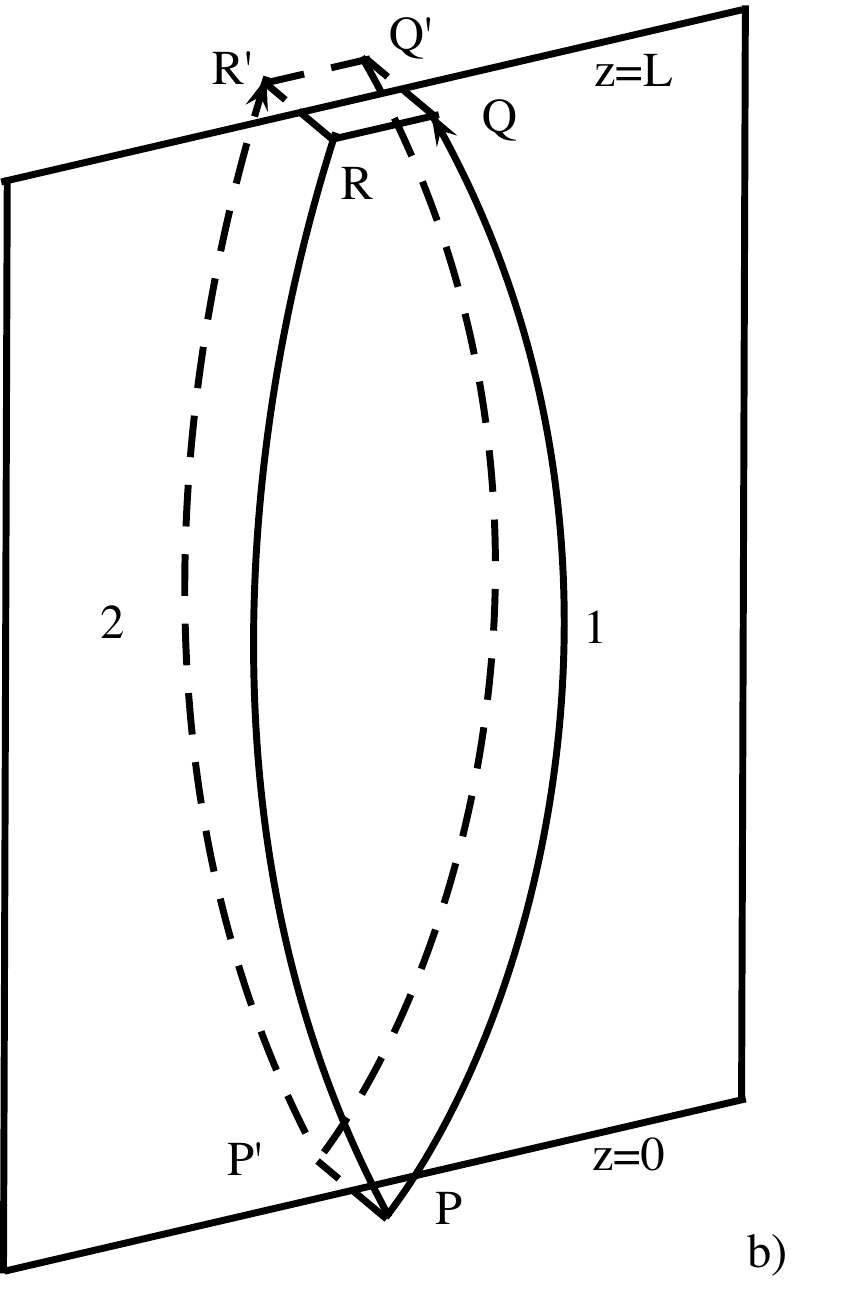}
\caption{\label{CLS}Illustrations for (a) a variant of van Ballegooijen's argument by \citet{Cowley1997} and (b) our counter argument to it with a finite shear in the magnetic field properly considered.}
\end{figure}

We first review the argument by \citet{Cowley1997}, which is based on force-free MHD. For the purpose of showing a contradiction, they assume that a singular current sheet exists, as shown in Fig.\,\ref{CLS}a. On the top of, yet infinitesimally close to the sheet, a magnetic field line starting from a point $P$ at $z=0$ connects to a point $Q$ at $z=L$. Label this field line 1. On the bottom of the sheet, another field line 2 connects a point $P'$, which is infinitesimally close to $P$, to a point $Q'$ at $z=L$. With the footpoint motions assumed to be smooth, $Q'$ must be infinitesimally close to $Q$ as well.

In a force-free equilibrium, the current density is parallel to the magnetic field, so that the current flows entirely along the current sheet and never leaves it \citep{Cowley1997}. Therefore, we have $\oint \mathbf{B}\cdot\mathrm{d}\mathbf{l}=0$ for any loop that lies on top or on the bottom of the sheet. For the path integral around the closed loop on the top of the sheet extending from P to Q along 1 and from Q to P along 2, as shown by the solid curves in Fig.\,\ref{CLS}a, we have
\begin{align}\label{top}
\oint \mathbf{B}\cdot\mathrm{d}\mathbf{l}=\int_1B_1(l)\,\mathrm{d}l-\int_2\mathbf{B}_1(l)\cdot\mathbf{b}_2(l)\,\mathrm{d}l=0.
\end{align}
Similarly, the path on the bottom of the sheet extending from $P'$ to $Q'$ along 2 and back to $P'$ along 1, as shown by the dashed curves in Fig.\,\ref{CLS}a, yields
\begin{align}\label{bottom}
\oint \mathbf{B}\cdot\mathrm{d}\mathbf{l}=\int_2B_2(l)\,\mathrm{d}l-\int_1\mathbf{B}_2(l)\cdot\mathbf{b}_1(l)\,\mathrm{d}l=0.
\end{align}
Here $\mathbf{B}_1$ and $\mathbf{B}_2$ are the magnetic fields on the top and the bottom of
the sheet, while $\mathbf{b}_1$ and $\mathbf{b}_2$ are the corresponding unit vectors, respectively.

The magnitude of the magnetic field must be continuous across the current sheet, that is, $B_1(l)=B_2(l)=B(l)$ \citep{Cowley1997}. We combine Eqs.\,\eqref{top} and \eqref{bottom} to obtain
\begin{align}\label{total}
\int_{1+2}B(l)\left[1-\mathbf{b}_2(l)\cdot\mathbf{b}_1(l)\right]\,\mathrm{d}l=0.
\end{align}
This requires the unit vectors $\mathbf{b}_1$ and $\mathbf{b}_2$ to coincide, which means that the current sheet has no net current, contradicting the assumption of the singular current sheet.

\citet{Ng1998} remarked that the argument given above relies strongly on the assumptions regarding the geometry of the current sheet. They demonstrated that the argument fails for a current sheet with a geometry less trivial than the simple one shown in Fig.\,\ref{CLS}a. Nonetheless, in this appendix, we show that even with such a simple geometry assumed, the above argument by \citet{Cowley1997} still does not stand.

A crucial assumption implied in the above argument is that the footpoints of field line 2, $P'$ and $Q'$, are the mirror images of the footpoints of field line 1, $P$ and $Q$, with regard to the current sheet. That is, $PP'$ and $QQ'$ are both orthogonal to the current sheet. This is almost always not true. More generally, the shear in the footpoint motions, or equivalently, the magnetic field, would induce separations tangentially along the sheet as well. The line-tied HKT problem considered in Chap.\,\ref{ch:3D}, among many others, is exactly a case of this kind.

We treat the effect of the finite shear more carefully in Fig.\,\ref{CLS}b. On the top of the sheet, the magnetic field line 1 still connects $P$ to $Q$. On the bottom of the sheet, we now choose $P'$ to be the mirror image of $P$ with regard to the sheet. The field line 2 starting from $P'$ connects to $R'$ at $z=L$, which is not necessarily the mirror image of $Q$. Instead, we take $R$ and $Q'$ to be the mirror images of $R'$ and $Q$, respectively. Assuming $PP'\sim QQ'\sim RR'\sim\epsilon$, which is infinitely small, we have $RQ\sim R'Q'\sim\epsilon s$, where $s$ stands for the finite shear across the sheet. When $\epsilon\rightarrow0$, $R'$ and $Q$ still coincide. However, as we shall show next, properly accounting for the finite shear $s$ can make a substantial difference.

For the path integral on the top of the sheet, along the solid curves in Fig.\,\ref{CLS}b, Eq.\,\eqref{top} now becomes
\begin{align}\label{topc}
\oint \mathbf{B}\cdot\mathrm{d}\mathbf{l}=\int_1B_1(l)\,\mathrm{d}l+\int_Q^R\mathbf{B}\cdot\mathrm{d}\mathbf{l}-\int_2\mathbf{B}_1(l)\cdot\mathbf{b}_2(l)\,\mathrm{d}l=0.
\end{align}
Similarly, along the path on the bottom of the sheet, as shown by the dashed curves in Fig.\,\ref{CLS}b, Eq.\,\eqref{bottom} becomes
\begin{align}\label{bottomc}
\oint \mathbf{B}\cdot\mathrm{d}\mathbf{l}=\int_2B_2(l)\,\mathrm{d}l+\int_{R'}^{Q'}\mathbf{B}\cdot\mathrm{d}\mathbf{l}-\int_1\mathbf{B}_2(l)\cdot\mathbf{b}_1(l)\,\mathrm{d}l=0.
\end{align}
Again, combining Eqs.\,\eqref{topc} and \eqref{bottomc}, Eq.\,\eqref{total} becomes
\begin{align}
\int_{1+2}B(l)\left[1-\mathbf{b}_2(l)\cdot\mathbf{b}_1(l)\right]\,\mathrm{d}l=\int^Q_R\mathbf{B}\cdot\mathrm{d}\mathbf{l}+\int^{R'}_{Q'}\mathbf{B}\cdot\mathrm{d}\mathbf{l}.
\end{align}
Since the magnetic field has no normal component across the current sheet, the RHS can be expressed as a loop integral along $R\rightarrow Q\rightarrow Q'\rightarrow R'\rightarrow R$, 
\begin{align}\label{totalc}
\int_{1+2}B(l)\left[1-\mathbf{b}_2(l)\cdot\mathbf{b}_1(l)\right]\,\mathrm{d}l=\oint_{\partial\Omega}\mathbf{B}\cdot\mathrm{d}\mathbf{l}=\int_{\Omega}\mathbf{j}\cdot\mathrm{d}\mathbf{S}.
\end{align}
Here $\Omega$ is the area enclosed by this loop at the top plane $z=L$. As $\epsilon\rightarrow0$, $\Omega\rightarrow0$, but the RHS of Eq.\,\eqref{totalc} can be finite if and only if $j_z$ contains a 2D delta function at ${z=L}$. Still, this means that it is possible that $\mathbf{b}_1$ and $\mathbf{b}_2$ do not coincide, unlike what \citet{Cowley1997} claimed.

We emphasize that our argument in this appendix does not confirm the conjecture by \citet{Parker1972}. We merely show that the argument due to \citet{Cowley1997} that allegedly disproved Parker's conjecture, which is a variant of that by \citet{VanBallegooijen1985}, does not really stand, and hence the Parker problem is still open.

In addition, the fact that $j_z$ needs to contain a 2D delta function at the footpoints in order for the singularity to exist suggests that the term ``current sheet" may not accurately describe the possible singular structure in 3D line-tied geometry. Therefore, in this thesis, we adopt the more general term of ``current singularity" instead.

\chapter{Numerical schemes\label{ch:scheme}}
In this appendix, the update schemes of our variational integrators for ideal MHD in cartesian coordinates in 1D, 2D, and 3D are presented, respectively. In 2D and 3D, the schemes are derived on unstructured simplicial complexes.

\section{1D scheme}\label{scheme:1D}
When projected into 1D, mass density and pressure become 1-forms. As for the magnetic field, a guide field is allowed. Its effect is exactly equivalent to a pressure with $\gamma=2$, so it can be treated as such. The Lagrangian in Lagrangian labeling \eqref{Lagrangian3} becomes
\begin{equation}
{L}[{x}]=\int\left[\frac{1}{2}\rho_0\dot{{x}}^2-\frac{p_0}{(\gamma-1)J^{\gamma-1}}\right]\mathrm{d}x_0,
\end{equation}
with the Jacobian $J=\partial x/\partial x_0$. And the Euler-Lagrange equation becomes
\begin{equation}
\rho_0\ddot{x}+
\frac{\partial }{\partial x_{0}}\left(\frac{p_0}{J^\gamma}\right)=0.\label{momentum1}
\end{equation}

Next we will spatially discretize the Lagrangian above. First, the manifold is discretized into $N+1$ grids $x_i\,(i=0,1,\cdots,N)$ together with $N$ line segments $(x_i,x_{i+1})$, where $x_i$ and $x_{i+1}$ are the left and right end of the segment respectively. The discrete 1-forms $\rho_i$ and $p_i$ are real numbers assigned to these segments. The initial grid size $x_{i+1}-x_{i}$ is denoted by a positive constant $h_i$. In this case, the Jacobian becomes $J_i=(x_{i+1}-x_i)/h_i$. The kinetic energy term is a product of discrete mass densities on the line segments and the square of the velocities on the grids, and we choose to average the latter when multiplying. The spatially discretized Lagrangian is therefore
\begin{equation}
L_d(x_i,\dot{x}_i)=\sum_i\left[\frac{1}{2}\rho_{i}\frac{\dot{x}_i^2+\dot{x}_{i+1}^2}{2}-\frac{p_{i}}{(\gamma-1)J_i^{\gamma-1}}\right]=\sum_i\frac{1}{2}M_{i}\dot{x}_i^2-W(x_i),
\end{equation}
where the subscript $0$ on the initial conditions $\rho_{0i}$ and $p_{0i}$ are dropped for convenience, and $M_i=(\rho_{i-1}+\rho_{i})/2$ is the effective mass of the $i$th vertex. Note that $\rho_i$, $p_i$, and $M_i$ are all constants. This Lagrangian has the form of a many-body Lagrangian, and the consequent Euler-Lagrange equation has the form of a Newton's equation
\begin{equation}
M_{i}\ddot{x}_i=-\partial W/\partial x_{i}=\frac{p_{i-1}/h_{i-1}}{(J_{i-1})^{\gamma}}-\frac{p_{i}/h_i}{(J_i)^{\gamma}}=F_i(x_{i},x_{i\pm1}).
\end{equation}

For temporal discretization, we refer to Sec.\,\ref{integrator:discretization}. Boundary conditions can be introduced as holonomic constraints. For instance, one can choose periodic boundary condition with $x_{N}=x_0+L$, or ideal walls where $x_1=0$ and $x_{N}=L$. Here the domain size $L=\sum^N_{i=0} h_i$. The effective mass needs to be adjusted accordingly for the boundary vertices as well. For periodic boundaries we have $M_0 = M_N=(\rho_0+\rho_{N-1})/2$, while for ideal walls $M_0=\rho_0/2$ and $M_N=\rho_{N-1}/2$.

\section{2D scheme}\label{scheme:2D}
When projected into 2D, mass density and pressure become 2-forms. The effective mass of a vertex $\sigma_0$ becomes $M(\sigma_0^0)=\sum_{\sigma_0^2\succ\sigma_0^0}\langle\rho_0,\sigma_0^2\rangle/3$. The guide field can effectively be treated as a magnetic pressure with $\gamma=2$. Hence we only need to consider the in-plane magnetic field, which becomes a 1-form. The Lagrangian and the Euler-Lagrange equation are basically the same as Eqs.~(\ref{Lagrangian3}) and (\ref{momentum3}), except in lower dimension. Note that the 1D derivation above is based on a structure grid, since any grid is naturally structured in 1D. But in 2D the mesh becomes genuinely unstructured, so we need to adopt a more abstract terminology.
As can be seen from Sec.\,\ref{integrator:discretization}, one only needs to calculate the discrete potential energy $W$ of an arbitrary volume simplex, and then derive its force on its $i$th vertex. Then one can iterate over all the volume simplices and calculate the total force on each vertex.

Following from Eq.~\eqref{Lagrangian3D}, the potential energy for a 2-simplex (0, 1, 2), namely a triangle, reads
\begin{align}
W=\frac{p}{(\gamma-1)(S/S_0)^{\gamma-1}}+\frac{1}{8S}\sum_{i=0}^2B^2_{i}I_i,
\end{align}
where $p$ is the initial pressure, $B_i$ is the initial magnetic field evaluated on the edge opposite to vertex $i$, and the area
\begin{align}
S=\left[\left(x_{i+1}-x_i\right)\left(y_{i+2}-y_i\right)-\left(x_{i+2}-x_i\right)\left(x_{i+2}-x_i\right)\right]/2.
\end{align}
Here $i=0,1,2$, and operation mod 3 is implied on the vertex indices (subscripts) $i+1$ and $i+2$.
$S_0$ is the initial area, and $p$, $B_i$ and $S_0$ are all constants. The inner product with respect to vertex $i$ reads
\begin{align}
I_i=\left(x_{i+1}-x_i\right)\left(x_{i+2}-x_i\right)+\left(y_{i+1}-y_i\right)\left(y_{i+2}-y_i\right).
\end{align}
Then, taking the derivative of $W$ then leads to the force on vertex $j$,
\begin{align}
-\frac{\partial W}{\partial x_j}=&\left[\frac{p/S_0}{2(S/S_0)^{\gamma}}+\frac{1}{16A^2}\sum_{i=0}^2B^2_{i}I_i\right]\left(y_{j+1}-y_{j+2}\right)\nonumber\\
&+\left[\left(B^2_{j+2}-B^2_{j+1}\right)\left(x_{j+1}-x_{j+2}\right)+B^2_{j}\left(2x_j-x_{j+1}-x_{j+2}\right)\right]/(8S),\\
-\frac{\partial W}{\partial y_j}=&\left[\frac{p/S_0}{2(S/S_0)^{\gamma}}+\frac{1}{16A^2}\sum_{i=0}^2B^2_{i}I_i\right]\left(x_{j+2}-x_{j+1}\right)\nonumber\\
&+\left[\left(B^2_{j+2}-B^2_{j+1}\right)\left(y_{j+1}-y_{j+2}\right)+B^2_{j}\left(2y_j-y_{j+1}-y_{j+2}\right)\right]/(8S).
\end{align}
Again, $j=0,1,2$, and the subscripts $j+1$ and $j+2$ are subject to operation mod 3. 

Boundary conditions can again be introduced as holonomic constraints. Besides periodic boundaries, there can also be ideal walls and no-slip boundaries. For ideal walls, the vertices on the boundaries are allowed to move tangentially along, but not normally to them. For no-slip boundaries, the vertices on the boundaries do not move.

\section{3D scheme}\label{scheme:3D}
For 3D MHD, we carry out the same procedure as in 2D. Following from Eq.~\eqref{Lagrangian3D}, the potential energy for a 3-simplex (0, 1, 2, 3), a tetrahedron, reads
\begin{align}
W=\frac{p}{(\gamma-1)(V/V_0)^{\gamma-1}}+\frac{1}{12V}\sum_{i=0}^{3}B^2_{i}\beta_i/S_i^2,
\end{align}
where $p$ is the initial pressure, $B_i$ is the initial magnetic field evaluated on the face opposite to vertex $i$. $V$ is the volume of the tetrahedron, while $V_0$ is its initial value. $p$, $B_i$ and $V_0$ are constants. $\beta_i$ is the barycentric coordinate with regard to vertex $i$,
\begin{align}
\beta_i=\sum_{k=0}^{2}l^2_{i,i+1+k}l_{i+1+(k+1),i+1+(k+2)}^2I_{i,k}-2l_{i+2,i+3}^2l_{i+1,i+3}^2l_{i+2,i+1}^2.
\end{align}
Here the squared length of the edge $(i,j)$ is
\begin{align}
l_{i,j}^2=(x_i-x_j)^2+(y_i-y_j)^2+(z_i-z_j)^2,
\end{align}
while $I_{i,k}$ denotes the doubled inner product with regard to vertex $i+1+k$ in the face opposite to vertex $i$,
\begin{align}
I_{i,k} =l_{i+1+k,i+1+(k+1)}^2+l_{i+1+k,i+1+(k+2)}^2-l_{i+1+(k+2),i+1+(k+1)}^2.
\end{align}
Note that here $(k+1),(k+2)$ are subject to operation mod 3, and all the vertex indices (subscripts), such as $i+1+k$, $i+1+(k+1)$, and $i+1+(k+2)$ are subject to operation mod 4. The same applies to the rest of this section.
Similarly, the sixteen-timed squared area of the face opposite to vertex $i$ is given by
\begin{align}
S_i^2&=\sum_{k=0}^{2}l_{i+1+(k+1),i+1+(k+2)}^2I_{i,k}.
\end{align}

Again, the force follows from taking the derivative of the potential energy,
\begin{align}
-\frac{\partial W}{\partial x_j}&=\left[\frac{p/V_0}{(V/V_0)^{\gamma}}+\frac{1}{12V^2}\sum_{i=0}^{3}B^2_{i}\beta_i/S_i^2\right]\frac{\partial V}{\partial x_j}
-\frac{1}{12V}\sum_{i=0}^{3}\frac{B^2_{i}}{S_i^2}\left(\frac{\partial \beta_i}{\partial x_j}-\frac{\beta_i}{S_i^2}\frac{\partial S_i^2}{\partial x_j}\right).
\end{align}
Here we only show the $x$ component, as the $y$ and $z$ components can be obtained by shuffling $x$, $y$, and $z$.
Note that
\begin{align}
\frac{\partial V}{\partial x_j}=\frac{(-1)^{j-1}}{6}\sum_{k=0}^{2}y_{j+1+k}[z_{j+1+(k+1)}-z_{j+1+(k+2)}],
\end{align}
and the volume can be calculated by
\begin{align}
V=\sum_{j=0}^{3}x_j\frac{\partial V}{\partial x_j}.
\end{align}
When calculating $\partial S_i^2/\partial x_{j}$ and $\partial \beta_i/\partial x_{j}$, one needs to differentiate between when $i=j$ and when $i\ne j$. For the doubled area, ${\partial S_i^2}/{\partial x_{i}}=0$, while
\begin{align}
{\partial S_i^2}/{\partial x_{i+1+k}}&=4[x_{i+1+k}-x_{i+1+(k+1)}]I_{i,(k+2)}+4[x_{i+1+k}-x_{i+1+(k+2)}]I_{i,(k+1)},
\end{align}
for $k=0,1,2$. Finally, the derivatives for the barycentric coordinates read
\begin{align}
{\partial \beta_i}/{\partial x_{i}}&=\sum_{k=0}^{2}2[x_{i}-x_{i+1+k}]l_{i+1+(k+1),i+1+(k+2)}^2I_{i,k},&\\
{\partial \beta_i}/{\partial x_{i+1+k}}&=2[x_{i+1+k}-x_{i+1+(k+1)}]\{[l^2_{i,i+1+k}-2l^2_{i+1+k,i+1+(k+2)}]l^2_{i+1+(k+2),i+1+(k+1)}\nonumber\\
&+l^2_{i,i+1+(k+1)}l^2_{i+1+k,i+1+(k+2)}+l^2_{i,i+1+(k+2)}[I^2_{i,(k+2)}-l^2_{i+1+k,i+1+(k+1)}]\}\nonumber\\
&+2[x_{i+1+k}-x_{i+1+(k+2)}]\{[l^2_{i,i+1+k}-2l^2_{i+1+k,i+1+(k+1)}]l^2_{i+1+(k+2),i+1+(k+1)}\nonumber\\
&+l^2_{i,i+1+(k+2)}l^2_{i+1+k,i+1+(k+1)}+l^2_{i,i+1+(k+1)}[I^2_{i,(k+1)}-l^2_{i+1+k,i+1+(k+2)}]\}\nonumber\\
&+2(x_{i+1+k}-x_{i})l^2_{i+1+(k+1),i+1+(k+2)}I_{i,k},
\end{align}
for $k=0,1,2$. The introduction of boundary conditions is equivalent to that in 2D, that is, as holonomic constraints.

\chapter{3D periodic geometry\label{ch:periodic}}
In this appendix, we generalize the 2D HKT problem discussed in Sec.\,\ref{2D:HKT} to 3D periodic geometry. Again, we consider the linear solution first. In 2D, the perturbation is assumed to have Fourier dependence $\chi(x_0,y_0)=\bar\chi(x_0)\sin{ky_0}$. In 3D, we generalize this assumption so that $\chi(x_0,y_0,z_0)=\bar\chi(x_0)\sin(k_yy_0+k_zz_0)$, with $k_y,k_z>0$. Accounting for the initial field $B_{0y}=x_0$ and adopting RMHD conventions ($B_{0z}=1$ and $\bm\xi = \nabla \chi\times \hat z $), the linear equilibrium equation $\mathbf{F}(\bm\xi )=0$ becomes  
\begin{align}\label{linearP}
(\partial_{x_0}^2-k_y^2)\left[\bar\chi\mathbf{k}\cdot\mathbf{B}_0(x_0)\right] = 0.
\end{align}
Here the wave vector $\mathbf{k}=k_y\hat y+k_z\hat z$. Note that this equation can also be obtained from Eq.\,\eqref{linear} by replacing the differential operator $\partial_{z_0}$ with $ik_z$.

Eq.\,\eqref{linearP} has two branches of solutions, $\bar\chi\sim\sinh(\mathbf{k}\cdot\mathbf{B}_0)/(\mathbf{k}\cdot\mathbf{B}_0)$ and $\bar\chi\sim\cosh(\mathbf{k}\cdot\mathbf{B}_0)/(\mathbf{k}\cdot\mathbf{B}_0)$. The latter is again discarded because it diverges where there is resonance, $\mathbf{k}\cdot\mathbf{B}_0(x_0)=0$, assuming that it exists in our domain of interest, $x_0\in[-a,a]$ (that is, $k_z/k_y<a$). If the boundary forcing at $x_0 = \pm a$ is assumed to be in the form of $x(\pm a, y_0,z_0) = \pm \{a - \delta \cos [k_yy(\pm a, y_0,z_0)+k_zz_0]\}$, the corresponding boundary condition for the linear solution becomes $\bar\chi(\pm a)=\mp \delta/k_y$. Consequently, the linear solution reads
\begin{align}\label{PS1}
\bar\chi=
\begin{dcases}
-\frac{\delta [\mathbf{k}\cdot\mathbf{B}_0(a)]\sinh [\mathbf{k}\cdot\mathbf{B}_0(x_0)]}{k_y\left[\mathbf{k}\cdot\mathbf{B}_0(x_0)\right]\sinh [\mathbf{k}\cdot\mathbf{B}_0(a)]},& \mathbf{k}\cdot\mathbf{B}_0(x_0)>0,\\
\frac{\delta [\mathbf{k}\cdot\mathbf{B}_0(-a)]\sinh [\mathbf{k}\cdot\mathbf{B}_0(x_0)]}{k_y\left[\mathbf{k}\cdot\mathbf{B}_0(x_0)\right]\sinh [\mathbf{k}\cdot\mathbf{B}_0(-a)]},& \mathbf{k}\cdot\mathbf{B}_0(x_0)<0.
\end{dcases}
\end{align}
As one can see, so far this is only a straightforward helical generalization of the 2D linear equation \eqref{HKE} and solution \eqref{HKS}. The location of the normal discontinuity in $\xi_x$ is generalized from the neutral line $x_0=0$ to $\mathbf{k}\cdot\mathbf{B}_0(x_0)=0$, where there is resonance.

Then, we consider a slightly more complicated case, where the boundary forcing at $x_0 = \pm a$ is assumed to be in the form of $x(\pm a, y_0,z_0) = \pm [a - \delta \cos k_yy(\pm a, y_0,z_0)\cos k_zz_0]$. In this setup, the perturbation contains two modes with $\mathbf{k}_{\pm}=k_y\hat y\pm k_z\hat z$. Accordingly, the linear solution $\chi=\chi_+ + \chi_-$, where $\chi_{\pm}=\bar\chi_{\pm}(x_0)\sin(k_yy_0\pm k_zz_0)$. With $\chi_{\pm}$ satisfying Eq.\,\eqref{linearP}, respectively, we have $\bar\chi_{\pm}\sim\sinh(\mathbf{k}_{\pm}\cdot\mathbf{B}_0)/(\mathbf{k}_{\pm}\cdot\mathbf{B}_0)$ (again, here the diverging branches are discarded). Accounting for the boundary condition $\chi(\pm a)=\mp (\delta/k_y)\sin k_yy_0 \cos k_zz_0$, we have the linear solution
\begin{align}\label{PS2}
\bar\chi_{\pm}=
\begin{dcases}
- \frac{\delta[\mathbf{k}_{\pm}\cdot\mathbf{B}_0(a)]\sinh [\mathbf{k}_{\pm}\cdot\mathbf{B}_0(x_0)]}{2k_y\left[\mathbf{k}_{\pm}\cdot\mathbf{B}_0(x_0)\right]\sinh [\mathbf{k}_{\pm}\cdot\mathbf{B}_0(a)]}, & \mathbf{k}_{\pm}\cdot\mathbf{B}_0(x_0)>0,\\
 \frac{\delta[\mathbf{k}_{\pm}\cdot\mathbf{B}_0(-a)]\sinh [\mathbf{k}_{\pm}\cdot\mathbf{B}_0(x_0)]}{2k_y\left[\mathbf{k}_{\pm}\cdot\mathbf{B}_0(x_0)\right]\sinh [\mathbf{k}_{\pm}\cdot\mathbf{B}_0(-a)]}, &\mathbf{k}_{\pm}\cdot\mathbf{B}_0(x_0)<0.
\end{dcases}
\end{align}
Note that $\bar\chi_{\pm}$ are identical to solution \eqref{PS1}, except for a factor of 2. 
The normal discontinuity in $\xi_x$ now appears at both locations of resonances, $\mathbf{k}_{\pm}\cdot\mathbf{B}_0(x_0)=0$. More generally, when the perturbation contains multiple modes (each with a different $\mathbf{k}$), the linear solution becomes a linear combination of the corresponding single-mode solutions \eqref{PS1}, and one should expect singularities to appear wherever there is resonance, $\mathbf{k}\cdot\mathbf{B}_0=0$.

Now, we study the latter case with two modes nonlinearly using the numerical method described in Chap.\,\ref{ch:integrator}. The domain is set up as $[-a,a]\times[-\pi/k_y,\pi/k_y]\times[0,2\pi/k_z]$. The update scheme in 3D is specified in Appx.\,\ref{scheme:3D}. The boundaries at $x_0=\pm a$ are constrained such that $x=\pm[a-\delta\cos ky\cos k_z z_0]$. The boundary conditions in $y$ and $z$ are both periodic. We use the same parameters as used in the line-tied problem in Chap.\,\ref{ch:3D}, namely $a=0.25$, $k_y=2\pi$, $\delta=0.05$. We choose $k_z=\pi/4$ so that resonances take place at $x_0=\pm k_z/k_y=\pm0.125$.

Similar to the line-tied problem, we adopt RMHD approximations, by setting $B_{0z}=1$ and $z=z_0$. We use moderate pressure to approximate incompressibility, by initializing with $p_0=0.1-x_0^2/2$ to balance the sheared field $B_{0y}=x_0$, and choosing $\gamma=5/3$. In our numerical solutions, we find $|J-1|\lesssim 0.02$. Again, we expect that $j_z$ should still be approximately constant along field lines.

We use a tetrahedral mesh where the vertices are arranged in a structured manner with resolution $N\times2N\times 2N$. The vertices are non-uniformly distributed in $x$ and to devote more resolution near the resonant surfaces at $x_0=\pm k_z/k_y$. We use a same profile of mesh packing while varying resolution $N$. The system starts from a smoothly perturbed configuration consistent with the boundary conditions and relaxes to equilibrium due to friction. 

\begin{figure}[h]
\centering
\includegraphics[scale=0.75]{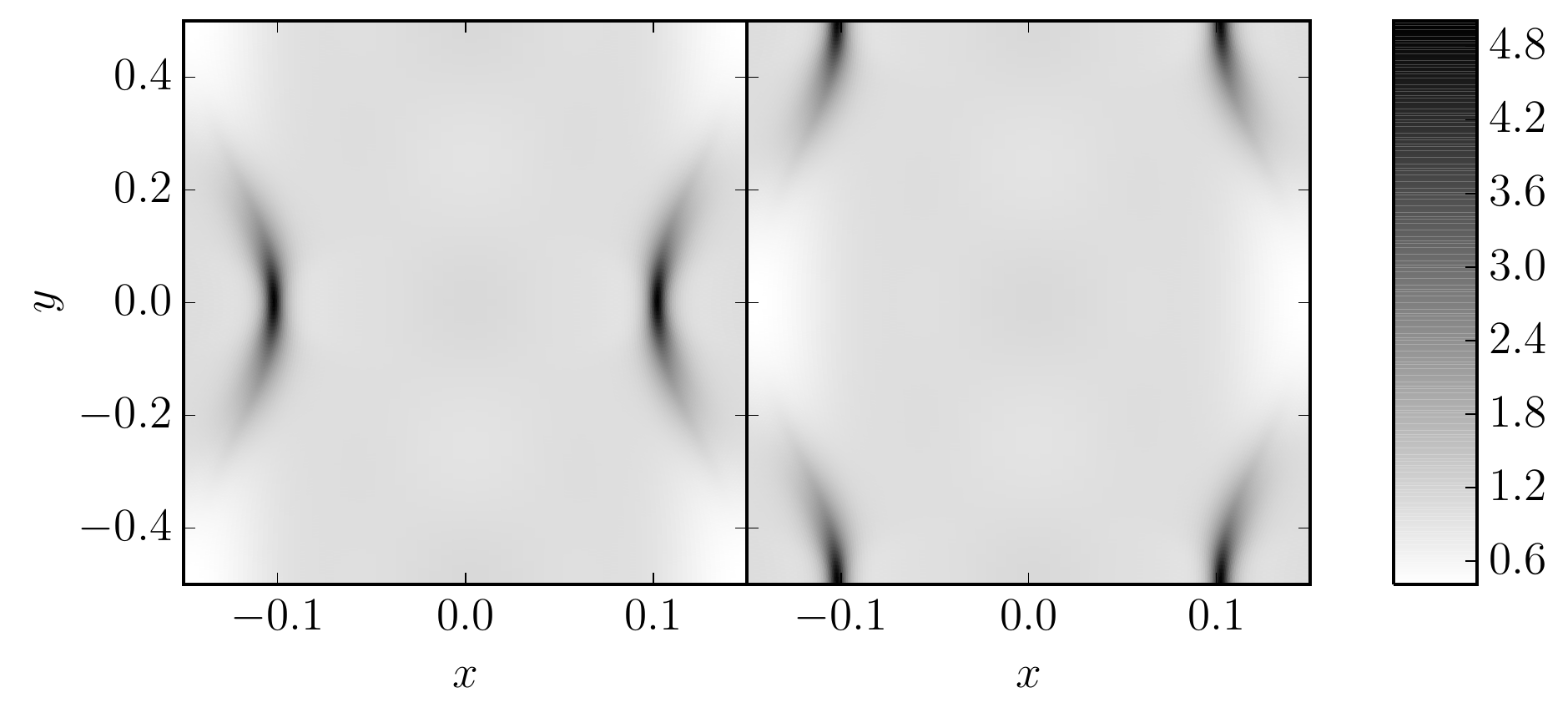}
\caption{\label{periodic}Distribution of current density $j_z(x,y)$ obtained with $N=160$ at $z=0$ (left) and $z=\pi/k_z$ (right) for the periodic setup.}
\end{figure}

In Fig.\,\ref{periodic}, we show the equilibrium current density distributions obtained with $N=160$ at $z=0$ and $z=\pi/k_z$. The current is concentrated to the resonant surfaces at $x_0=\pm k_z/k_y$, unlike the $z$ axis in the line-tied problem, as shown in Fig.\,\ref{current}. Notably, the peaks of the current density do follow the field lines from $(\pm k_z/k_y,0,0)$ to $(\pm k_z/k_y,\pm\pi/k_y,\pm\pi/k_z)$, respectively.

\begin{figure}[h]
\centering
\includegraphics[scale=0.75]{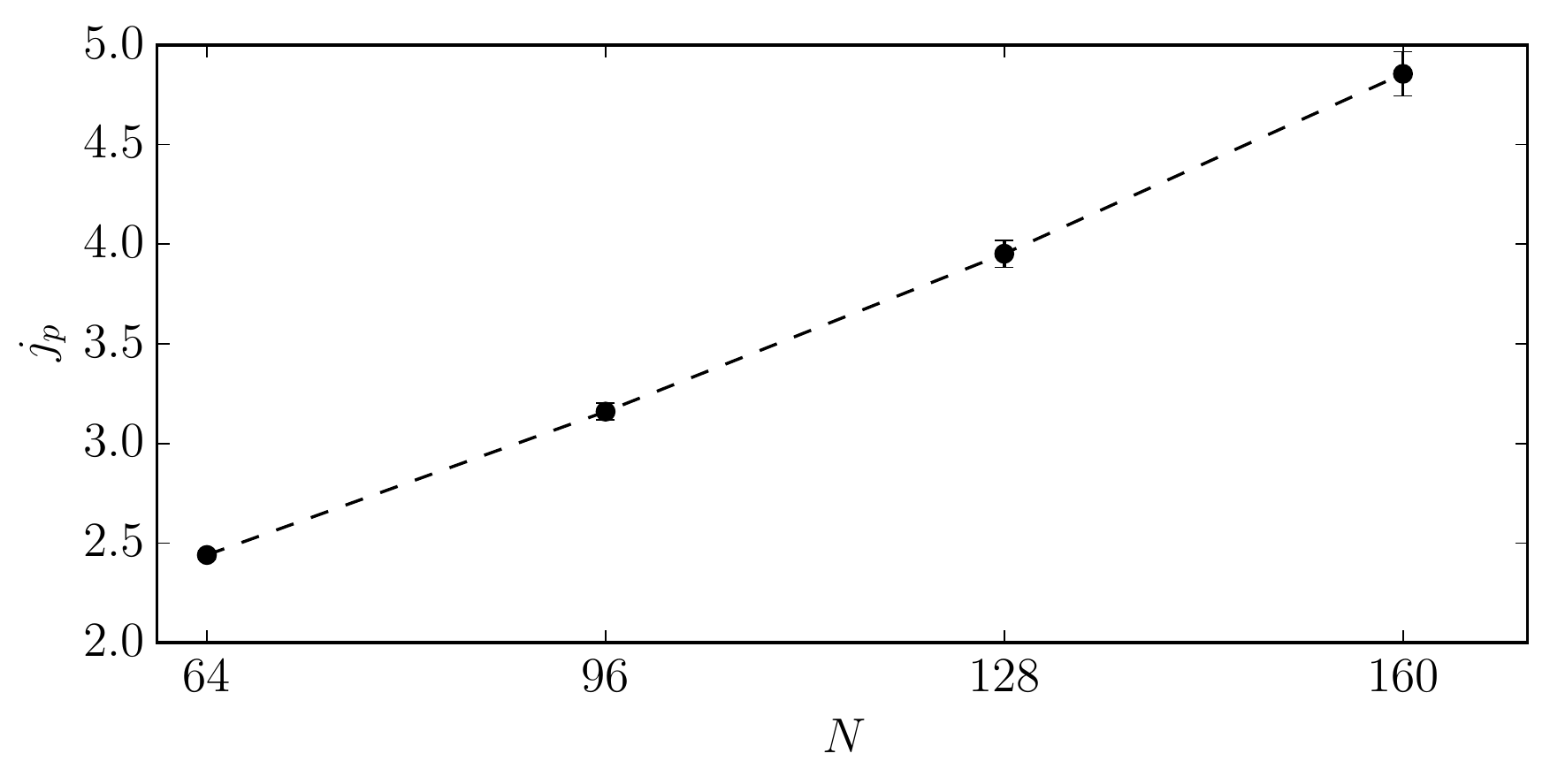}
\caption{\label{diverge}The mean ($j_p$) and the standard deviation (error bars) of the in-plane maximum of the current density $j_z$ are shown to diverge with increasing resolution $N$. }
\end{figure}

Moreover, our convergence study shows that at the resonant surfaces, the solutions do not converge. We obtain the in-plane maximum of the current density $j_z$ as a function of $z$, and then show its mean and standard deviation in Fig.\,\ref{diverge}. For resolutions $N=64,96,128,160$, no sign of convergence can be observed. In contrast, the solutions in the line-tied problem shown in Fig.\,\ref{resolution} are shown to converge. These results suggest that the current singularity that shows up in the 2D HKT problem survives in 3D periodic geometry.

\singlespacing
\bibliographystyle{apalike2}
\setlength{\bibsep}{1.5pt plus 0.3ex}

\cleardoublepage
\ifdefined\phantomsection
  \phantomsection  
\else
\fi
\addcontentsline{toc}{chapter}{Bibliography}

\bibliography{thesis_Zhou}

\end{document}